\documentclass{article}
\usepackage[utf8]{inputenc}
\usepackage{amsfonts}
\usepackage{amssymb}
\usepackage{amsmath}
\usepackage{amsthm}
\usepackage{graphicx}
\usepackage{setspace}
\usepackage[T1]{fontenc}
\usepackage{natbib}
\usepackage{tikz}
\usepackage{mathbbol}
\usepackage[toc,page]{appendix}
\usepackage{stackengine}
\usepackage{subcaption}
\usepackage{hyperref}

\usetikzlibrary{arrows.meta}
\usetikzlibrary{calc}

\graphicspath{{./images}}
\DeclareGraphicsExtensions{.png,.pdf}
\DeclareMathOperator{\plim}{plim}

\DeclareMathOperator{\diag}{diag}

\title{\Large Dynamic Spatial Interaction Models for a Resource Allocator’s Decisions and Local Agents’ Multiple Activities\thanks{%
I would like to thank Jieun Lee, Lung-fei Lee, Konrad Menzel, Eric Renault, Shuping Shi, Bruce Weinberg and Yiran Xie for their valuable comments and suggestions. I also thank Aureo de Paula and Pedro Souza for sharing codes to generate the U.S. states' economic neighbors. Further, I thank the seminar participants at Monash University, University of Melbourne, University of Sydney, University of Queensland, and the audience from 2024 Australasia Econometric Society Meeting, Melbourne, Australia and 33rd Australian New Zealand Econometric Study Group (ANZESG) Meeting, Melbourne, Australia}}
\author{Hanbat Jeong\thanks{%
Department of Economics, Macquarie Business School, Macquarie University. E-mail: hanbat.jeong@mq.edu.au}}

\theoremstyle{definition}

\newtheorem{theorem}{Theorem}[section]

\newtheorem{assumption}{Assumption}[section]
\newtheorem{remark}{Remark}[section]

\newtheorem{definition}{Definition}[section]

\begin{document}

\maketitle

\begin{abstract}
This paper introduces a novel spatial interaction model to explore the decision-making processes of a resource allocator and local agents, with central and local governments serving as empirical representations. The model captures two key features: (i) resource allocations from the allocator to local agents and the resulting strategic interactions, and (ii) local agents' multiple activities and their interactions. We develop a network game for the micro-foundations of these processes. In this game, local agents engage in multiple activities, while the allocator distributes resources by monitoring the externalities arising from their interactions. The game's unique Nash equilibrium establishes our econometric framework. To estimate the agent payoff parameters, we employ the quasi-maximum likelihood (QML) estimation method and examine the asymptotic properties of the QML estimator to ensure robust statistical inference. Empirically, we study interactions among U.S. states in public welfare and housing and community development expenditures, focusing on how federal grants influence these expenditures and the interdependencies among state governments. Our findings reveal significant spillovers across the states' two expenditures. Additionally, we detect positive effects of federal grants on both types of expenditures, inducing a responsive grant scheme based on states' decisions. Last, we compare state expenditures and social welfare through counterfactual simulations under two scenarios: (i) responsive intervention by monitoring states' decisions and (ii) autonomous transfers. We find that responsive intervention enhances social welfare by leading to an increase in the states' two expenditures. However, due to the heavy reliance on autonomous transfers, the magnitude of these improvements remains relatively small compared to the share of federal grants in total state revenues.
\end{abstract}

\begin{quote}
\end{quote}

\section{Introduction}

Spatial autoregressive (SAR) models are increasingly popular tools for analyzing the interdependent decision-making processes of economic agents, each situated in a specific location. These local agents interact with one another, with the intensity of their interactions shaped by geosocial proximity. For example, SAR models can characterize the best response functions of local governments competing over welfare spillovers and movements of welfare-motivated residents \citep{Caseetal1993}. Much of the existing literature focuses either on strategic interactions related to local agents' (local governments) activities or extends to their multiple activities, examining their complementarity or substitutability. In other words, existing SAR-type models formulate local governments' decisions, and their spillovers originate from flexible environments of demographic and economic movements. However, another distinct feature of local governments' decisions has been disregarded in the SAR model's framework: the role of intergovernmental grants from the federal government. When local agents utilize their resources across various activities over multiple periods and engage in strategic interactions, a central resource allocator—tasked with national resource distribution—monitors the resulting externalities from competing local agents. This gives rise to two important questions: How are the allocator’s decisions related to the multiple activities of interdependent local agents? And how can these elements be incorporated into the SAR framework?

To address these questions, this paper introduces a new dynamic spatial interaction model, informed by panel data that record local agents’ multiple activities and the corresponding resource allocations. The model delineates the interactions between a central allocator and \( n (\geq 2) \) local agents, who allocate the received resources across various activities. In our empirical application, the U.S. federal government is treated as the "resource allocator," while state governments serve as "local agents." The model highlights two pivotal dimensions: (i) resource allocations from the allocator to local agents and their interaction—vertical interaction, and (ii) the local agents’ multiple activities and their interaction—horizontal interaction, particularly when these activities draw upon common resource types. Through this lens, the model enables an examination of how federal resource distributions affect states’ expenditures across multiple activities and how these state-level expenditure decisions, in turn, shape future federal allocations over time.

As the model's microfoundation, we develop a dynamic network game involving the allocator and local agents. The game unfolds over an infinite sequence of decision-making periods, each consisting of two stages. In the first stage, the benevolent allocator chooses grants for local agents. We define the allocator's objective as social welfare, represented by the sum of the local agents' payoffs (Jackson and Zenou (\citeyear{JacksonZenou2015}), Ch.4.1.3; \cite{HsiehLin2021}). Accordingly, the allocator internalizes the externalities arising from the local agents’ (non-cooperative) decisions. In the second stage, after observing the allocator’s actions, the $n$ local agents simultaneously choose their activity levels. This structure generates a directional influence from the allocator to the local agents. However, by assuming forward-looking behavior on the part of the agents, the model also allows for a feedback channel through which local decisions affect the allocator’s future grant allocations. Thus, the allocator acts not only as a financial distributor but also as a strategic leader, guiding local behavior to enhance long-term social welfare. This hierarchical structure and the strategic interactions it entails provide a general framework for settings in which multiple agents engage in diverse activities under centralized resource allocation.

As the payoff specification, we assume that each local agent’s decision is governed by a linear-quadratic (LQ) parametric payoff function. The LQ specification is the only functional form that allows for feasible computation in a dynamic model with shared state variables induced by network interactions, as it yields closed-form solutions for optimal activities. Through numerical experiments, we verify that the LQ specification approximates optimal behavior well under a non-LQ payoff, provided a contraction mapping generates the equilibrium. This payoff function incorporates both current activities and agent-specific characteristics. These characteristics—comprising relevant demographic and economic variables for each local agent—serve as state variables alongside past activity levels in the dynamic system. We assume that each demographic or economic variable evolves according to a first-order Markov process. When local agents’ activities and the allocator’s grants are interpreted as “efforts,” this structure captures the cumulative effect of past efforts on current demographic and economic characteristics.

Four categories of parameters characterize each local agent’s incentive structure. First, parameters governing the relationships among a local agent’s multiple activities capture either substitutability or complementarity. Second, interactions among local agents arise not only within a single activity but also across different activities. These inter-agent interactions are modeled through a network (spatial weighting matrix), denoted by \(\mathbf{W} = [w_{ij}]\), and are associated with specific payoff parameters linked to the network weights \(w_{ij}\). Third, the influence of the allocator on local agents’ activities is incorporated into their payoffs through dedicated parameters. Fourth, the effects of local agents’ demographic and economic characteristics on their payoffs are captured by another set of parameters. Beyond the sum of local agents’ payoffs, the allocator’s payoff also includes economic indicators reflecting the current economic environment, as well as fixed-effect components representing incentives for autonomous transfers. If these autonomous transfer components account for a large share of the variation in allocations, the allocator becomes less responsive to local agents’ decisions regarding their activities.

The econometric model is grounded in a Markov Perfect Nash Equilibrium (MPNE), which is characterized by recursive value functions. The resulting estimating equations extend the spatial dynamic panel simultaneous equations (SDPSE) framework by incorporating an additional endogenous variable that captures the allocator’s intervention. SDPSE models represent a panel-data extension of SAR models, accommodating multiple activities and their intertemporal dynamics. By deriving explicit payoff structures that rationalize the SDPSE framework, our model yields welfare implications beyond standard estimates of strategic interactions and substitutability or complementarity among activities. We establish conditions on the model parameters and the network matrix $\mathbf{W}$ that ensure the uniqueness of the MPNE, thereby enabling point identification of the payoff parameters. A key condition for equilibrium uniqueness involves the stability of spatial-dynamic influences inherent in the system. Unlike existing spatial or network interaction models, the network spillovers in our setting emerge from the nuanced interplay between the local agents’ non-cooperative behaviors and the allocator’s altruistic decisions.

Leveraging the uniquely characterized MPNE, this study introduces two measures to interpret the model’s implications. First, we derive the marginal effect of changes in a local agent’s characteristics on their activities, operating through the allocator’s decision rule. Second, by the SDPSE model's dynamic nature, we construct a measure that captures the cumulative dynamic impact of a characteristic over time. The first measure corresponds to the \textit{short-run impact}, while the second reflects the \textit{long-run impact}. Because the allocator’s intervention responds to changes in local characteristics and, in turn, influences agents’ decisions, there exists an indirect effect whereby a local agent’s characteristics affect their own activities via the allocator’s grant allocations.

The primary objective of our estimation procedure is to identify and estimate the agents’ payoff parameters. To this end, we construct a log-likelihood function based on the uniquely derived MPNE-implied activities and grant allocations. Drawing on likelihood theory as formalized by \citet{Rothenberg1971}, we derive identification conditions for the key structural parameters. The MPNE induces a structural vector autoregressive (SVAR) representation, which provides sources of identification through the conditional mean and variance of agents’ activities. For estimation, we employ the quasi-maximum likelihood (QML) method, combined with a direct approach to estimating fixed-effect parameters. This strategy yields robust statistical inference even in the presence of non-normal errors, time-specific shocks, and unobserved agent heterogeneity. We also examine the large-sample properties of the QML estimator. Monte Carlo simulations confirm that the estimator performs well in finite samples.

In the empirical application, we explore two key aspects: (i) interactions among U.S. states in terms of public welfare (PWE) and housing and community development expenditures (HCDE), and (ii) the impact of intergovernmental revenue from the federal government on PWE and HCDE. Our model assumes that each state cares only about the utility of its own residents, although this utility may be influenced by neighboring states’ decisions regarding PWE and HCDE. Meanwhile, the federal government allocates grants to states with the objective of promoting nationwide welfare. One motivation for this study lies in the shared role of PWE and HCDE in supporting residents, which makes it important to examine their potential complementarity. These expenditures may also be interrelated across states due to policy spillovers. Another motivating factor is the states’ substantial reliance on intergovernmental revenue, which can be regarded as direct federal grants. On average, these grants account for approximately 38\% of total state revenue, underscoring the considerable influence of federal policy decisions. Through our analysis, we examine how the federal government intervenes in state-level decisions, each driven by local interests, to promote nationwide social welfare through grant allocations.

Among the possible specifications of $\mathbf{W}$, the Akaike weight—representing model probability—selects a geographic network, supporting the notion of benefit spillovers from the two expenditures \citep{SOLEOLLE200632}. Our findings yield several key insights. First, federal grants have a significant positive effect on both types of expenditures, resulting in a responsive grant scheme shaped by states’ decisions. Second, we find strong evidence of spillovers in both types of state expenditures. Third, PWE and HCDE function as complements within a state. Finally, the fixed-effect components—representing the main part of the autonomous transfer—account for approximately 74\% of the variation in federal grant allocations.

Further analysis shows that states’ tax revenues, demographic characteristics, and economic conditions significantly affect their public expenditures. For instance, a \$1,000 increase in a state’s tax revenue per capita is associated with a \$61.28 rise in PWE per capita and a \$2.96 decline in HCDE per capita. In contrast, a \$1,000 increase in federal grants per capita leads to a \$23.35 rise in PWE per capita and a marginal \$0.05 increase in HCDE per capita. These comparatively smaller effects of federal grants highlight their limited flexibility relative to state tax revenues. This finding is consistent with the structural composition of revenue sources: federal grants are largely explained by fixed-effect components—the main elements of autonomous transfers—whereas state tax revenues are more responsive to local conditions. Moreover, a 1\% increase in a state’s population growth results in a \$63.24 decline in PWE per capita and an \$8.01 rise in HCDE per capita. Finally, a 1\% increase in income inequality within a state leads to a \$0.52 increase in HCDE per capita.

Finally, we compare state-level expenditures and social welfare under our estimated responsive grant scheme with those under a regime consisting solely of autonomous transfers, using counterfactual simulations. The results show that responsive federal intervention leads to a per capita increase of \$67.17 in PWE and \$1.21 in HCDE for each year, contributing to a 7.27\% overall improvement in social welfare. These findings suggest that federal resource allocation can enhance social welfare by increasing state-level PWE and HCDE. However, the magnitude of these welfare gains is modest relative to the share of federal grants in total state revenue (38\%), implying that the federal government’s ability to improve welfare through the adjustment of state-level policy spillovers is constrained by the system’s reliance on autonomous transfers—primarily fixed-effect components. Variance decomposition supports this interpretation: the responsive component explains only 1.24\% of the variation in federal grants, with 94.93\% attributed to autonomous transfers—highlighting the structural limits on federal responsiveness. In contrast, for state tax revenue, responsiveness explains 12.15\% of the variation, which is much larger than the contribution of the responsive part for the federal grant. These findings highlight the importance of designing more responsive federal grant mechanisms to achieve more effective policy outcomes.

\subsection{Related literature}

In this study, we advance the literature on spatial and network interactions by developing a model that captures (i) resource allocations by a central agent, (ii) local agents’ decisions across multiple activities, and (iii) their dynamic interactions. Methodologically, we integrate multivariate spatial econometric modeling with hierarchical decision-making, positioning our framework within the structural branch of spatial modeling.\footnote{For comprehensive reviews of structural models in fiscal competition and local public goods, see \cite{HOLMES201569}. For spatial models in urban economic issues, refer to \cite{GIBBONS2015115}.}

First, our study bridges a gap between macroeconomic analyses of federal decision-making and spatial econometric research on local government behavior. \citet{Agrawaletal2022} emphasize that state governments consider cross-border demographic and economic flows but face neither strict border controls nor linguistic barriers, resulting in substantial policy spillovers across states. Spatial econometric studies have traditionally focused on local governments’ interdependence and strategic responses \citep{Caseetal1993, Brueckner2003, BAICKER2005529, Revelli2005, SOLEOLLE200632, DiPortoRevelli2013, HanLee2016, JeongLee2020, JeongLee2021, dePaulaetal2024}. In particular, SAR models and their extensions estimate best response functions in the context of welfare competition and policy spillovers, often within a game-theoretic framework.

Another key aspect of state-level decision-making is the significant reliance on intergovernmental grants from the federal government \citep{Agrawaletal2022}. While existing spatial econometric studies often treat grant levels as exogenous explanatory variables, they rarely capture the strategic interaction between local and federal governments. In this study, we incorporate (i.e., endogenize) federal behavior (e.g., \cite{Barro1990}) into the SAR framework to represent the interaction between the allocator’s decisions and state policy choices. This idea is related to several existing studies. \citet{Knight2002} develops a bargaining model in which the federal government’s grant decisions and the state’s highway spending are jointly determined. Building on this theoretical foundation, he also accounts for the endogeneity of federal grants in a reduced-form estimation framework. \citet{REVELLI2005JUE} examines a two-tier fiscal system that captures both horizontal and vertical interactions, but assumes that each level of government provides similar public services, rather than allocating resources hierarchically. Extending SAR models in this direction enables us to capture multiple forms of policy interdependence—such as fiscal competition, yardstick competition, and expenditure spillovers—while embedding the model in a network-based framework with broader welfare implications.

In contrast to previous studies—most notably those summarized in the seminal work by \citet{Agrawaletal2022}, which primarily examine local government interactions in isolation—this study introduces a hierarchical framework that incorporates both federal influence and local government responses. By dynamically modeling federal–state interactions, we offer new insights into multi-level policy coordination, highlighting federal intervention not only as a form of financial support but also as strategic guidance. This approach captures the complexity of intergovernmental relationships and illustrates how federal resources shape decentralized policy actions over time. Our extended SAR model thus provides a structural explanation of the relationship between federal resource allocation and state-level policy interdependence, and it can be adapted to broader contexts involving interactions between central resource allocators and strategically responsive agents.

Second, from a modeling perspective, we extend existing network interaction models to a dynamic hierarchical setting to describe interactions between a central allocator and strategically responsive local agents. As a result, our framework captures both horizontal and vertical interactions, highlighting how the allocator’s decisions influence the activity choices of local agents. \citet{HelsleyZenou2014} and \citet[Ch.4.1.3]{JacksonZenou2015} emphasize that Nash equilibrium outcomes in network games are typically inefficient in terms of social welfare, and that welfare can be evaluated as the sum of all agents’ payoffs. We adopt this welfare specification to formulate the benevolent allocator’s objective in determining grant allocations. Moreover, this formulation corresponds to the highest level of altruism in \citet{HsiehLin2021}. Recognizing the repeated nature of the allocator’s decisions and the local agents’ activities in panel data, we model their interaction as a dynamic game. To do so, we employ the structure of a dynamic Stackelberg game \citep[Ch.19]{LjungqvistSargent2012}, in which a large agent (the allocator) interacts with smaller agents (the local agents).\footnote{\citet{Diazetal2021} also study interactions between a large agent and follower agents in a network setting, but their model features a single type of activity (crime) and prescriptive guidance rather than strategic incentives.}

Third, we extend linear-quadratic (LQ) network interaction models with multiple activities to a dynamic game setting. LQ payoff functions have gained popularity in network models because they yield linear best responses \citep{Ballester2006, Calvoetal2009, Bramoulleetal2014, Blumeetal2015, Boucher2016}. This structure can be explored within the linear-in-means model \citep{Manski1993, Moffitt2001, Bramoulleetal2009} or the SAR model \citep{CliffandOrd1973, Ord1975, Lee2004}, both of which offer analytical tractability and facilitate comparative static analysis. As an extension, several studies examine network interactions across multiple activities in static game environments \citep{BelhajDeroian2014, Liu2014, Cohen-Coleetal2017, Chenetal2018}. These works, relying on cross-sectional data, primarily investigate the complementarity or substitutability of multiple activities. In contrast, our framework emphasizes (i) repeated decision-making by a central allocator and local agents, and (ii) multiple activities that draw on a shared type of resource.\footnote{We focus on multiple activities that utilize a common type of resource—e.g., local governments’ various expenditure types funded by central government grants, as in our application. In contrast, for some multiple-activity contexts (e.g., hours of study and smoking frequency), it is difficult to conceptualize the role of a central allocator.}

Fourth, recall that our model belongs to the class of spatial dynamic panel simultaneous equations (SDPSE) models, augmented by an additional endogenous variable representing the allocator’s intervention. As such, our framework provides an economic foundation for SDPSE models, allowing for structural interpretation and welfare analysis. Recent studies on SAR models with multiple activities have primarily addressed statistical properties such as identification, estimation, and asymptotic behavior \citep{KelejianPrucha2004, Baltageetal2015, YangLee2017, YangLee2019, LiuSaraiva2019, Zhuetal2020, YangLee2021, Ambaetal2023, Drukkeretal2023, Lu2023}. Within our framework, these models can be viewed as representing the best responses of local agents under myopic behavior in the absence of allocator intervention. By building on existing notions of equilibrium measures in network and spatial econometric models \citep{Katz1953, Bonacich1987, LeSagePace2008, Blochetal2023}, we characterize how a change in a local agent’s characteristics leads to immediate adjustments in their own or others’ activities—capturing what we refer to as \textit{short-run impacts}. Furthermore, our model, along with other SDPSE models, can be interpreted in conjunction with structural vector autoregressive (SVAR) models, as discussed by \citet{Elhorstetal2021}, who outline the commonalities and differences between spatial dynamic and SVAR frameworks. As a multivariate time-series model, our framework also allows us to derive measures of \textit{long-run impacts} based on cumulative dynamic responses.

The remainder of the paper is organized as follows. Section 2 introduces the model and its economic foundations. Section 3 derives key measures for interpreting the model. Section 4 presents the quasi-maximum likelihood (QML) estimation method and its asymptotic properties. Section 5 evaluates the finite-sample performance of the QML estimator through Monte Carlo simulations. Section 6 applies the model to U.S. federal resource allocations and state-level PWE and HCDE. Section 7 concludes. Additional derivations and asymptotic analyses are provided in the Appendices and supplementary material.

\textbf{Notations}. The following notations will be used throughout this paper.
\begin{itemize}
    \item $e_{n,i}$: $n$-dimensional unit vector with its $i$-th element equal to 1 and all other elements equal to 0,
    \item $l_{n}$: $n$-dimensional vector consisting of ones, $\mathbf{I}_{n}$: $n$-dimensional identity matrix,
    \item $(\cdot)$ denotes a vector, while $\left[\cdot\right]$ represents a matrix. Note that $\left[\mathbf{A} \right]_{ij}$ denotes the $(i,j)$-element of $\mathbf{A}$.
\end{itemize}
In general, we use boldface and uppercase letters to represent matrices, while lowercase letters represent vectors. Sets are represented using calligraphic font. To avoid notational complexity, we will mostly refrain from using double array notations frequently used in spatial econometrics.

\section{Model}

\label{modelspecification}

\setcounter{theorem}{0}
\setcounter{assumption}{0}
\setcounter{remark}{0}
\setcounter{definition}{0}
\renewcommand{\thetheorem}{2.\arabic{theorem}}
\renewcommand{\theassumption}{2.\arabic{assumption}}
\renewcommand{\theremark}{2.\arabic{remark}}
\renewcommand{\thedefinition}{2.\arabic{definition}}%

\subsection{Basic setup and payoff specifications}
\label{payoffspec}

\begin{figure}[!htbphtbp]
\centering
\caption{Basic description}
    \label{vardesc}
\begin{tikzpicture}[>=stealth, font=\small]

\fill[blue!10] (-4,-2) -- (4,-2) -- (5,-3) -- (-3,-3) -- cycle;

\node[above] (allocator) at (0,2) {\textbf{Resource allocator}};
\node[below] (agent1) at (-2,-2.5) {\textbf{Local agent 1}};
\node[below] (agent2) at (2,-2.5) {\textbf{Local agent 2}};

\draw[thick,->] (allocator) -- node[left=2pt] {$g_{1,t}$} (agent1);
\draw[thick,->] (allocator) -- node[right=2pt] {$g_{2,t}$} (agent2);

\node[draw, fill=green!20, rounded corners, minimum width=2.6cm] (a11) at (-4,1.0) {Activity 1 ($y_{1,t,1}$)};
\node[draw, fill=red!20, rounded corners, minimum width=2.6cm] (a12) at (-4,-0.3) {Activity 2 ($y_{1,t,2}$)};
\draw[<->] (a11) -- (a12);

\node[draw, fill=green!20, rounded corners, minimum width=2.6cm] (a21) at (4,1.0) {Activity 1 ($y_{2,t,1}$)};
\node[draw, fill=red!20, rounded corners, minimum width=2.6cm] (a22) at (4,-0.3) {Activity 2 ($y_{2,t,2}$)};
\draw[<->] (a21) -- (a22);

\draw[dotted, thick,<->] (agent1) to[bend left=15] (agent2);

\draw[dashed, ->] (agent1) -- (a11);
\draw[dashed, ->] (agent1) -- (a12);

\draw[dashed, ->] (agent2) -- (a21);
\draw[dashed, ->] (agent2) -- (a22);
\end{tikzpicture}
\end{figure}

We begin by introducing the notation and basic setup, which is briefly summarized in Figure~\ref{vardesc}. The index \( 0 \) denotes the resource allocator, while indices \( i = 1, 2, \cdots, n \) refer to the local agents. Each local agent $i$ is located at $\ell(i) \in \mathbb{R}^{d}$ ($d \geq 1$). Given the set of locations $\lbrace \ell(i) \rbrace_{i=1}^{n}$, the interactions among local agents are characterized by an \( n \times n \) spatial weighting matrix \( \mathbf{W} =\left[ w_{ij}\right] \). To allow richer forms of proximity, the weights $w_{ij}$ may also incorporate region-specific characteristics.\footnote{For example, population or capital mobility measures—possibly time-varying—can be used to construct spatial weights. In our framework, a long-run (steady-state) relationship can serve as a practical approximation. See Section~\ref{Application} for an illustrative application. A time-varying or endogenous network structure, as discussed by \citet{JeongLee2021}, entails additional technical challenges, including the possibility of multiple equilibria due to nonlinearities. Since our focus lies in integrating vertical and horizontal interactions, we maintain the simplifying assumption that $\mathbf{W}$ is exogenous and time-invariant.} In line with the spatial econometric literature, we assume that \( w_{ij} \geq 0 \) and \( w_{ii} = 0 \) for all \( i = 1, \cdots, n \), implying that no local agent interacts with itself.

The allocator's primary role is to allocate resources to local agents. In each period $t$, the allocator chooses grant levels $g_{i,t}$ for $i = 1, \ldots, n$. Given $g_{i,t}$, each local agent $i$ decides on \( m^{*} \geq 2 \) continuous types of activities, denoted by $y_{i,t,l}$ for $l = 1, \ldots, m^{*}$. We interpret both the local agents' activities and the allocator's grants as forms of “effort,” which implies that they must be strictly positive in our framework. Accordingly, we assume $y_{i,t,1}, \ldots, y_{i,t,m^{*}} \in (0, \infty)$ and $g_{i,t} \in (0, \infty)$ for all $i = 1, \ldots, n$ and all $t$.\footnote{If zero values are observed in practice, an alternative model specification accommodating corner solutions or truncated distributions would be required. Such extensions are beyond the scope of this study.} Since the allocator's decisions are also treated as endogenous, the total number of endogenous variables in the system is given by \( m = m^{*} + 1 \). In the theoretical model, the number of local agents $n$ is fixed, and the decision-making horizon is infinite. Based on this setup, we assume that the econometrician observes the panel data $\left\{ y_{i,t,1}, \ldots, y_{i,t,m^*}, g_{i,t} \right\}_{i=1}^{n}\vert_{t=0}^{T}$. In Section~\ref{econometricmodel}, we adopt a large-$n$, large-$T$ asymptotic framework. Consequently, for estimation, agent- and time-specific characteristics are treated as fixed effects, even though some of these—particularly time-specific ones—are interpreted as random shocks in the theoretical model.

\subsubsection{Local agent's payoff function}
\label{followerpayoffspec}

A local agent \(i\)'s objective is to maximize their lifetime payoff by choosing multiple activities, \(y_{i,t} = (y_{i,t,1}, \dots, y_{i,t,m^{*}})^\prime\). The per-period payoff for local agent \(i\) at time \(t\) is specified as follows:
\begin{equation}
    \begin{split}
        & U_{i}\left( y_{i,t}, \mathbf{y}_{-i,t}, g_{i,t}, \mathbf{Y}_{t-1}, x_{i,t}, u_{i,t}\right) \\
= & \sum_{l=1}^{m^{*}} \left( \sum_{k=1}^{K}\pi_{kl}x_{i,t,k} + u_{i,t,l} + \phi_{l}g_{i,t}
+ \sum_{l^{\prime}=1}^{m^{*}}\sum_{j=1}^{n}\rho_{l^{\prime }l}w_{ij}y_{j,t-1,l^{\prime}}
+ \sum_{l^{\prime }=1}^{m^{*}}\sum_{j=1}^{n}\lambda_{l^{\prime
}l}w_{ij}y_{j,t,l^{\prime}}\right) y_{i,t,l} \\
& - \underset{\text{Cost I}}
{\underbrace{\frac{1}{2}\sum_{l=1}^{m^{*}}\sum_{l^{\prime }=1}^{m^{*}}p_{l^{\prime}l}\left( y_{i,t,l^{\prime}} - y_{i,t-1,l^{\prime}}\right) \left(
y_{i,t,l} - y_{i,t-1,l}\right)}}
- \underset{\text{Cost II}}{\underbrace{\frac{1}{2}\sum_{l=1}^{m^{*}}\sum_{l^{\prime}=1}^{m^{*}}\psi_{l^{\prime
}l}y_{i,t,l^{\prime}}y_{i,t,l}}}
\label{m2-2-fpayoff}
    \end{split}
\end{equation}
where \(\mathbf{y}_{-i,t} = \left(y_{1,t}^\prime,\dots,y_{i-1,t}^\prime,y_{i+1,t}^\prime,\dots,y_{n,t}^\prime\right)^\prime\) denotes the collection of reactions from all other local agents to \(i\)'s actions, \(x_{i,t} = \left(x_{i,t,1}, \dots, x_{i,t,K} \right)^\prime\) is a \(K \times 1\) vector of observable characteristics, and \(u_{i,t} = \left(u_{i,t,1},\dots,u_{i,t,m^{*}} \right)^\prime\) is an \(m^{*} \times 1\) vector of characteristics unobserved by econometricians.

Payoff (\ref{m2-2-fpayoff}) is a dynamic extension of the linear-quadratic (LQ) payoff used in static models, as presented in prior works by \cite{BelhajDeroian2014}, \cite{Liu2014}, \cite{Cohen-Coleetal2017}, and \cite{Chenetal2018}. In those static network interaction models, the LQ payoff specification can generate a linear best response, which offers advantages in comparative static analysis as well as linkage to econometric models. The LQ payoff setting in dynamic models, where agents optimize their lifetime payoffs, holds additional significance. Due to network interactions, agents in a dynamic model share common state variables, leading to a large dimensionality of these variables (proportional to \(n\)). The LQ payoff specification is the only functional form allowing feasible computations in dynamic network interaction models. In Appendix~B, we present a sensitivity analysis that supports the LQ payoff specification.  When the actual payoff function deviates moderately from the LQ form and the contraction-mapping conditions hold, the resulting equilibrium activities and grant allocations remain close to those obtained under the baseline LQ model.

First, the choice-specific benefit for \(y_{i,t,l}\) from characteristics is given by \(\left(\sum_{k=1}^{K}\pi_{kl}x_{i,t,k} + u_{i,t,l} \right)y_{i,t,l}\), where \(\pi_{kl}\) illustrates the influence of \(x_{i,t,k}\) on the \(l\)th activity-related payoff. Explicit assumptions for characteristics \(x_{i,t,k}\) and \(u_{i,t,l}\) are provided in Assumption \ref{asmproch}. Second, the allocator's grant (\(g_{i,t}\)) impacts local agent \(i\)'s marginal payoff through the parameters \(\phi_{l}\) (in detail, $\phi_{l} = \frac{\partial^{2}U_{i}(\cdot)}{\partial g_{i,t} \partial y_{i,t,l}}$). Third, the influence of preceding neighboring activities is represented by \(\sum_{l^{\prime }=1}^{m^{*}}\sum_{j=1}^{n}\rho_{l^{\prime} l}w_{ij}y_{j,t-1,l^{\prime}}\), where \(\rho_{l^\prime l}\) characterizes the dynamic strategic substitutability (\(\rho_{l^\prime l}w_{ij} = \frac{\partial^{2}U_{i}(\cdot)}{\partial y_{j,t-1,l^\prime} \partial y_{i,t,l}} < 0\)) or complementarity (\(\rho_{l^\prime l}w_{ij} = \frac{\partial^{2}U_{i}(\cdot)}{\partial y_{j,t-1,l^\prime} \partial y_{i,t,l}} > 0\)) between \(i\)'s \(l\)th activity and the \(l^\prime\) activities of $i$'s neighbors. Fourth, the contemporaneous spatial interactions among local agents are depicted by \(\sum_{l^{\prime }=1}^{m^{*}}\sum_{j=1}^{n}\lambda_{l^{\prime} l}w_{ij}y_{j,t,l^{\prime}}\), with the parameter \(\lambda_{l^\prime l}\) describing the contemporaneous strategic substitutability (\(\lambda_{l^\prime l}w_{ij} = \frac{\partial^{2}U_{i}(\cdot)}{\partial y_{j,t,l^\prime} \partial y_{i,t,l}} < 0\)) or complementarity (\(\lambda_{l^\prime l}w_{ij} = \frac{\partial^{2}U_{i}(\cdot)}{\partial y_{j,t,l^\prime} \partial y_{i,t,l}} > 0\)) between \(i\)'s \(l\)th activity and the \(l^\prime\)th activities of $i$'s neighbors.

The cost function is divided into two components: (i) Cost I includes LQ dynamic adjustment costs, and (ii) Cost II consists of quadratic costs associated with selecting activity levels (corresponding to the model presented by \cite{Cohen-Coleetal2017}). In addition to the dynamic strategic substitutability/complementarity term, the dynamic adjustment cost drives the dynamics of the local agents' decisions by characterizing the persistence of their decision-making. For interpretational purposes, we assume that both \(P = \left[p_{l^\prime l}\right]\) and \(\Psi = \left[ \psi_{l^\prime l}\right]\) in the cost function are nonnegative definite. To facilitate the identification of the main parameters, we normalize all the diagonal elements of \(\Psi\) to one.\footnote{In the context of linear simultaneous equations models, this normalization is analogous to normalizing the coefficient matrix for endogenous variables. Theoretically, this can be interpreted as an affine transformation of the utility function. For example, in a single-variable model like the one presented by \cite{JeongLee2020}, one can normalize the cost parameter for two different types of costs (see footnote 6 in \cite{JeongLee2020}). In their model, the parameter \(\gamma\) represents the dynamic adjustment cost, while \(1 - \gamma\) is assigned for determining the activity level. Normalizing \(\Psi\) is computationally advantageous for effective parameter searching. Despite our model's highly nonlinear parametric structure, I have found that searching for the optimizer of the statistical objective function becomes challenging when \(\Psi\) is not normalized.} The diagonal elements of \(P\), \(p_{ll}\), characterize the persistence of the agent's \(l\)th activity. The off-diagonal elements of \(P\) and \(\Psi\) represent the interactions among activities within the same agent's decision-making process: they indicate complementarity if \(-(p_{l^\prime l} + \psi_{l^\prime l}) = \frac{\partial^{2}U_{i}(\cdot)}{\partial y_{i,t,l^\prime} \partial y_{i,t,l}} > 0\), and substitutability if \(-(p_{l^\prime l} + \psi_{l^\prime l}) = \frac{\partial^{2}U_{i}(\cdot)}{\partial y_{i,t,l^\prime} \partial y_{i,t,l}} < 0\).

Under the complete information setting with myopia, the first-order conditions yield the local agents' best responses: 
\begin{equation}
\mathbf{Y}_{t}\left( P + \Psi \right) = \mathbf{W} \mathbf{Y}_{t}\Lambda + \mathbf{Y}_{t-1}P + \mathbf{W} \mathbf{Y}_{t-1}\boldsymbol{\rho} 
+ \mathbf{g}_{t}\phi^\prime + \mathbf{X}_{t}\Pi + \mathbf{U}_{t},
\text{ for }t=1,\cdots,T,  \label{m2-2-example}
\end{equation}
where $\mathbf{Y}_{t} = \left[y_{i,t,l} \right]$ is an $n \times m^{*}$ matrix, $\mathbf{g}_{t} = \left(g_{1,t}, \cdots, g_{n,t} \right)^\prime$ is an $n \times 1$ vector, $\mathbf{X}_{t} = \left[x_{1,t}, \cdots, x_{n,t} \right]^\prime$ is an $n \times K$ matrix, and $\mathbf{U}_{t} = [u_{i,t,l}]$ is an $n \times m^{*}$ matrix. For the parameters, $\phi = \left( \phi_{1},\cdots,\phi_{m^{*}}\right)^\prime$ is an $m^{*}\times 1$ vector, $\Lambda = \left[ \lambda_{l^\prime l}\right]$ and $\boldsymbol{\rho} = \left[\rho_{l^\prime l}\right]$ are $m^{*}\times m^{*}$ square matrices, and $\Pi = [\pi_{kl}]$ denotes a $K\times m^{*}$ coefficient matrix for $\mathbf{X}_{t}$.

The positive definiteness of \(P + \Psi\) is required to obtain a unique best response vector. When $\mathbf{I}_{nm^{*}} - \left(P + \Psi \right)^{-1}\Lambda^\prime \otimes \mathbf{W}$ is invertible, this myopic model has a unique Nash equilibrium. Indeed, (\ref{m2-2-example}) follows a spatial dynamic panel simultaneous equations (SDPSE) model \citep{KelejianPrucha2004, Baltageetal2015, YangLee2017, YangLee2019, LiuSaraiva2019, Zhuetal2020, YangLee2021, Ambaetal2023, Drukkeretal2023, Lu2023}. $\mathbf{g}_{t}\phi^\prime$ is a new term in our specification, and this term highlights how the allocator directly affects the local agents' activities. The SDPSE model itself can provide evidence of strategic interactions and substitutability/complementarity among multiple activities, but it has limitations in providing welfare implications. The payoff structure \eqref{m2-2-fpayoff} generating the SDPSE model can produce welfare implications.

$\mathbf{X}_{t}$ contains observable demographic/economic consequences induced by previous efforts, while the unobservable matrix $\mathbf{U}_{t}$ possibly includes agent- or time-specific components and contemporaneous shocks. Assumption \ref{asmproch} describes the detailed processes for $\mathbf{X}_{t}$ and $\mathbf{U}_{t}$, which are essential for agents to form conditional expectations regarding future characteristics.

\begin{assumption}
\label{asmproch} 
(i) For $k=1,\cdots,K$, $\mathbf{x}_{t,k} = \left(x_{1,t,k}, \cdots, x_{n,t,k} \right)^\prime$ follows a stable Markov process: 
\begin{equation}
\mathbf{x}_{t,k} = \mathbf{A}_{k}^{x}\mathbf{x}_{t-1,k} + \mathbf{B}_{k}^{x}\operatorname{vec}(\mathbf{Y}_{t-1}) + C_{k}^{x}\mathbf{g}_{t-1} +
\underset{=\text{agent/time specific components}}{\underbrace{\mathbf{c}_{k}^{x} + \alpha_{t,k}^{x}l_{n}}}
+ \mathbf{e}_{t,k}^{x},
\label{processx}
\end{equation}
where $\mathbf{A}_{k}^{x} = \gamma_{k}\mathbf{I}_{n} + \varrho_{k}\mathbf{W}$, $\mathbf{B}_{k}^{x} = [\mathbf{B}_{k1}^{x},\cdots,\mathbf{B}_{km^{*}}^{x}]$ with $\mathbf{B}_{kl}^{x} = \gamma_{kl}\mathbf{I}_{n} + \varrho_{kl}\mathbf{W}$ for $l=1,\cdots,m^{*}$, $C_{k}^{x} = \gamma_{k}^{g}\mathbf{I}_{n} + \varrho_{k}^{g}\mathbf{W}$, $\mathbf{c}_{k}^{x}$ is an $n\times 1$ vector of local agent fixed effects, $\alpha_{t,k}^{x}$ is a time specific component with zero mean, and $\mathbf{e}_{t,k}^{x}$ is an $n\times 1$ vector of idiosyncratic components.  

Let $\mathbf{E}_{t}^{x} = [\mathbf{e}_{t,1}^{x},\cdots,\mathbf{e}_{t,K}^{x}]$ for each $t$. Assume $\operatorname{vec}(\mathbf{E}_{t}^{x}) \sim i.i.d.\left( \mathbf{0},\diag\{\sigma_{k}^{2}\} \otimes \mathbf{I}_{n}\right)$ with $\sigma_{k}^{2} > 0$ for $k = 1,\cdots,K$.

(ii) For each $t$,
\begin{equation*}
   \mathbf{U}_{t} = \underset{=\text{agent/time specific components}}{\underbrace{\boldsymbol{\eta} + \widetilde{\alpha}_{t}^{\prime}\otimes l_{n}}}  + \mathbf{E}_{t},
\end{equation*}
where $\boldsymbol{\eta}$ is an $n\times m^{*}$ matrix of agent-specific components, $\widetilde{\alpha}_{t} = \left(\alpha_{t,1},\cdots,\alpha_{t,m^{*}}\right)^\prime$ is a vector of time-specific components with zero means at time $t$, and $\mathbf{E}_{t} = [e_{i,t,l}]$ is an $n\times m^{*}$ matrix of idiosyncratic components. We assume $\operatorname{vec}(\mathbf{E}_{t}) \sim i.i.d.\left( \mathbf{0}, \Sigma \otimes \mathbf{I}_{n}\right) $ with $\Sigma > 0$.

(iii) $\mathbf{E}_{t}$, $\alpha_{t,1}^{x}$, $\cdots$, $\alpha_{t,K}^{x}$, and $\mathbf{E}_{t}^{x}$ are independently generated over $t$.
\end{assumption}

Assumption \ref{asmproch} (i) states that \(\mathbf{x}_{t+1,k}\) is influenced by the local agents' activities (\(\mathbf{Y}_{t}\)) and grant allocations ($\mathbf{g}_{t}$). Equation \eqref{processx} can be written as
\begin{equation}
    \mathbf{x}_{t,k} = \sum_{s=1}^{\infty}\left(\mathbf{A}_{k}^{x} \right)^{s-1}\left(\mathbf{B}_{k}^{x} \operatorname{vec}(\mathbf{Y}_{t-s}) + C_{k}^{x}\mathbf{g}_{t-s} + \mathbf{c}_{k}^{x} + \alpha_{t-s,k}^{x}l_{n} + \mathbf{e}_{t-s,k}^{x} \right)
    \label{processx_alter}
\end{equation}
since $\Vert \mathbf{A}_{k}^{x} \Vert_2 < 1$ is assumed for stability. Then, \eqref{processx_alter} reflects the cumulative feature of the previous efforts on the current outcomes (refer to \cite{ToddWolpin2003}). Although we focus on continuous $\mathbf{x}_{t,k}$ for ease of exposition, the same framework accommodates discrete-value processes by representing $\mathbf{x}_{t,k} \vert \mathbf{Y}_{t-1}, \mathbf{g}_{t-1}, \mathbf{x}_{t-1,k}$ as a Markov chain.\footnote{Our estimation procedure is based on the joint density $f\bigl(\{\mathbf{Y}_{t}, \mathbf{g}_{t}, \mathbf{X}_{t}\}_{t=1}^{T}\bigr)$. If $\mathbf{x}_{t,k}$ is discrete, we represent
$\mathbf{x}_{t,k} \mid \mathbf{Y}_{t-1}, \mathbf{g}_{t-1}, \mathbf{x}_{t-1,k}$ as a finite-state Markov chain.
Hence, the main derivations remain valid under either a continuous or discrete specification of $\mathbf{x}_{t,k}$ (see Section 3.2 of the supplement for details). For details on the relationship between discrete-valued Markov chains and continuous Markov processes, see \cite{Tauchen1985}.} For further analysis, let $\mathbf{C}_{k}^{x} = \left[\frac{1}{m^{*}}C_{k}^{x}, \cdots, \frac{1}{m^{*}}C_{k}^{x} \right]$ to have $\mathbf{C}_{k}^{x}\left(l_{m^{*}} \otimes \mathbf{g}_{t} \right) = C_{k}^{x}\mathbf{g}_{t}$ for $k = 1,\cdots, K$.

Assumption \ref{asmproch} (ii) establishes that \(\mathbf{U}_{t}\) contains \(\boldsymbol{\eta}\), \(\widetilde{\alpha}_{t}\), and \(\mathbf{E}_{t}\). Assumption \ref{asmproch} (iii) ensures that future shocks and time-specific components are unpredictable at time \(t\). This implies that the conditional expectations of future shocks and time-specific components are zero, simplifying the derivation of the agents' optimal activities.\footnote{This assumption can be extended to nonzero expected values of \(\widetilde{\alpha}_{t+s}\) and \(\alpha_{t+s,k}^{x}\) for \(k = 1,\dots,K\) conditional on the \(t\)th period information set. However, this extension does not affect the derivation of the agents' optimal activities but adds additional complications since those constants will be absorbed in the intercept terms in the model. Hence, this extension also does not affect the estimation procedure.} In the estimation, we will treat \(\widetilde{\alpha}_{t}\) and \(\alpha_{t,k}^{x}\) as time fixed-effect components.

\subsubsection{Resource allocator's payoff function}
\label{leaderpayoffspec}

A benevolent allocator aims to allocate grants to local agents, with the objective function geared towards maximizing social welfare. Consequently, the allocator's per-period payoff for allocating $g_{1,t},\cdots,g_{n,t}$ is formulated as a summation of the local agents' payoffs and associated costs:
\begin{equation}
U_{0}\left(\mathbf{g}_{t}, \mathbf{Y}_{t},\mathbf{Y}_{t-1}, \mathbf{X}_{t}, \mathbf{U}_{t}, \boldsymbol{\tau}_{t}\right) = \underset{\text{Social benefits}}{\underbrace{\sum_{i=1}^{n}U_{i}\left( \mathbf{Y}_{t}, g_{i,t}, \mathbf{Y}_{t-1}, x_{i,t}, u_{i,t}\right) }} + \underset{\text{Autonomous transfer}}{\underbrace{\sum_{i=1}^{n}\tau_{i,t} g_{i,t}}} - \underset{\text{Cost}}{\underbrace{\frac{1}{2}\sum_{i=1}^{n}g_{i,t}^{2}}}.  
\label{m2-2-lpayoff}
\end{equation}
Here, \( \boldsymbol{\tau}_{t} = \left(\tau_{1,t},\cdots,\tau_{n,t} \right)^\prime \) represents autonomous transfer factors associated with the allocator's decisions at time \( t \). The term \( \sum_{i=1}^{n}\tau_{i,t}g_{i,t} \) in \eqref{m2-2-lpayoff} acts as an incentive (or a cost) for the allocator to make autonomous transfers. Specifically, when \(\phi_{l} = 0\) for all \( l = 1, \cdots, m^{*} \), the allocator's optimal grant for local agent \( i \) simplifies to \( g_{i,t}^{*} = \tau_{i,t} \). We will introduce a specification of $\boldsymbol{\tau}_{t}$ in Assumption \ref{asmproctau}. The quadratic cost in \eqref{m2-2-lpayoff} streamlines the cost or budget constraint incurred by grant allocation. If relevant covariates $x_{i,t}^{c}$ are available, this cost function can be extended as $- \frac{1}{2}\sum_{i=1}^{n}\left(g_{i,t} - G(x_{i,t}^{c}) \right)^2$ for some function $G(\cdot)$ representing the constraint $g_{i,t} = G(x_{i,t}^c)$.

Following Jackson and Zenou (\citeyear{JacksonZenou2015}, Ch.4.1.3), the first part of \( U_{0}\left(\cdot\right) \) represents social benefits, defined as the summation of the local agents' payoffs.\footnote{\cite{Diazetal2021} also consider a hierarchical decision-making process for crime activities; that is, two-type agents in their model produce the same type activity. \cite{HelsleyZenou2014} employ a similar idea in a cross-sectional linear-quadratic model for the agent's effort choice problem. A social planner knows that the Nash equilibrium outcome is inefficient in their setting. The first-best outcome can be achieved by giving subsidies (Pigouvian subsidies) to support individuals' activities. However, they do not combine a planner's optimization problem with agents' optimization problems exactly to their paper's specifications. For details, see Helsley and Zenou (\citeyear{HelsleyZenou2014}, Sec.5). In contrast, our model's allocator aims to maximize social welfare by choosing grants $\mathbf{g}_{t}$.} Adopting the perspective of \cite{HsiehLin2021}, the allocator embodies the highest level of altruism. In the model, $\mathbf{g}_{t}$ influences each local agent's payoff, \( U_{i}(\cdot) \), in two ways: directly and indirectly. The direct effect manifests as \( \sum_{l=1}^{m^{*}}\phi_{l}g_{i,t}y_{i,t,l} \) within \( U_{i}(\cdot) \). The indirect effect comes into play through changes in the elements of \( \mathbf{Y}_{t} \): the allocator chooses \( \mathbf{g}_{t} \) with the understanding that this choice will impact \( \mathbf{Y}_{t} \) and, consequently, the collective payoffs \( \sum_{i=1}^{n}U_{i}\left( \cdot \right) \). Hence, this specification positions the allocator not only as a financial allocator but also as a strategic leader aiming to guide local agents' activities toward social welfare objectives.

It should be noted that \( \tau_{i,t} \) encapsulates local agents' innate characteristics and temporary economic conditions (or shocks) but are unrelated to the local agents' decisions once \( \mathbf{g}_{t} \) is determined. The assumption below imposes a structure on $\boldsymbol{\tau}_{t}$ for estimation.

\begin{assumption}
\label{asmproctau} 
(i) For each $t$, assume 
\begin{equation*}
    \boldsymbol{\tau}_{t} = \boldsymbol{\tau} + \mathbf{X}_{t}^{\tau}\beta + \alpha_{t}^{\tau} l_{n} + \mathbf{e}_{t}^{\tau},
\end{equation*}
where \(\boldsymbol{\tau} > 0 \) is a vector of local agents' innate characteristics, \(\mathbf{X}_{t}^{\tau} = [ \mathbf{x}_{t,1}^{\tau}, \cdots, \mathbf{x}_{t,Q}^{\tau}]\) with \(\mathbf{x}_{t,q}^{\tau} = (x_{1,t,q}^{\tau}, \cdots, x_{n,t,q}^{\tau})^{\prime}\) for \(q=1, \cdots, Q\) is an \(n \times Q\) matrix of (mean zero) indicators for the allocator's decisions, $\alpha_{t}^{\tau}$ is a
time-specific component with zero mean, $\beta = \left( \beta_{1},\cdots,\beta_{Q}\right)^{\prime}$ is a parameter vector, and $\mathbf{e}_{t}^{\tau} = \left( e_{1,t}^{\tau},\cdots,e_{n,t}^{\tau}\right)^{\prime}$ be a vector of idiosyncratic components. Assume $e_{i,t}^{\tau} \sim i.i.d.\left( 0,\sigma^{2}\right) $ with $\sigma^{2} > 0$ across $i$ and $t$. 

(ii) $\mathbf{X}_{t}^{\tau}$, $\alpha_{t}^{\tau}$ and $\mathbf{e}_{t}^{\tau}$ are independently generated over $t$.
\end{assumption}

Assumption \ref{asmproctau} specifies that $\boldsymbol{\tau}_{t}$ comprises of the positive autonomous payments ($\boldsymbol{\tau}$) and the $t$-period specific mean-zero components $\mathbf{X}_{t}^{\tau}$, $\alpha_{t}^{\tau}$, and $\mathbf{e}_{t}^{\tau}$. This ensures that the baseline level of $\tau_{i,t}$ remains positive, so that even if $\phi_{l} = 0$ for all $l$, the allocator’s optimal grant $g_{i,t}^{*}$ does not become negative. Of course, one could further impose constraints on the error term $\mathbf{e}_{t}^{\tau}$ or its variance to maintain positivity in all realizations, but we keep our framework simpler by assuming such non-negativity is guaranteed in expectation. \(\mathbf{X}_{t}^{\tau}\) contains cyclical components, which account for contemporaneous fluctuations in local agents' characteristics (see Section \ref{Application} for the example). As $\widetilde{\alpha}_{t}$ and $\alpha_{t,k}^{x}$, the random component assumption in Assumption \ref{asmproctau} (ii) also simplifies the derivation of the agents' optimal activities. In the estimation part, \(\alpha_{t}^{\tau}\) will act as a time-fixed effect in the equation for $\mathbf{g}_{t}$.

\subsection{Two-stage decision-making processes}
\label{networkgame}

Based on the setting in Section \ref{payoffspec}, this section characterizes the allocator's optimal grants ($\mathbf{g}_{t}^{*}$) and the local agents' optimal activities ($\mathbf{Y}_{t}^{*}$). Our model's decision-making process is organized into two stages within each period, adapting the dynamic Stackelberg game's idea to outline strategic interactions between the allocator and the local agents.\footnote{Unlike the traditional dynamic Stackelberg game, our model's allocator does not produce any output for itself but allocates resources to maximize social welfare.} Our model's two stages can also be related to the two stages (federal and state stages) in the bargaining model of \cite{Knight2002}. We introduce a time-discounting factor \( \delta \in (0,1) \) to differentiate between future and present payoffs.

In the first stage, the allocator chooses $\mathbf{g}_{t}$ for local agents. The $t$th-period his/her information set is
\begin{equation*}
\mathcal{B}_{t,1} = \sigma \left( \left\{ \mathbf{g}_{s},\mathbf{Y}_{s}, \widetilde{\alpha}_{s} \right\}_{s=-\infty}^{s=t-1}, \left\{\mathbf{X}_{s}\right\}_{s=-\infty}^{t}, \boldsymbol{\eta}, \left\{\mathbf{E}_{s}, \mathbf{E}_{s}^{x}, \boldsymbol{\tau}_{s}\right\}_{s=-\infty}^{s=t}\right),
\end{equation*}
where $\sigma \left( \cdot \right)$ denotes a $\sigma$-field generated by the arguments inside it. Note that the $t$th-period time-specific components for the local agents' activities, $\widetilde{\alpha}_{t}$, are unpredictable to the resource allocator when making the grant allocation decision. The local agents' information set at time $t$, denoted by $\mathcal{B}_{t,2}$, includes both chosen grants $\mathbf{g}_{t}^{\ast}$ and $\widetilde{\alpha}_{t}$ in addition to the components of $\mathcal{B}_{t,1}$. Based on this setting, we define the conditional expectations $\mathbb{E}_{t,s}\left( \cdot \right)$ for stages $s=1$ and $2$.\footnote{This information structure is introduced to achieve the additive individual/time effect specifications in the estimation part. Theoretically, it represents uncertainty in the first stage's decision-making relative to the second stage.}

In the second stage, local agents choose their optimal activities ($\mathbf{Y}_{t}^{\ast}$) after observing $\mathbf{g}_{t}^{*}$. Once the variables $\left\{ \mathbf{g}_{t}^{\ast},\mathbf{Y}_{t}^{\ast}\right\}$ are set for the $t$-th period, the first stage of the $(t+1)$-th period becomes active. Consequently, the local agents' chosen activities $\mathbf{Y}_{t}^{\ast}$ are reflected for the allocator's decision-making in the following period ($\mathbf{g}_{t+1}$). This leads to an intertemporal linkage.

We formalize the lifetime optimization problems for the allocator and local agents. The Nash equilibrium (NE) concept appropriate for our econometric model is a Markov perfect Nash equilibrium (MPNE), which we will uniquely characterize later. Consequently, the agents' optimal activities can be characterized by the recursive relation of the value functions. Below, we introduce notations for state variables that are essential for characterizing the MPNE:
\begin{equation}
\begin{split}
    &\text{State var. for the local agents: }\mathbf{z}_{t}^{A} = \left( \operatorname{vec}\left( \mathbf{Y}_{t-1}\right)^\prime, \left(l_{m^{*}} \otimes \mathbf{g}_{t}^{\ast}\right)^\prime, \operatorname{vec}(\mathbf{X}_{t})', \operatorname{vec}\left( \mathbf{U}_{t}\right)^\prime \right)^\prime \\
    &\text{State var. for the resource allocator: } \mathbf{z}_{t}^{R} = \left( \operatorname{vec}\left( \mathbf{Y}_{t-1}\right)^\prime, \operatorname{vec}(\mathbf{X}_{t})^\prime,\operatorname{vec}\left( \mathbf{E}_{t}\right)^{\prime},\left( l_{m^{*}}\otimes \boldsymbol{\tau}_{t}\right)^\prime \right)^\prime 
    \label{state_var}
\end{split}
\end{equation}
for each time $t$. In the aspect of observed characteristics, note that the allocator and local agents make decisions by reacting to $\mathbf{Y}_{t-1}$ and $\mathbf{X}_{t}$ as common characteristics, while the allocator additionally utilizes information in $\mathbf{X}_{t}^\tau$ in his/her decision-making.

Now, we describe the two-stage decision-making process. The detailed value function forms are in Appendix B.

\paragraph*{Stage 1: Resource allocator's decision-making.}
Given $\mathbf{z}_{t}^{R}$, the allocator's value
function is
\begin{equation}
V_{0} \left(\mathbf{z}_{t}^{R}\right)  = \max_{
\mathbf{g}_{t}}\mathbb{E}_{t,1}\left\lbrace
{U}_{0}\left( \mathbf{g}_{t}, \mathbf{Y}_{t}^{\ast}, \mathbf{Y}_{t-1}, \mathbf{X}_{t}, \mathbf{U}_{t}, \boldsymbol{\tau}_{t} \right) 
 + \delta V_{0} \left( \mathbf{z}_{t+1}^{R} \right)
\right\rbrace  
\label{leadervalue}
\end{equation}
with expecting the local agents' optimal activities as 
\begin{equation}
\operatorname{vec}\left( \mathbf{Y}_{t}^{\ast}\right) = \mathbf{\bar{c}}^{y} + \mathbb{L}^{y}\mathbf{z}_{t}^{A},\text{ where }\mathbb{L}^{y}
=
\begin{bmatrix}
\mathbf{A}^{y}, & \mathbf{B}^{y}, & \mathbf{C}_{1}^{y}\boldsymbol{\Pi}_{1}, & \cdots, & 
\mathbf{C}_{K}^{y}\boldsymbol{\Pi}_{K}, & \mathbf{C}_{u}^{y}
\end{bmatrix}.
\label{constraint_followeract}
\end{equation}
Here, $\mathbf{\bar{c}}^{y}$ is an $nm^{*} \times 1$ vector of constants, and $\mathbf{A}^{y}$, $\mathbf{B}^{y}$, $\left\{ \mathbf{C}_{k}^{y}\right\}_{k=1}^{K}$, and $\mathbf{C}_{u}^{y}$ are $nm^{*}\times nm^{*}$ matrices derived from the local agents' value functions $\left\{ V_{i}\left( \cdot \right) \right\}_{i=1}^{n}$ in the second stage. Note that $\boldsymbol{\Pi}_{k} = \Pi^\prime e_{K,k}\otimes \mathbf{I}_{n}$ for $k=1,\cdots,K$ to have $\operatorname{vec}(\mathbf{X}_{t}\Pi) = \left(\mathbf{I}_{m^{*}} \otimes \mathbf{X}_{t}\right) \operatorname{vec}\left(\Pi\right) = \sum_{k=1}^{K}(\Pi^\prime e_{K,k} \otimes \mathbf{x}_{t,k}) = \sum_{k=1}^{K}\boldsymbol{\Pi}_{k}\mathbf{x}_{t,k}$, which is useful to deal with Riccati equations.

Observe that $\mathbf{Y}_{t}^{*}$ is a function of $\mathbf{z}_{t}^{A}$ that contains $\mathbf{g}_{t}$: the allocator chooses $\mathbf{g}_{t}$ by considering $\mathbf{g}_{t}$ impacts on $\mathbf{Y}_{t}^{*}$. Since the allocator's per period payoff (\ref{m2-2-lpayoff}) is LQ in $\mathbf{g}_{t}$, it follows that $\mathbf{g}_{t}^{\ast}$ is a linear function of $\mathbf{z}_{t}^{R}$, represented by 
\begin{equation}
\left( l_{m^{*}}\otimes \mathbf{g}_{t}^{\ast}\right) = \mathbf{\bar{c}}^{g} + \mathbb{L}^{g}\mathbf{z}_{t}^{R},\text{ where }\mathbb{L}^{g} =
\begin{bmatrix}
\mathbf{A}^{g}, & \mathbf{C}_{1}^{g} \boldsymbol{\Pi}_{1}, \cdots , \mathbf{C}_{K}^{g} \boldsymbol{\Pi}_{K}, & \mathbf{C}_{e}^{g}, & \mathbf{B}^{g}
\end{bmatrix}.
\label{m2-2-laction}
\end{equation}
Here, $\mathbf{\bar{c}}^{g}$ is an $nm^{*} \times 1$ vector, and $\mathbf{A}^{g}$, $\mathbf{B}^{g}$, $\lbrace \mathbf{C}_{k}^{g} \rbrace_{k=1}^{K}$ and $\mathbf{C}_{e}^{g}$ are $nm^{*} \times nm^{*}$ matrices of which all come from $V_{0}(\cdot)$.\footnote{Having $nm^{*}\times nm^{*}$ matrices, $\mathbf{A}^{g}$, $\mathbf{B}^{g}$, $\left\{ \mathbf{C}_{k}^{g}\right\}_{k=1}^{K}$ and $\mathbf{C}_{e}^{g}$ for $l_{m}^{*}\otimes \mathbf{g}_{t}^{*}$ gives a better connection between a theoretical model specification and regularity conditions for asymptotic analysis. For example, we need to impose restrictions on the matrix governing spatial dependence for spatial stability. Those assumptions are based on the uniform boundedness of a square matrix in its row sum norm.}

\paragraph*{Stage 2: Local agents' decision-making.}

The local agent $i$'s lifetime problem given $\mathbf{y}_{-i,t}^{*}$ (other local agents' activities) and $\mathbf{z}_{t}^{A}$ is 
\begin{equation}
V_{i}\left(\mathbf{z}_{t}^{A}\right) = \max_{y_{i,t}}\left\lbrace 
U_{i}\left( y_{i,t},\mathbf{y}_{-i,t}^{*}, g_{i,t}^{*}, \mathbf{Y}_{t-1}, x_{i,t}, u_{i,t}\right)
+\delta \mathbb{E}_{t,2}\left( V_{i}\left( \mathbf{z}_{t+1}^{A}\right) \right)
\right\rbrace.  \label{m2-2-flpv}
\end{equation}
Observe $\mathbf{z}_{t+1}^{A}$ includes $\mathbf{g}_{t+1}$. It implies each local agent decides $y_{i,t}$ given $g_{i,t}^{*}$ but expects $\mathbf{g}_{t+1}^{*}$ based on $\mathcal{B}_{t,2}$.

This structure leads to a directional impact from the allocator to local agents, but the forward-looking agent assumption generates a channel through which the local agents' decisions affect the allocator's next-period grant allocations. Hence, this specification positions the allocator not only as an allocator but also as a strategic leader aiming to guide local agents' activities toward long-term social welfare represented by $V_{0}(\mathbf{z}_{t}^{L})$, thus highlighting its role in shaping long-term policy outcomes. This hierarchical structure and the strategic intergovernmental interactions modeled here provide a framework that can be generalized to a broad range of contexts where economic agents engage in multiple activities under a resource allocation from a central authority.

\subsection{Uniqueness of the MPNE}

Before elaborating on the uniqueness conditions for the MPNE, let us define two parameter vectors: \( \theta_{P} \) and \( \theta_{E} \). The vector \( \theta_{P} \) contains the payoff parameters, including \( \Lambda \), \( \boldsymbol{\rho} \), \( P \), \( \Psi \), \( \phi \), and \( \Pi \), while \( \theta_{E} \) is a parameter vector governing the processes of \( \mathbf{x}_{t,k}  \) for $k = 1,\cdots,K$ as described in equation \eqref{processx}.

Given that the agents' payoff functions, (\ref{m2-2-fpayoff}) and (\ref{m2-2-lpayoff}), are LQ in activities, their corresponding value functions will also be LQ:
\begin{equation*}
    \begin{split}
        V_{i}(\mathbf{z}_{t}^{A}) & = \mathbf{z}_{t}^{A\prime} \mathbb{Q}_{i} \mathbf{z}_{t}^{A} + \mathbf{z}_{t}^{A\prime}\mathbb{L}_{i} + c_{i} \text{ for }i=1,\cdots,n,\text{ and} \\
        V_{0}(\mathbf{z}_{t}^{R}) & = \mathbf{z}_{t}^{R\prime}\mathbb{Q}_{0} \mathbf{z}_{t}^{R} + \mathbf{z}_{t}^{R\prime}\mathbb{L}_{0} + c_{0}.
    \end{split}
\end{equation*}
The detailed forms of $\lbrace \mathbb{Q}_{i} \rbrace_{i=0}^{n}$ and $\lbrace \mathbb{L}_{i} \rbrace_{i=0}^{n}$ can be found in \eqref{QF_form}, \eqref{LF_form}, \eqref{QL_form}, and \eqref{LL_form} of Appendix B.\footnote{We do not provide the forms of constants $\lbrace c_{i} \rbrace_{i=0}^{n}$ since they are well-defined if $\lbrace \mathbb{Q}_{i}, \mathbb{L}_{i} \rbrace_{i=0}^{n}$ and the variance parameters are well-defined. Those constant terms do not affect the agents' optimal decisions.} The key elements of the optimal decisions, $\mathbb{L}^{y}$ and $\mathbb{L}^{g}$, come from the quadratic components, $\lbrace \mathbb{Q}_{i} \rbrace_{i=0}^{n}$, while the linear components, $\lbrace \mathbb{L}_{i} \rbrace_{i=0}^{n}$, form the intercept terms, $\overline{\mathbf{c}}^{y}$ and $\overline{\mathbf{c}}^{g}$.

The MPNE can be characterized by a set of quadruples \(\lbrace (\theta_{P}, \theta_{E}, \delta, \mathbf{W}) \rbrace \) (specific levels of $\mathbf{z}_{t}^{A}$ and $\mathbf{z}_{t}^{R}$ do not affect the MPNE uniqueness).\footnote{Since Assumption \ref{asmproctau} (ii) implies $\mathbb{E}_{t,1}(\boldsymbol{\tau}_{t+1}) = \boldsymbol{\tau}$, $\beta$ does not affect the MPNE uniqueness condition.} We will specify a set \( \mathcal{M} \) that contains such quadruples. That is, if \( (\theta_{P}, \theta_{E}, \delta, \mathbf{W}) \in \mathcal{M} \), \( \mathbf{Y}_{t}^{*} \) and \( \mathbf{g}_{t}^{*} \) will be unique. The key feature of $\mathcal{M}$ is that it restricts $(\theta_{P}, \theta_{E}, \delta, \mathbf{W})$ to ensure that any exogenous change to the economy fades over space and over time. Derivations can be found in Appendix B.

$\mathcal{M}$ consists of two parts: invertibility ($\mathcal{I}$) and stability ($\mathcal{S}$), i.e., $\mathcal{M} = \mathcal{I} \cap \mathcal{S}$. $\mathcal{I}$ and $\mathcal{S}$ play exactly the roles of the two convergence requirements for the infinite continued-fraction expansion when there is a single local agent with his/her single activity (see Sec.1.3.3 of the Supplement). $\mathcal{I}$ ensures that each denominator in the nested fractions stays strictly positive, while $\mathcal{S}$ guarantees that the feedback terms at each deeper level are geometrically damped, so that successive corrections shrink. Together, these conditions imply that the full, untruncated nesting converges to a unique, finite solution.

\textbf{Invertibility condition ($\mathcal{I}$)}. The first-order conditions of the local agents' problems \eqref{m2-2-flpv} can be represented by
\begin{equation*}
    \mathbf{R}_{1:n}\operatorname{vec}(\mathbf{Y}_{t}^{*}) = \ddot{\overline{\mathbf{c}}}^{y} + \ddot{\mathbb{L}}^{y}\mathbf{z}_{t}^{A},
    \text{ for some }\ddot{\overline{\mathbf{c}}}^{y}\text{ and }\ddot{\mathbb{L}}^{y},
\end{equation*}
where $\mathbf{R}_{1:n} = \mathbf{S} - \delta\mathbf{Q}_{1:n}$ with $\mathbf{S} = \left( P + \Psi \right) \otimes \mathbf{I}_{n} - \left( \Lambda^{\prime} \otimes \mathbf{W} \right)$, and $\mathbf{Q}_{1:n}$ is an $nm^{*} \times nm^{*}$ matrix showing the expected future influences among activities (the form of $\mathbf{Q}_{1:n}$ can be found in \eqref{BQF_form}). Hence, invertibility of $\mathbf{R}_{1:n}$ is required for uniqueness of $\mathbf{Y}_{t}^{*}$. When $\mathbf{R}_{1:n}$ is invertible, $\overline{\mathbf{c}}^{y} = \mathbf{R}_{1:n}^{-1} \ddot{\overline{\mathbf{c}}}^{y}$ and $\mathbb{L}^{y} = \mathbf{R}_{1:n}^{-1} \ddot{\mathbb{L}}^{y}$. Let 
\begin{equation*}
    \mathbf{T}_{1:n} = (P + \Psi)^{-1}\Lambda^\prime \otimes \mathbf{W} + \delta\left((P + \Psi)^{-1} \otimes \mathbf{I}_{n} \right)\mathbf{Q}_{1:n}
\end{equation*}
to have $\mathbf{R}_{1:n} = \left((P + \Psi) \otimes \mathbf{I}_{n} \right)\left(\mathbf{I}_{nm^{*}} - \mathbf{T}_{1:n} \right)$. Then, $\Vert \mathbf{T}_{1:n} \Vert_{2} < 1$ is a sufficient condition for invertibility of $\mathbf{R}_{1:n}$. This condition implies that an exogenous change from $\ddot{\overline{\mathbf{c}}}^{y}$ or $\ddot{\mathbb{L}}^{y}$ in the second stage eventually disappears over space (relevant results in a static model can be found in Lemma 4 of \cite{Chenetal2018}). Under this condition, $\mathbf{R}_{1:n}^{-1} = \left(\sum_{k=0}^{\infty}\mathbf{T}_{1:n}^{k} \right)\cdot (P + \Psi)^{-1}\otimes \mathbf{I}_{n}$.

Similarly, the first-order conditions of the allocator's problem \eqref{leadervalue} can form the following equation:
\begin{equation*}
    \left(\mathbf{I}_{m^{*}} \otimes \mathbf{R}_{0}  \right)\left(l_{m^{*}} \otimes \mathbf{g}_{t}^{*} \right) = \ddot{\overline{\mathbf{c}}}^{g} + \ddot{\mathbb{L}}^{g}\mathbf{z}_{t}^{R}\text{ for some }\ddot{\overline{\mathbf{c}}}^{g}\text{ and }\ddot{\mathbb{L}}^{g},
\end{equation*}
where $\mathbf{R}_{0} = \mathbf{I}_{n} - \mathbf{T}_{0}$ and the form of $\mathbf{T}_{0}$ can be found in \eqref{T0_form}. Hence, invertibility of $\mathbf{R}_{0}$ is required for uniqueness of $\mathbf{g}_{t}^{*}$. If $\mathbf{R}_{0}$ is invertible,  $\overline{\mathbf{c}}^{g} =  \left(\mathbf{I}_{m^{*}} \otimes \mathbf{R}_{0}^{-1}  \right)\ddot{\overline{\mathbf{c}}}^{g}$ and $\mathbb{L}^{g} = \left(\mathbf{I}_{m^{*}} \otimes \mathbf{R}_{0}^{-1}  \right)\ddot{\mathbb{L}}^{g}$. Then, $\Vert \mathbf{T}_{0} \Vert_{2} < 1$ is a sufficient condition for invertibility of $\mathbf{R}_{0}$. Under this condition, $\mathbf{R}_{0}^{-1} = \sum_{k=0}^{\infty}\mathbf{T}_{0}^{k}$. Hence, this condition also means that an exogenous change from $\ddot{\overline{\mathbf{c}}}^{g}$ or $\ddot{\mathbb{L}}^{g}$ in the first stage fades over space. In consequence, we establish the invertibility condition:
\begin{equation*}
    \mathcal{I} = \left\lbrace (\theta_{P}, \theta_{E}, \delta, \mathbf{W}):\Vert \mathbf{T}_{1:n} \Vert_{2} < 1 \text{ and }\Vert \mathbf{T}_{0} \Vert_{2} < 1  \right\rbrace.
\end{equation*}
Remark \ref{remark_invertibility} in Appendix B delves into the invertibility condition, particularly when \(\delta = 0\) (i.e., myopic agents).

\textbf{Stability condition ($\mathcal{S}$)}. The additional condition for the uniqueness of the MPNE, stemming from the model's dynamic nature, pertains to dynamic stability. Intuitively, each agent's optimized lifetime payoff should be a bounded infinite discounted summation to have its recursive structure. By its recursive nature, note that \(\mathbb{Q}_{i}\) (for $i = 0,1,\cdots,n$) is the unique solution to the discrete \textit{Lyapunov equation} under some stability conditions: \(\mathbb{Q}_{0} = \widetilde{\mathbb{Q}}_{0} + \delta \mathbb{A}_{0}^\prime \mathbb{Q}_{0} \mathbb{A}_{0}\) and \(\mathbb{Q}_{i} = \widetilde{\mathbb{Q}}_{i} + \delta \mathbb{A}_{1:n}^\prime \mathbb{Q}_{i} \mathbb{A}_{1:n}\) for \(i = 1, \dots, n\), where \(\widetilde{\mathbb{Q}}_{i}\) (for \(i = 0,1, \dots, n\)) denotes the per-period payoff components. Here, \(\mathbb{A}_{0}\) and \(\mathbb{A}_{1:n}\) are transition matrices governing the dynamic evolution of $\mathbf{z}_{t}^{R}$ and $\mathbf{z}_{t}^{A}$, respectively (see \eqref{AF_form} and \eqref{AL_form} for the detailed forms of $\mathbb{A}_{1:n}$ and $\mathbb{A}_{0}$), i.e., \(\mathbb{E}_{t,1}(\mathbf{z}_{t+1}^{R}) = \mathbb{c}_{0} + \mathbb{A}_{0}\mathbf{z}_{t}^{R}\) and \(\mathbb{E}_{t,2}(\mathbf{z}_{t+1}^{A}) = \mathbb{c}_{1:n} + \mathbb{A}_{1:n}\mathbf{z}_{t}^{A}\) for some constant vectors \(\mathbb{c}_{0}\) and \(\mathbb{c}_{1:n}\). For stability, we should have $\Vert \mathbb{A}_{0} \Vert_{2} < 1$ and $\Vert \mathbb{A}_{1:n} \Vert_{2} < 1$. Let
\begin{equation*}
    \mathcal{S} = \left\lbrace (\theta_{P}, \theta_{E}, \delta, \mathbf{W}): \Vert \mathbb{A}_{1:n} \Vert_{2} < 1 \text{ and }\Vert \mathbb{A}_{0} \Vert_{2} < 1 \right\rbrace.
\end{equation*}
If \((\theta_{P}, \theta_{E}, \delta, \mathbf{W}) \in \mathcal{S}\), each component in \(\lbrace \mathbb{Q}_{i}, \mathbb{L}_{i} \rbrace_{i=0}^{n}\) has a unique infinite discounted summation representation (e.g., \(\mathbb{Q}_{i} = \sum_{k=0}^{\infty} \delta^{k} (\mathbb{A}_{1:n}^\prime)^k \widetilde{\mathbb{Q}}_{i} \mathbb{A}_{1:n}^{k}\)).\footnote{To obtain the unique solutions to \(\lbrace \mathbb{Q}_{i} \rbrace_{i=0}^{n}\), sufficient conditions are \(\Vert \sqrt{\delta} \mathbb{A}_{0} \Vert_{2} < 1\) and \(\Vert \sqrt{\delta} \mathbb{A}_{1:n} \Vert_{2} < 1\). The same argument applies to \(\mathbb{L}_{i}\) (for \(i = 0,1, \dots, n\)), but stricter conditions in \(\mathcal{S}\) are introduced for the unique representations of \(\lbrace \mathbb{L}_{i} \rbrace_{i=0}^{n}\) by achieving the well-definedness of \(\sum_{k=0}^{\infty}\mathbb{A}_{0}^{k}\) and \(\sum_{k=0}^{\infty}\mathbb{A}_{1:n}^{k}\).}

\begin{theorem}[Uniqueness of the MPNE]
    \label{uniquempne}
    Assume $P + \Psi$ is diagonally dominant. If $(\theta_{P}, \theta_{E}, \delta, \mathbf{W}) \in \mathcal{M} = \mathcal{I} \bigcap \mathcal{S}$, we obtain (i) the unique value functions, $\lbrace V_{i}(\cdot) \rbrace_{i=0}^{n}$, and (ii) the unique optimal activities, $\overline{\mathbf{c}}^{y}$, $\mathbb{L}^{y}$, $\overline{\mathbf{c}}^{g}$, and $\mathbb{L}^{g}$.
\end{theorem}

The positive definiteness of \(P + \Psi\) ensures the strict concavity of \(U_{i}(\cdot)\), which is essential for the interpretations detailed in Sec.1.2 of the supplement. While the spatial-dynamic influences manifested by our model share similar statistical features with those of conventional models in terms of weak dependence, the model's dependence structure is more intricate. This complexity primarily arises from (i) the interplay between the local agents' non-cooperative behaviors and the allocator's benevolent actions and (ii) their forward-looking behaviors.

\subsection{The derived equilibrium system}

Combining equations (\ref{constraint_followeract}) and (\ref{m2-2-laction}), we have a system of equations with \(m = m^{*} + 1\) endogenous variables:
\begin{equation}
\begin{split}
    \begin{bmatrix}
    \mathbf{R}_{1:n}   & - (\phi \otimes \mathbf{I}_{n}) \\
    \mathbf{0} & \mathbf{R}_{0}
    \end{bmatrix}
    \begin{pmatrix}
        \operatorname{vec}(\mathbf{Y}_{t}^{*}) \\
        \mathbf{g}_{t}^{*}
    \end{pmatrix}
    = & \begin{bmatrix}
        (P \otimes \mathbf{I}_{n}) + (\boldsymbol{\rho}^\prime \otimes \mathbf{W}) \\
        \mathbf{R}_{0}A^{g}
    \end{bmatrix}
    \operatorname{vec}(\mathbf{Y}_{t-1}) \\
    + & \sum_{k=1}^{K}
    \begin{bmatrix}
        \mathbf{I}_{nm^{*}} + \delta \mathbf{L}_{1:n,k}(\mathbf{I}_{m^{*}} \otimes \mathbf{A}_{k}^{x}) \\
        \mathbf{R}_{0}C_{k}^{g}
    \end{bmatrix}
    \boldsymbol{\Pi}_{k}\mathbf{x}_{t,k} 
    + \begin{bmatrix}
        \mathbf{0} \\
        \mathbf{I}_{n}
    \end{bmatrix}
    \mathbf{X}_{t}^{\tau}\beta 
    + \mathbf{c}
    + \widetilde{\boldsymbol{\alpha}}_{t} \otimes l_{n}
    + \mathbf{v}_{t}.
\label{m2-2-fd1}
\end{split}
\end{equation}
Here, 
\begin{itemize}
   \item \(\mathbf{c} = \begin{pmatrix}
    \mathbf{c}^{y} \\
    c^{g}
\end{pmatrix}\) with \(\mathbf{c}^{y} = \mathbf{R}_{1:n}\bar{\mathbf{c}}^{y} + \operatorname{vec}\left( \boldsymbol{\eta} \right)\) and \(c^{g} = \mathbf{R}_{0}\bar{c}^{g}\), \(\widetilde{\boldsymbol{\alpha}}_{t} = \begin{pmatrix}
    \widetilde{\alpha}_{t} \\
    \alpha_{t}^{\tau}
\end{pmatrix}\), and \(\mathbf{v}_{t} = \begin{bmatrix}
    \mathbf{I}_{nm^{*}} \\
    \mathbf{R}_{0} C_{e}^{g}
\end{bmatrix}\operatorname{vec}(\mathbf{E}_{t}) + \begin{bmatrix}
    \mathbf{0} \\
    \mathbf{I}_{n}
\end{bmatrix}\mathbf{e}_{t}^{\tau}\).
\end{itemize}
Matrices \(A^{g}\), \(C_{k}^{g}\) for \(k=1,\dots,K\), \(C_{e}^{g}\), and \(\bar{c}^{g}\) satisfy \(\mathbf{A}^{g} = l_{m^{*}} \otimes A^{g}\), \(\mathbf{C}_{k}^{g} = l_{m^{*}} \otimes C_{k}^{g}\) for \(k=1,\dots,K\), \(\mathbf{C}_{e}^{g} = l_{m^{*}} \otimes C_{e}^{g}\), and \(\bar{\mathbf{c}}^{g} = l_{m^{*}} \otimes \bar{c}^{g}\) in \eqref{m2-2-laction}. In the estimation procedure, \(\mathbf{c}\) and \(\widetilde{\boldsymbol{\alpha}}_{t}\) will be treated as fixed-effect components. Note that \(\mathbf{v}_{t}\) is a vector of the model's errors, which is a linear combination of \(\operatorname{vec}(\mathbf{E}_{t})\) and \(\mathbf{e}_{t}^{\tau}\).

The system of equations \eqref{m2-2-fd1} can be viewed as a structural vector autoregressive (SVAR) model. A zero block, which represents the lack of contemporaneous influence of $\mathbf{Y}_{t}^{*}$ on $\mathbf{g}_{t}^{*}$, serves as an exclusion restriction originating from the hierarchical decision-making structure. Then, the first part of equation \eqref{m2-2-fd1} characterizes a vector of the local agents' best responses, which is an extension of \eqref{m2-2-example}. The term \(\delta \mathbf{L}_{1:n,k}\left(\mathbf{I}_{m^{*}} \otimes \mathbf{A}_{k}^{x} \right)\), reflecting the influences from the expected future characteristics, includes the core part \(\mathbf{L}_{1:n,k}\) discounted by \(\mathbf{A}_{k}^{x} = \frac{\partial \mathbf{x}_{t+1,k}}{\partial \mathbf{x}_{t,k}^{\prime}}\) from Assumption \ref{asmproch} and the time discounting factor \(\delta\). The second part of equation \eqref{m2-2-fd1} is a vector of the allocator's optimal grants. Its detailed structure can be decomposed into:
\begin{equation}
    \mathbf{g}_{t}^* = \underset{\equiv \mathbf{g}_{t}^{L} \text{ (Responsive part)}}{\underbrace{A^g \text{vec}(\mathbf{Y}_{t-1}) + \sum_{k=1}^{K}C_{k}^{g}\boldsymbol{\Pi}_{k}\mathbf{x}_{t,k}}}  + C_{e}^{g}\text{vec}(\mathbf{E}_{t}) + \underset{\equiv \mathbf{g}_{t}^{A} \text{ (Autonomous transfer part)}}{\underbrace{\mathbf{R}_{0}^{-1}\left( \mathbf{X}_{t}^\tau \beta + c^g + \alpha_{t}^\tau \otimes l_{n} \right)}}  + \mathbf{R}_{0}^{-1}\mathbf{e}_{t}^\tau.
    \label{optimal_grant}
\end{equation}
Here, $\mathbf{g}_{t}^{L}$ represents the part of reacting to the local agents' past activities and their observable characteristics. When $\phi = \mathbf{0}$, $\mathbf{g}_{t}^L = \mathbf{0}$ since $A^g = \mathbf{0}$ and $C_{1}^{g} = \cdots C_{K}^{g} = C_{e}^{g} = \mathbf{0}$ (see Appendix B). On the other hand, $\mathbf{g}_{t}^{A}$ can illustrate the systematic component of the autonomous transfers. After identifying the key parameters, one can evaluate the contribution of each part. For example, if the variation of $\mathbf{g}_{t}^L$ is a major part of the variation of $\mathbf{g}_{t}^*$, the allocator is responsive to the local agents' decisions and characteristics. From this view, another exclusion restriction is evident on the right-hand side of \eqref{m2-2-fd1}, where the temporal economic indicators ($\mathbf{X}_{t}^{\tau}$) — denoting the allocator's autonomous transfers — do not directly influence the local agents' activities.

\paragraph*{Examples.}

\begin{figure}[!htbp]
    \centering
    \caption{An example of $\mathbf{W}$}
    \label{networkexample}
    \begin{tikzpicture}[baseline={(current bounding box.center)}]
  \node[circle, draw=black, fill=blue!10, minimum size=0.6cm, inner sep=1pt] (n1) at (0,1) {1};
  \node[circle, draw=black, fill=blue!10, minimum size=0.6cm, inner sep=1pt] (n2) at (-0.9,0) {2};
  \node[circle, draw=black, fill=blue!10, minimum size=0.6cm, inner sep=1pt] (n3) at (0.9,0) {3};

  \draw[thick] (n1) -- (n2);
  \draw[thick] (n1) -- (n3);
\end{tikzpicture}
\hspace{1.5cm}
\raisebox{0.5ex}{$
\mathbf{W} =
\begin{bmatrix}
0 & 0.5 & 0.5 \\
1 & 0 & 0 \\
1 & 0 & 0
\end{bmatrix}
$}
\end{figure}

\begin{table}[htbp]
\caption{Equilibrium activities}
\label{equilactexample}
\begin{center}
\footnotesize
\begin{tabular}{l|c|ccccccccc}
\hline																					
	&	Case 0	&	Case1.1	&	Case 1.2	&	Case 1.3	&	Case 2.1	&	Case 2.2	&	Case 2.3	&	Case 3.1	&	Case 3.2	&	Case 3.3	\\
\hline																					
$y_{1,t,1}$	&	3.00	&	4.29	&	4.72	&	5.08	&	5.56	&	5.95	&	12.29	&	3.78	&	4.24	&	4.21	\\
$y_{2,t,1}$	&	3.00	&	4.29	&	4.69	&	5.03	&	5.56	&	5.94	&	11.27	&	3.78	&	4.23	&	4.21	\\
$y_{3,t,1}$	&	3.00	&	4.29	&	4.69	&	5.03	&	5.56	&	5.94	&	11.21	&	3.78	&	4.23	&	4.22	\\
$y_{1,t,2}$	&	2.00	&	2.86	&	2.86	&	3.65	&	4.44	&	4.55	&	11.16	&	1.78	&	1.65	&	2.19	\\
$y_{2,t,2}$	&	2.00	&	2.86	&	2.86	&	3.60	&	4.44	&	4.56	&	10.15	&	1.78	&	1.65	&	2.18	\\
$y_{3,t,2}$	&	2.00	&	2.86	&	2.86	&	3.59	&	4.44	&	4.56	&	10.08	&	1.78	&	1.65	&	2.18	\\
$g_{1,t}$	&	1.00	&	1.00	&	1.55	&	2.61	&	1.00	&	1.27	&	10.20	&	1.00	&	1.52	&	1.91	\\
$g_{2,t}$	&	1.00	&	1.00	&	1.39	&	2.41	&	1.00	&	1.21	&	7.32	&	1.00	&	1.41	&	1.87	\\
$g_{3,t}$	&	1.00	&	1.00	&	1.39	&	2.41	&	1.00	&	1.21	&	7.01	&	1.00	&	1.41	&	1.88	\\
\hline																					
											
\end{tabular}
\end{center}
\footnotesize
In this experiment, all local agents' given characteristics are time-invariant and identical across all entities (Case 0). Hence, the resulting activities are also time-invariant. Below are the descriptions of different cases concerning the parameters in $\boldsymbol{\rho}$. Additionally, three subcases of $\phi$ are considered: (i) subcase 1 with $\phi = (0, 0)^\prime$, representing no responsive intervention, (ii) subcase 2 with $\phi = (0.2, 0)^\prime$, indicating responsive intervention on the first activity, and (iii) subcase 3 with $\phi = (0.2, 0.2)^\prime$ illustrating responsive intervention on both activities.

\begin{itemize}
\item Case 1 (Positive spillovers within activities): $\rho_{11} = \rho_{22} = 0.3$ and $\rho_{12} = \rho_{21} = 0$.
\item Case 2 (Positive spillovers within activities + Strategic complementarity): $\rho_{11} = \rho_{22} = 0.3$ and $\rho_{12} = \rho_{21} = 0.2$.
\item Case 3 (Positive spillovers within activities + Strategic substitutability): $\rho_{11} = \rho_{22} = 0.3$ and $\rho_{12} = \rho_{21} = -0.2$.
\end{itemize}
\end{table}

To understand the equilibrium activities, we generate data under specific key parameter values. The model's essential parameters---\(\Lambda\), \(\boldsymbol{\rho}\), and \(\phi\)---characterize the local agents' interactions and the allocator's interventions. Since the results from various combinations of \(\Lambda\) and \(\phi\) are similar to those of \(\boldsymbol{\rho}\) and \(\phi\), we report activity values across \(\boldsymbol{\rho}\) and \(\phi\). Positive spillovers within activities result in local agent 1 exerting the highest effort in both activities, with the allocator allocating the largest grant to this agent. When activities are strategic complements, local agent 1 increases efforts more than in other scenarios, prompting the allocator to raise grant levels for all local agents, particularly focusing on local agent 1. Responsive interventions by the allocator amplify efforts in both activities across all agents. Conversely, if activities are strategic substitutes, all agents reduce their efforts compared to the case of strategic complements. Without any responsive intervention (\(\phi = (0, 0)^\prime\)), the allocator issues uniform autonomous transfers to all local agents. If intervention targets only the first activity, efforts increase for that activity but decrease for the second. When responsive interventions target both activities, efforts increase for both, although the increments are less pronounced than in the case of strategic complements.

\section{Equilibrium measures}
\label{sec_eqmeasure}

Informed by equation \eqref{m2-2-fd1}, we define equilibrium measures for the model's interpretation. The standard errors of these equilibrium measures can be evaluated using the delta method.

\subsection{Short-run impacts (Marginal effects)}
\label{sec_me}

Leveraging ideas from \cite{Katz1953, Bonacich1987, LeSagePace2008, Blochetal2023} and equation (\ref{m2-2-fd1}), the marginal effects of \(\mathbf{x}_{t,k}\) on \(\mathbf{Y}_{t}^{*}\) elucidate the short-run consequences of a change of local agent's characteristic. We derive two summary measures for these effects:
\begin{equation*}
    \begin{split}
        \mathsf{ADI}_{x_{k} \to y_{l}} & = \frac{1}{n}\sum_{i=1}^{n}\frac{\partial y_{i,t,l}^{*}}{\partial x_{i,t,k}},\text{ and }\mathsf{ATSI}_{x_{k} \to y_{l}}
        = \frac{1}{n}\sum_{i=1}^{n}\sum_{j=1}^{n}\frac{\partial y_{j,t,l}^{*}}{\partial x_{i,t,k}},
    \end{split}
\end{equation*}
where \( \mathsf{ADI} \) denotes the average direct impact, and \( \mathsf{ATSI} \) encapsulates the average total spillover impact, highlighting the cumulative effects on all local agents' decisions. The differential, $\mathsf{ATSI} - \mathsf{ADI}$, thus portrays the net externalities. The aforementioned measures stem from the derivative:
\begin{equation*}
    \begin{split}
       \frac{\partial \operatorname{vec}(\mathbf{Y}_{t}^{*})}{\partial \mathbf{x}_{t,k}^{\prime}} & =
       \underset{\text{Direct channel}}{\underbrace{\left(
\mathbf{R}_{1:n} \right)^{-1} \left(\mathbf{I}_{nm^{*}} + \delta \mathbf{L}_{1:n,k}(\mathbf{I}_{m^{*}} \otimes \mathbf{A}_{k}^{x}) \right)\boldsymbol{\Pi}_{k}}}
+ \underset{\text{Indirect channel through }\mathbf{g}_{t}^{*}}{\underbrace{\left(\mathbf{R}_{1:n} \right)^{-1}\left( \phi \otimes \mathbf{I}_{n}\right) C_{k}^{g}\boldsymbol{\Pi}_{k}}}.
    \end{split}
\end{equation*}

Note that each parameter $\phi_{l}$ ($l = 1,\cdots,m^*$) does not represent the marginal effect of $g_{i,t}$ on $y_{i,t,l}$. For this, we can also define the marginal effects of $\mathbf{g}_{t}^*$ on $\mathbf{Y}_{t}^*$ based on $\frac{\partial \operatorname{vec}(\mathbf{Y}_{t}^{*})}{\partial \mathbf{g}_{t}^{*\prime}} = \left(\mathbf{R}_{1:n} \right)^{-1}\left( \phi \otimes \mathbf{I}_{n}\right)$: $\mathsf{ADI}_{g \to y_{l}} = \frac{1}{n}\sum_{i=1}^{n}\frac{\partial y_{i,t,l}^{*}}{\partial g_{i,t}^*}$ and $\mathsf{ATSI}_{g \to y_{l}} = \frac{1}{n}\sum_{i=1}^{n}\sum_{j=1}^{n}\frac{\partial y_{j,t,l}^{*}}{\partial g_{i,t}^*}$.

\subsection{Long-run impacts (Cumulative spatial dynamic effects)}
\label{sec_drf}

Since the model described by equation \eqref{m2-2-fd1} can be regarded as an SVAR model, we can evaluate the effect of $x_{i,t,k}$ on $y_{j,t+h,l}$ for $h = 0,1,2\cdots$, which corresponds to the impulse response function:
\begin{equation*}
    \begin{split}
        \mathsf{ADDI}_{x_{k} \to y_{l}}(h) = & \frac{1}{n}\sum_{i=1}^{n}\frac{\partial y_{i,t+h,l}^{*}}{\partial x_{i,t,k}}\text{, and }
        \mathsf{ADTSI}_{x_{k} \to y_{l}}(h) 
= \frac{1}{n}\sum_{i=1}^{n}\sum_{j=1}^{n}\frac{\partial y_{j,t+h,l}^{*}}{\partial x_{i,t,k}}.
    \end{split}
\end{equation*}
Here, $\mathsf{ADDI}(h)$ describes the average direct impact of an intervention on the \( h \)-period ahead local agent's activity, and $\mathsf{ADTSI}(h)$ shows the average total spillovers from an intervention on the \( h \)-period ahead local agent's activity. Using the measures above, we can define the cumulative dynamic effects (long-run impacts) to capture the full effects of changing a characteristic ($x_{i,t,k}$) over time: $\mathsf{CADDI}_{x_{k} \to y_{l}} = \sum_{h=0}^{\infty}\mathsf{ADDI}_{x_{k} \to y_{l}}(h)$ and $\mathsf{CADTSI}_{x_{k} \to y_{l}} = \sum_{h=0}^{\infty}\mathsf{ADTSI}_{x_{k} \to y_{l}}(h)$.

\section{Statistical inference}
\label{econometricmodel}

\setcounter{theorem}{0}
\setcounter{assumption}{0}
\setcounter{remark}{0}
\setcounter{definition}{0}
\renewcommand{\thetheorem}{4.\arabic{theorem}}
\renewcommand{\theassumption}{4.\arabic{assumption}}
\renewcommand{\theremark}{4.\arabic{remark}}
\renewcommand{\thedefinition}{4.\arabic{definition}}%

\subsection{Log-likelihood function}

This section constructs a log-likelihood function for estimation based on equation (\ref{m2-2-fd1}). Henceforth, we omit the superscript `\(*\)' in \(\mathbf{Y}_{t}^{*}\) and \(\mathbf{g}_{t}^{*}\) because we assume that the observed activities are optimally realized. Following Assumptions \ref{asmproch} and \ref{asmproctau}, we assume errors with a zero mean and finite variances without specifying their distributions. Therefore, we apply a quasi-maximum likelihood (QML) estimation approach. For the fixed-effect parameters, we adopt the direct estimation approach. These approaches produce robust statistical inferences concerning error distributions, agents' unobserved characteristics, and time-specific components.

The main purpose is to estimate the main structural parameters, denoted by \( \theta \). The parameter vector \( \theta \) includes: (i) local agent's payoff parameters (\( \theta_{P} \)), (ii) coefficients $\beta$ for \( \mathbf{X}_{t}^\tau \) in the allocator's payoff, and (iii) variance parameters for shocks (\( \Sigma, \sigma^{2} \)). For the theoretical analysis presented in the main draft, we assume that the parameters \( \theta_{E} \) and $\sigma_{1}^{2}, \cdots, \sigma_{K}^{2}$ governing the processes of \(  \mathbf{X}_{t}  \) are known. This assumption is made for notational simplicity.\footnote{For practical uses (application), we consider utilizing the joint likelihood of \( \left\{ \mathbf{Y}_{t}, \mathbf{g}_{t}, \mathbf{X}_{t} \right\} \) in estimating $\theta$, $\theta_{E}$, and $\sigma_{1}^{2}, \cdots, \sigma_{K}^{2}$. More details are provided in the supplement file (Section 3.2).}
From this section onward, we use the subscript "\(0\)" to denote the true parameter values (e.g., \( \theta_{P,0} \) would signify the true value of \( \theta_{P} \)). Note that the Riccati matrices from the value functions rely on the payoff parameters $\theta_{P}$. The argument $\theta_{P}$ of the Riccati matrices, for example $\mathbf{R}_{1:n}(\theta_{P})$, represents $\mathbf{R}_{1:n}$ evaluated at $\theta_{P}$. When $\theta_{P} = \theta_{P,0}$, we will omit its argument, e.g., $\mathbf{R}_{1:n} = \mathbf{R}_{1:n}(\theta_{P,0})$. Observe that $\mathbb{E}\left( \mathbf{v}_{t}\right) = \mathbf{0}$, and 
\begin{equation}
\mathbb{Var}(\mathbf{v}_{t}) \underset{\text{let}}{=} \boldsymbol{\Delta} =
\begin{bmatrix}
\Sigma_{0}\otimes \mathbf{I}_{n} & \left( \Sigma_{0} \otimes \mathbf{I}_{n}\right)
C_{e}^{g\prime} \mathbf{R}_{0}^{\prime} \\ 
\mathbf{R}_{0} C_{e}^{g} \left( \Sigma_{0} \otimes \mathbf{I}_{n}\right) & 
\mathbf{R}_{0} C_{e}^{g}\left( \Sigma_{0} \otimes \mathbf{I}_{n}\right)
C_{e}^{g\prime} \mathbf{R}_{0}^{\prime } + \sigma_{0}^{2}\mathbf{I}_{n}
\end{bmatrix}
.  \label{m3-3-varu}
\end{equation}
$\boldsymbol{\Delta}$ involves variance parameters $\left(\Sigma_{0}, \sigma_{0}^{2}\right)$ as well as $\theta_{P,0}$. For each tuple $(\theta_{P}, \Sigma, \sigma^{2})$, $\boldsymbol{\Delta}(\theta_{P}, \Sigma, \sigma^{2})$ denotes the variance of $\mathbf{v}_{t}$ evaluated at $(\theta_{P}, \Sigma, \sigma^{2})$.

Given \( \left(\mathbf{Y}_{t-1}, \mathbf{X}_{t}\right) \), the joint density of \( \left( \mathbf{Y}_{t}, \mathbf{g}_{t}\right) \) for \( t=1,\cdots,T \) is derived as follows: 
\begin{equation*}
f\left(\mathbf{Y}_{t}, \mathbf{g}_{t}|\mathbf{Y}_{t-1}, \mathbf{X}_{t}\right) = \left(2 \pi \right)^{-\frac{nm}{2}}
\left\vert \boldsymbol{\Delta} \right\vert^{-\frac{1}{2}}\exp \left( -\frac{1}{2}\mathbf{v}_{t}^\prime \boldsymbol{\Delta}^{-1}\mathbf{v}_{t}\right) \cdot
\vert \mathbf{R}_{1:n}\vert \cdot \vert\mathbf{R}_{0} \vert.
\end{equation*}
The additive individual and time effects can be concentrated out, implying that the concentrated log-likelihood is solely a function of \( \theta \). The concentrated log-likelihood function for \( \theta \) is denoted by \( \ell_{mnT}^{c}\left( \theta \right) \):
\begin{equation}
\begin{split}
    \ell_{mnT}^{c}\left( \theta \right) = &
- \frac{ mnT}{2}\ln
2\pi 
+ T\ln \vert \mathbf{R}_{1:n} \left( \theta_{P} \right) \vert 
+ T\ln \vert \mathbf{R}_{0} \left( \theta_{P} \right) \vert
- \frac{T}{2}\ln \vert \boldsymbol{\Delta} \left( \theta_{P}, \Sigma, \sigma^{2} \right)
\vert \\
& - \frac{1}{2}\sum_{t=1}^{T}\mathbf{\tilde{v}}_{t}\left( \theta \right)^{\prime} \mathbf{J}_{n,m} \boldsymbol{\Delta}^{-1}\left( \theta_{P}, \Sigma, \sigma^{2} \right) \mathbf{J}_{n,m}
\mathbf{\tilde{v}}_{t}\left( \theta \right),
\label{m3-3-cll_chap3}
\end{split}
\end{equation}
where $\mathbf{J}_{n,m} = \mathbf{I}_{m}\otimes \mathbf{J}_{n}$ with $\mathbf{J}_{n} = \mathbf{I}_{n} - \frac{1}{n}
l_{n}l_{n}^{\prime}$, $\mathbf{\tilde{v}}_{t}\left( \theta \right) = \mathbf{v}_{t}\left( \theta \right) - \frac{1}{T}\sum_{s=1}^{T}\mathbf{v}_{s}\left( \theta \right)$, 
\begin{equation*}
    \mathbf{v}_{t}\left( \theta \right) =
    \mathbf{R} \left( \theta_{P} \right) 
\begin{pmatrix}
\operatorname{vec}\left( \mathbf{Y}_{t}\right) \\ 
\mathbf{g}_{t}
\end{pmatrix}
-
\begin{bmatrix}
\left( P \otimes \mathbf{I}_{n}\right) + \left( \boldsymbol{\rho}^\prime \otimes
\mathbf{W} \right) \\ 
\mathbf{R}_{0}\left( \theta_{P} \right) A^{g}\left( \theta_{P} \right)
\end{bmatrix}
\operatorname{vec}\left( \mathbf{Y}_{t-1}\right)
-\sum_{k=1}^{K}
\begin{bmatrix}
\mathbf{I}_{nm} + \delta\mathbf{L}_{1:n, k}\left( \theta_{P} \right)(\mathbf{I}_{m^{*}} \otimes \mathbf{A}_{k}^{x}) \\ 
\mathbf{R}_{0} \left( \theta_{P} \right) C_{k}^{g}\left( \theta_{P} \right)
\end{bmatrix}
\boldsymbol{\Pi}_{k} \mathbf{x}_{t,k}-\begin{bmatrix}
            \mathbf{0} \\
            \mathbf{I}_{n}
        \end{bmatrix}
        \mathbf{X}_{t}^{\tau}\beta,
\end{equation*}
where $\mathbf{R}(\theta_{P}) = \begin{bmatrix}
    \mathbf{R}_{1:n}(\theta_{P}) & - \phi \otimes \mathbf{I}_{n} \\
    \mathbf{0} & \mathbf{R}_{0}(\theta_{P})
\end{bmatrix}$. Then, the QMLE is given by $\hat{\theta}_{nmT}=\arg \max_{\theta \in \Theta}\ell_{mnT}^{c}\left( \theta \right)$, where $\Theta$ denotes a parameter space of $\theta$.

\subsection{Regularity assumptions for large sample properties}

Some regularity conditions are required to derive consistency and asymptotic normality of $\hat{\theta}_{nmT}$. For those conditions, we define $\mathbf{A}_{\mathbf{z}} = \begin{bmatrix}
    \mathbf{R}_{1:n} & - (\phi_{0} \otimes \mathbf{I}_{n}) & -(\mathbf{I}_{nm^{*}} + \delta \mathbf{L}_{1:n, 1}(\mathbf{I}_{m^{*}} \otimes \mathbf{A}_{1}^{x}))\boldsymbol{\Pi}_{1,0}  & \cdots & - (\mathbf{I}_{nm^{*}} + \delta \mathbf{L}_{1:n, K}(\mathbf{I}_{m^{*}} \otimes \mathbf{A}_{K}^{x} ))\boldsymbol{\Pi}_{K,0} \\
    \mathbf{0} & \mathbf{R}_{0} & - \mathbf{R}_{0}C_{1}^{g}\boldsymbol{\Pi}_{1,0} & \cdots & - \mathbf{R}_{0}C_{K}^{g}\boldsymbol{\Pi}_{K,0} \\
    \mathbf{0} & \mathbf{0} & \mathbf{I}_{n} & \cdots & \mathbf{0} \\
    \vdots & \vdots & \vdots & \ddots & \vdots \\
    \mathbf{0} & \mathbf{0} & \mathbf{0} & \cdots & \mathbf{I}_{n}
    \end{bmatrix}$ and
\begin{equation*}
    \mathbf{B}_{\mathbf{z}} = 
\begin{bmatrix}
    (P_{0} \otimes \mathbf{I}_{n}) + (\boldsymbol{\rho}_{0}' \otimes \mathbf{W}) & \mathbf{0} & \mathbf{0} \\
    \mathbf{R}_{0}A^{g} & \mathbf{0} & \mathbf{0} \\
    \begin{matrix}
\mathbf{B}_{1}^{x} \\
\vdots \\
\mathbf{B}_{K}^{x}
    \end{matrix}
    & \begin{matrix}
        C_{1}^{x} \\
        \vdots \\
        C_{K}^{x}
    \end{matrix} 
    & \diag\left(\lbrace\mathbf{A}_{k}^{x} \rbrace \right)
\end{bmatrix},
\end{equation*}
where \( \diag\left(\lbrace\mathbf{A}_{k}^{x} \rbrace \right) \) denotes a diagonal block matrix with blocks \( \mathbf{A}_{k}^{x} \) for \( k = 1, \cdots, K \). Then, the system of equations, \eqref{processx} and \eqref{m2-2-fd1}, can be represented by $\mathbf{A}_{\mathbf{z}}\mathbf{z}_{t} = \mathbf{B}_{\mathbf{z}}\mathbf{z}_{t-1} + \text{other term}$, where $\mathbf{z}_{t} = \left(\operatorname{vec}(\mathbf{Y}_{t})^\prime, \mathbf{g}_{t}^\prime, \operatorname{vec}(\mathbf{X}_{t})^\prime \right)^\prime$.

\begin{assumption}
\label{asmw} 
$\mathbf{W}$ is time-invariant, strictly exogenous and uniformly bounded (UB) in row and column sums in absolute value.
\end{assumption}

\begin{assumption}
\label{asmeps} Let $e_{i,t} = \left(e_{1,i,t},\cdots,e_{m,i,t} \right)$ denote the $i$th row of $\mathbf{E}_{t}$. For all $i$ and $t$, we assume $e_{i,t}\sim i.i.d.\left(\mathbf{0}, \Sigma_{0}\right)$ and $e_{i,t}^{\tau} \sim i.i.d.\left(0,\sigma_{0}^{2}\right)$ with $\Sigma_{0}$ being positive definite, $\sigma_{0}^{2} > 0$, $\sup_{i,t}\max_{k,l,p,q}\mathbb{E}\left\vert e_{i,t,k}e_{i,t,l}e_{i,t,p}e_{i,t,q}\right\vert^{1 + \eta}<\infty$, $\sup_{i,t}\mathbb{E}\left\vert e_{i,t}^{\tau} \right\vert^{4 + \eta}< \infty$ for some $\eta > 0$. Also, we assume $\left\{e_{i,t}\right\}$ and $\left\{e_{i,t}^{\tau}\right\}$ are independent.
\end{assumption}

\begin{assumption}
\label{asmps} 
$\Theta$ is a compact parameter space with $\theta_{0} \in \operatorname{int}\left( \Theta \right)$. For $\theta \in \Theta$, $\boldsymbol{\Delta}(\theta)$ is positive definite.
\end{assumption}

\begin{assumption}
\label{asmreg1} 
(i) For some $\eta >0$, $\sup_{i,t}\max_{k}\mathbb{E}\left\vert x_{i,t,k}\right\vert^{4+\eta} < \infty$ and $\sup_{i,t}\max_{q}\mathbb{E}\left\vert x_{i,t,q}^{\tau} \right\vert^{4+\eta} < \infty$.

(ii) We assume that the fixed effects $\mathbf{c}_{0}$,
and $\left\{ \widetilde{\boldsymbol{\alpha}}_{t,0} \right\}$ are conditionally nonstochastic constants. Further, we assume $\sup_{T}\frac{1}{T}\sum_{t=1}^{T}\left\vert \alpha_{t,l,0}\right\vert^{4+\eta} < \infty$, and $\sup_{n}\frac{1}{n}\sum_{i=1}^{n}\left\vert c_{i,l,0}\right\vert ^{4+\eta}<\infty $ for $l=1,\cdots,m$, where $c_{i,l,0}$ denotes the individual fixed-effect for $i$'s $l$th activity and $\alpha_{t,l,0}$ represents the time fixed-effect for the period $t$ and $l$th activity.
\end{assumption}

\begin{assumption}
\label{asmmodel} 
(i) Let $\Theta_{P}$ denote the parameter space for $\theta_{P}$. For $\theta_{P} \in \Theta_{P}$, $\mathbf{R}\left(\theta_{P} \right) $ is nonsingular. The Riccati matrices (e.g., $\mathbf{R}\left(\theta_{P} \right) $) are UB in both row and column sum norms, uniformly in $\theta_{P} \in \Theta_{P}$.

(ii) For $\theta_{P} \in \operatorname{int}\left(\Theta_{P} \right) $, the first, second,
and third derivatives of the Riccati matrices with respect to $\theta_{P}$ exist and are UB in both row and column sum norms, uniformly in $\theta_{P} \in \Theta_{P}$.

(iii) $\sum_{h=1}^{\infty }\operatorname{abs}\left( \left( \mathbf{A}_{\mathbf{z}}^{-1}\mathbf{B}_{\mathbf{z}}\right) ^{h}\right)$ is UB in both row and column sum
norms, where $\left[ \operatorname{abs}\left(\mathbf{A}_{\mathbf{z}}^{-1}\mathbf{B}_{\mathbf{z}}\right)\right]_{ij}=\left\vert \left[ \mathbf{A}_{\mathbf{z}}^{-1}\mathbf{B}_{\mathbf{z}}\right]_{ij}\right\vert$.
\end{assumption}

\begin{assumption}
\label{asmnT} $n$ is an increasing function of $T$ with $T\rightarrow
\infty $.
\end{assumption}

Assumptions \ref{asmw}-\ref{asmreg1}, and \ref{asmnT} are standard in the spatial econometrics literature, as discussed in \citep{YangLee2017, YangLee2019}. Assumption \ref{asmmodel} (i) ensures that the invertibility of $\mathbf{R}\left(\theta_{P} \right)$ guarantees the existence and uniqueness of solutions for (\ref{m2-2-fd1}) for each possible $\theta_{P}$. Its second part describes the weak space-time dependencies of dependent variables as implied by the model for any possible value of $\theta_{P}$. Assumption \ref{asmmodel} (ii) addresses technical considerations, enabling the application of uniform laws of large numbers. For Assumption \ref{asmmodel} (iii), it serves as a sufficient condition to ensure time-space stability of  $\mathbf{z}_{t} = \left(\operatorname{vec}(\mathbf{Y}_{t})^\prime, \mathbf{g}_{t}^\prime, \operatorname{vec}(\mathbf{X}_{t})^\prime \right)^\prime$ for the asymptotic analysis.

\subsection{Consistency}

The primary aim of this subsection is to demonstrate that $\plim_{n,T\rightarrow \infty }\hat{\theta}_{nmT} = \theta_{0}$. The proof proceeds in three steps. First, we establish uniform convergence: $\sup_{\theta \in \Theta}\left\vert \frac{1}{nT}\ell_{nmT}^{c}\left( \theta \right) - \mathbb{E}\left(\frac{1}{nT}\ell_{nmT}^{c}(\theta) \right)  \right\vert \overset{p} \to 0$ when $n,T \to \infty $. Second, we prove the uniform equicontinuity of $\mathbb{E}\left(\frac{1}{nT}\ell_{nmT}^{c}(\theta) \right)$. Lastly, identification conditions are introduced. For this, we define the generated regressors as follows: for each $t$ and for each $(\theta_{P}, \beta)$,
\begin{equation}
\boldsymbol{\Xi}_{t}^{*} \left(\theta_{P},\beta \right) 
= \boldsymbol{\Xi}_{t}(\theta_{P}) 
+ \sum_{q=1}^{Q}(\beta_{q,0} - \beta_{q})\left(l_{m} \otimes \mathbf{x}_{t,q}^{\tau}\right),
\label{Xi_form}
\end{equation}
where $\boldsymbol{\Xi}_{t}(\theta_{P}) = \mathbf{F}^{y}\left( \theta_{P}\right) 
\operatorname{vec}(\mathbf{Y}_{t-1}) + \sum_{k=1}^{K}\mathbf{F}_{k}^{x}\left(\theta_{P}\right) \mathbf{x}_{t,k} + \sum_{q=1}^{Q}\mathbf{F}_{q}^{\tau} \left(\theta_{P},\beta_{q,0}\right) \left(l_{m} \otimes \mathbf{x}_{t,q}^{\tau}\right) + \mathbf{R} \left( \theta_{P}\right) \mathbf{R}^{-1} 
\widetilde{\boldsymbol{\alpha}}_{t,0}$ with
\begin{equation*}
    \begin{split}
        \mathbf{F}^{y}\left(\theta_{P}\right) & = \mathbf{R}\left(
\theta_{P}\right) \mathbf{R}^{-1}
\begin{bmatrix}
    \left( P_{0}\otimes \mathbf{I}_{n}\right) + \left( \boldsymbol{\rho}_{0}^{\prime} \otimes \mathbf{W} \right) \\
    \mathbf{R}_{0} A^{g}
\end{bmatrix}
-
\begin{bmatrix}
    \left( P \otimes \mathbf{I}_{n}\right) + \left( 
\boldsymbol{\rho}^{\prime} \otimes \mathbf{W} \right) \\
    \mathbf{R}_{0}\left(\theta_{P}\right) A^{g} \left( \theta_{P}\right)
\end{bmatrix}, \\
      \mathbf{F}_{k}^{x}\left(\theta_{P}\right) & = \mathbf{R}\left(\theta_{P}\right)\mathbf{R}^{-1}
\begin{bmatrix}
    \mathbf{I}_{nm^{*}} + \delta\mathbf{L}_{1:n, k}(\mathbf{I}_{m^{*}} \otimes \mathbf{A}_{k}^{x}) \\
    \mathbf{R}_{0}C_{k}^{g}
\end{bmatrix}
\boldsymbol{\Pi}_{k,0} -
\begin{bmatrix}
    \mathbf{I}_{nm^{*}} + \delta\mathbf{L}_{1:n, k}\left(\theta_{P}\right)(\mathbf{I}_{m^{*}} \otimes \mathbf{A}_{k}^{x}) \\
    \mathbf{R}_{0}\left(\theta_{P}\right)C_{k}^{g}\left( \theta_{P}\right)
\end{bmatrix}
\boldsymbol{\Pi}_{k}\text{ for }k=1,\cdots,K,\text{ and}\\
      \mathbf{F}_{q}^{\tau}\left(\theta_{P},\beta_{q}\right) & = \beta_{q,0}\mathbf{R} \left(\theta_{P}\right)\mathbf{R}^{-1}-\beta_{q}\mathbf{I}_{nm}\text{ for }q=1,\cdots,Q.
    \end{split}
\end{equation*}

Hence, $\boldsymbol{\Xi}_{t}(\theta_{P})$ represents the source of variations to identify $\theta_{P,0}$ stemming from $\text{vec}(\mathbf{Y}_{t-1})$, $\mathbf{x}_{t,1}, \cdots, \mathbf{x}_{t,K}$ and their linear transformations derived by $\mathbf{W}$ and $\theta_{P}$. $\sum_{q=1}^{Q}(\beta_{q,0} - \beta_{q})(l_{m} \otimes \mathbf{x}_{t,q}^\tau)$ is for identifying $\beta_{q,0}$ for $q = 1,\cdots,Q$ which are the exclusive parameters for the allocator's decisions. Observe that $\mathbf{F}^{y}(\theta_{P,0}) = \mathbf{0}$, $\mathbf{F}_{k}^{x}(\theta_{P,0}) = \mathbf{0}$ for $k=1,\cdots,K$, $\mathbf{F}_{q}^{\tau}(\theta_{P,0}, \beta_{q,0}) = \mathbf{0}$ and $\mathbf{F}_{q}^{\tau}(\theta_{P,0}, \beta_{q}) = (\beta_{q,0} - \beta_{q})\mathbf{I}_{nm}$ for $q=1,\cdots,Q$.\footnote{For example, $\mathbf{F}^{y}(\theta_{P})\operatorname{vec}(\mathbf{Y}_{t-1})$ in \eqref{Xi_form} represents the misspecification of $\theta_{P}$ from the part of $\operatorname{vec}(\mathbf{Y}_{t-1})$. Also, note that $\mathbf{F}_{q}^{\tau}(\theta_{P}, \beta_{q,0}) = \beta_{q,0}\left(\mathbf{R}(\theta_{P})\mathbf{R}^{-1} - \mathbf{I}_{nm} \right)$ for $q = 1,\cdots,Q$.} Define 
\begin{equation*}
    \begin{split}
        \boldsymbol{\Xi}_{nmT}^{*}(\theta_{P}, \beta) = & \left[\boldsymbol{\Xi}_{1}^{*\prime}(\theta_{P}, \beta),\cdots,\boldsymbol{\Xi}_{T}^{*\prime}(\theta_{P}, \beta) \right]^\prime\text{ for each }(\theta_{P}, \beta), \text{ and }\\
        \mathbf{M} \left(\theta_{P}, \Sigma, \sigma^{2} \right) = & \boldsymbol{\Delta}^{\frac{1}{2}}\mathbf{R}^{-1\prime}\mathbf{R}\left(\theta_{P}\right)
^{\prime}\boldsymbol{\Delta}^{-1}\left(\theta_{P}, \Sigma, \sigma^{2} \right) \mathbf{R}\left(\theta_{P}\right) \mathbf{R}^{-1}\boldsymbol{\Delta}^{\frac{1}{2}}
\text{ for each }(\theta_{P}, \Sigma, \sigma^{2}).
    \end{split}
\end{equation*}

Assumption \ref{asmiden} delineates the identification assumption under a large sample. The conditions below are derived from the information inequality (\cite{Rothenberg1971}) and the distribution of $\lbrace \mathbf{Y}_{t}, \mathbf{g}_{t} \rbrace$ induced by the MPNE.

\begin{assumption}[Sufficient conditions for identification]
\label{asmiden} 

(i) For $\left(\theta_{P}, \beta \right) \neq \left(\theta_{P,0}, \beta_{0}\right)$, and for all $\Sigma > 0$ and $\sigma^{2} > 0$,
\begin{equation}
    \lim_{n,T \to \infty}\frac{1}{nT}\mathbb{E}\left[ \boldsymbol{\Xi}_{nmT}^{*} \left(\theta_{P}, \beta \right)^{\prime}\left( \mathbf{J}_{T}\otimes
\mathbf{J}_{n,m}\boldsymbol{\Delta}^{-1}\left( \theta_{P}, \Sigma, \sigma^{2} \right) \mathbf{J}_{n,m}\right) \boldsymbol{\Xi}_{nmT}^{\ast}\left(\theta_{P}, \beta \right) \right] > 0,
\label{asmiden_cond1}
\end{equation}
where $\mathbf{J}_{T} = \mathbf{I}_{T} - \frac{1}{T}l_{T}l_{T}^\prime$. 

(ii)  Alternatively, we assume 
\begin{equation*}
\lim_{n\rightarrow \infty }\frac{1}{nm}\sum_{i}\left[ \ln \varphi_{i}\left( \mathbf{M} \left(\theta_{P}, \Sigma, \sigma^{2} \right) \right) - \left(\varphi_{i}\left( \mathbf{M} \left(\theta_{P}, \Sigma, \sigma^{2} \right) \right) - 1\right) \right] < 0 \text{ unless }(\theta_{P}, \Sigma, \sigma^{2}) = (\theta_{P,0}, \Sigma_{0}, \sigma_{0}^{2}).
\end{equation*}
To identify $\beta_{0}$, we assume $\plim_{n,T\to \infty}\frac{1}{nT} \mathbf{X}_{nmT}^{\tau \prime}\left(\mathbf{J}_{T}\otimes \mathbf{J}_{n,m}\boldsymbol{\Delta}^{-1} \mathbf{J}_{n,m}\right) \mathbf{X}_{nmT}^{\tau} > 0$, where $\mathbf{X}_{nmT}^{\tau} = \left[\mathbf{x}_{nmT, 1}^{\tau},\cdots,\mathbf{x}_{nmT, Q}^{\tau} \right]$ with $\mathbf{x}_{nmT,q}^{\tau} = \left(l_{m}^\prime \otimes \mathbf{x}_{1,q}^{\tau\prime}, \cdots, l_{m}^\prime \otimes \mathbf{x}_{T,q}^{\tau\prime} \right)^\prime$ for $q = 1,\cdots,Q$.
\end{assumption}

The main focus of Assumption \ref{asmiden} (i) is identifying $\theta_{P,0}$ and $\beta_{0}$. This pinpoints the mean-part misspecification errors that emerge when $\left(\theta_{P}, \beta \right)$ is inaccurately chosen and when the two-way fixed effects are eliminated using $\mathbf{J}_{T}$ and $\mathbf{J}_{n}$. The quadratic variation in \eqref{asmiden_cond1} represents the expected value of the misspecification quantity, so it becomes zero when $(\theta_{P}, \beta) = (\theta_{P,0}, \beta_{0})$.\footnote{In the linear regression model for cross-section data, $\mathbf{y} = \mathbf{X}\beta_{0} + \mathbf{e}$, the corresponding condition is $\left(\beta_{0} -\beta\right)^\prime \mathbb{E}\left(\frac{\mathbf{X}^\prime \mathbf{X}}{n} \right) \left(\beta_{0} -\beta\right) > 0$ if $\beta \neq \beta_{0}$. Notably, this assumption parallels Assumption 8 in \cite{Lee2004} and Assumption 5 in \cite{YangLee2017}.} Here is a brief explanation of Condition (i):
\begin{itemize}
    \item Under the hierarchical decision-making structure, there is only a directional influence of $\mathbf{g}_{t}$ on $\mathbf{Y}_{t}$. Note that $\mathbf{g}_{t}$ is an endogenous variable in the first part of equation~\eqref{m2-2-fd1}. Since the structure of $\mathbf{g}_{t}$ is described by the second part of equation~\eqref{m2-2-fd1}, its conditional expectation can be expressed as a function of $\mathbf{Y}_{t-1}$ and $\mathbf{X}_{t}$.

    \item Exogenous variation necessary for identifying $\theta_{P,0}$ can thus be achieved. A sufficient condition is that the set of the exogenous variables and conditional expectations of the endogenous variables in the first part of equation~\eqref{m2-2-fd1}, for varying values of $\theta_{P,0}$, is linearly independent. Intuitive explanations based on the myopic agents’ model (equation~\eqref{m2-2-example}) can be found in Sec.~3.3.1 of the supplement.

    \item Once $\theta_{P,0}$ is identified from the first part of equation~\eqref{m2-2-fd1}, the parameters $\beta_{0}$ can be subsequently identified using the second part of equation~\eqref{m2-2-fd1}. That is, $\beta_{0}$ is identified when there are sufficient spatial/time variations in $\lbrace x_{i,t,q}^{\tau} \rbrace$ after removing the two-way fixed effects.
\end{itemize}

Given that \(\mathbb{Var}\left(\operatorname{vec}\left(\mathbf{Y}_{t}\right)^{\prime}, \mathbf{g}_{t}^{\prime}\right)^{\prime} = \mathbf{R}^{-1}\boldsymbol{\Delta}\mathbf{R}^{-1\prime}\), Assumption \ref{asmiden} (ii) characterizes the variance-part misspecification errors that may arise when \((\theta_{P}, \Sigma, \sigma^{2})\) is incorrectly chosen. Specifically, we can observe that \(\mathbf{M}\left( \theta_{P,0}, \Sigma_{0}, \sigma_{0}^{2}\right) = \mathbf{I}_{nm}\) implying \(\varphi_{i}\left( \mathbf{M}\left(\theta_{P,0}, \Sigma_{0}, \sigma_{0}^{2} \right) \right) = 1\) for all \(i\). Intuitively, some $\theta_{P}$, $\Sigma$, and $\sigma^{2}$ with $(\theta_{P}, \Sigma, \sigma^{2}) \neq (\theta_{P,0}, \Sigma_{0}, \sigma_{0}^{2})$, $\mathbf{M}(\theta_{P}, \Sigma, \sigma^{2})$ becomes far from $\mathbf{I}_{nm}$. Then, the geometric mean of $\varphi_{1}\left(\mathbf{M}(\theta_{P}, \Sigma, \sigma^{2}) \right), \cdots, \varphi_{nm}\left(\mathbf{M}(\theta_{P}, \Sigma, \sigma^{2}) \right)$ and their arithemetic mean are not close.\footnote{In detail, $\bar{x}^{G} \leq \bar{x}^{A}$, where $\bar{x}^{G} = \left(x_{1}x_{2}\cdots x_{nm} \right)^{\frac{1}{nm}}$ and $\bar{x}^{A} = \frac{1}{nm}\sum_{i=1}^{nm}x_{i}$ with $x_{i} > 0$ for all $i$. Without the case of $x_{1} = x_{2} = \cdots = x_{nm}$, $\bar{x}^{G} < \bar{x}^{A} \Rightarrow \ln \bar{x}^{G} < \ln \bar{x}^{A} < \bar{x}^{A} - 1$.} Explanations for this condition using a simpler model can be found in Sec.3.3.2 of the
supplement. Since this variance structure does not account for \(\beta\), an additional condition is needed to identify \(\beta_{0}\).

Here is the result stating the consistency of $\hat{\theta}_{nmT}$.

\begin{theorem}
\label{thmconsistency} Under Assumptions \ref{asmw} - \ref{asmiden}, $\plim_{n,T \rightarrow \infty }\hat{\theta}_{nmT}=\theta _{0}$.
\end{theorem}

\subsection{Asymptotic normality}

Applying conventional arguments from the spatial econometrics literature \citep{Yuetal2008, YangLee2017}, we derive the asymptotic distribution of \( \hat{\theta}_{nmT} \). In this subsection, we provide an overview of this derivation. Detailed arguments and formulas are available in Appendix C and Section 3.4 of the supplementary file.

Note that the asymptotic distribution of $\hat{\theta}_{nmT}$ is characterized by $\frac{1}{\sqrt{nT}}\frac{\partial \ell_{nmT}^{c}(\theta_{0})}{\partial \theta}$. For the \( j\)-th component \( \theta_j \) of the vector \( \theta \), let \( \mathbf{\tilde{s}}_{nmT}^{\theta_j} = \frac{1}{\sqrt{nT}}\frac{\partial \ell_{nmT}^{c}(\theta_0)}{\partial \theta_j} \). Each component of \( \mathbf{\tilde{s}}_{nmT}^{\theta_j} \) is an LQ form of \( \lbrace \mathbf{v}_{t} \rbrace \). Due to the existence of fixed effects, there exist asymptotic biases in \( \frac{1}{\sqrt{nT}}\frac{\partial \ell_{nmT}^{c}(\theta_0)}{\partial \theta} \), and consequently in \( \hat{\theta}_{nmT} \), because \( \mathbb{E}(\mathbf{\tilde{s}}_{nmT}^{\theta_j}) \neq 0 \).

For each $\mathbf{\tilde{s}}_{nmT}^{\theta _{j}}$, we have the following decomposition:
\begin{equation*}
    \mathbf{\tilde{s}}_{nmT}^{\theta_{j}} = \mathbf{s}_{nmT}^{\theta_{j}} - \textbf{bias}_{1,nmT}^{\theta_{j}} - \textbf{bias}_{2,nmT}^{\theta_{j}}.
\end{equation*}
Here, $\mathbf{s}_{nmT}^{\theta
_{j}}$ satisfies $\mathbb{E}\left(\mathbf{s}_{nmT}^{\theta
_{j}}\right) = 0$ for all $\theta_{j}$, $\textbf{bias}_{1,nmT}^{\theta_{j}} = \sqrt{\frac{n}{T}}\mathbf{a}_{nm,1}^{\theta_{j}} + o_{p}\left(1\right) $, and $\textbf{bias}_{2,nmT}^{\theta_{j}} = \sqrt{\frac{T}{n}}\mathbf{a}_{nm,2}^{\theta_{j}}$, where $\mathbf{a}_{nm,1}^{\theta_{j}}\left(\theta \right)$ and $\mathbf{a}_{nm,2}^{\theta_{j}}\left(\theta \right)$ denote the non-stochastic approximations for the asymptotic bias terms evaluated at $\theta \in \Theta$, and $\mathbf{a}_{nm,1}^{\theta_{j}} = \mathbf{a}_{nm,1}^{\theta_{j}}(\theta_{0})$ and $\mathbf{a}_{nm,2}^{\theta_{j}} = \mathbf{a}_{nm,2}^{\theta_{j}}(\theta_{0})$. Using the decomposition scheme above, we obtain $\frac{1}{\sqrt{nT}}\frac{\partial \ell_{nmT}^{c}\left( \theta_{0}\right) }{\partial \theta } = \frac{1}{\sqrt{nT}}\frac{\partial \ell_{nmT}^{c,\left(u\right) }\left( \theta_{0}\right) }{\partial \theta} 
 - \textbf{bias}_{1,nmT} - \textbf{bias}_{2,nmT}$ from above summarized in the vector form. Based on $\textbf{bias}_{1,nmT}$ and $\textbf{bias}_{2,nmT}$, the corresponding vectors $\mathbf{a}_{nm,1}$ and $\mathbf{a}_{nm,2}$, which are finite in their limits, can also be defined.

\begin{assumption}
\label{asminfo}
For each \( \theta \in \Theta \), we define \( \boldsymbol{\Sigma}_{nmT}(\theta) = -\mathbb{E}\left( \frac{1}{nT}\frac{\partial^{2}\ell_{nmT}^{c}(\theta)}{\partial \theta \partial \theta^{\prime}} \right) \) and \( \boldsymbol{\Omega}_{nmT}(\theta) = \mathbb{E}\left( \frac{1}{nT}\frac{\partial \ell_{nmT}^{c,(u)}(\theta)}{\partial \theta }\frac{\partial \ell_{nmT}^{c,(u)}(\theta)}{\partial \theta^{\prime}}  \right) \). \(\liminf_{n,T \to \infty} \varphi_{\min}\left( \boldsymbol{\Sigma }_{nmT}\right) > 0\) and \(\liminf_{n,T \to \infty} \varphi_{\min}\left( \boldsymbol{\Omega}_{nmT}\right) > 0\), where \(\varphi_{\min}\left(\mathbf{A}\right)\) denotes the smallest eigenvalue of \(\mathbf{A}\), $\boldsymbol{\Sigma }_{nmT} = \boldsymbol{\Sigma }_{nmT}(\theta_{0})$ and $\boldsymbol{\Omega}_{nmT} = \boldsymbol{\Omega}_{nmT}(\theta_{0})$.
\end{assumption}

Assumption \ref{asminfo} ensures that \(\boldsymbol{\Sigma} = \lim_{n,T \to \infty} \boldsymbol{\Sigma}_{nmT}\) and \(\boldsymbol{\Omega} = \plim_{n,T \to \infty} \boldsymbol{\Omega}_{nmT}\) are positive definite constant matrices.

\begin{theorem}
\label{thmasymdist} 
Under Assumptions \ref{asmw} - \ref{asmps}, \ref{asmreg1} (i), \ref{asmmodel} - \ref{asminfo},
\begin{equation*}
\sqrt{nT}\left( \hat{\theta}_{nmT} - \theta_{0}\right) 
+ \sqrt{\frac{n}{T}}\boldsymbol{\Sigma}_{nmT}^{-1} \mathbf{a}_{nm,1} 
+ \sqrt{\frac{T}{n}}\boldsymbol{\Sigma}_{nmT}^{-1}\mathbf{a}_{nm,2} 
\overset{d}{\to} N\left( \mathbf{0}, \boldsymbol{\Sigma}^{-1} \boldsymbol{\Omega}\boldsymbol{\Sigma}^{-1}\right) 
\text{ as } n, T \rightarrow \infty.
\end{equation*}
\end{theorem}

By Theorem \ref{thmasymdist}, we can define a bias-corrected QMLE:
\begin{equation*}
\hat{\theta}_{nmT}^{c} = \hat{\theta}_{nmT} - \frac{1}{T}
\left( -\boldsymbol{\Sigma}_{nmT}^{-1}(\hat{\theta}_{nmT})\mathbf{a}_{nm,1}
\left( \hat{\theta}_{nmT}\right) \right) -\frac{1}{n}\left( -\boldsymbol{\Sigma}_{nmT}^{-1}(\hat{\theta}_{nmT})\mathbf{a}_{nm,2}\left( \hat{\theta}_{nmT}\right) \right).
\end{equation*}
This feasible bias-corrected estimator can be asymptotically equivalent to the infeasible bias-corrected QMLE (the ideal bias-corrected one), if Assumption \ref{asmae} is satisfied.

\begin{assumption}
\label{asmae} 
(i) $\frac{n}{T} \to c \in (0, \infty)$.

(ii) In a neighborhood of $\theta_{P,0}$, $\sum_{h=0}^{\infty}\left( 
\mathbf{A}_{\mathbf{z}}^{-1}\left(\theta_{P}\right) \mathbf{B}_{\mathbf{z}}\left( \theta_{P}\right) \right)^{h}$ and $\sum_{h=1}^{\infty}h\left( \mathbf{A}_{\mathbf{z}}^{-1}\left( \theta_{P}\right) \mathbf{B}_{\mathbf{z}}\left(\theta_{P}\right)
\right)^{h-1}$ are UB in row and column sums.
\end{assumption}

Assumption \ref{asmae} imposes additional space-time stability conditions for the asymptotic equivalence. For relevant discussions on Assumption \ref{asmae}, refer to Theorem 7 in \cite{LeeYu2010}, and Assumption 4.5
and Theorem 3 in \cite{YangLee2019}. Consequently, under Assumption \ref{asmae}, we have 
\begin{equation*}
    \sqrt{nT}\left(\hat{\theta}_{nmT}^{c} - \theta
_{0}\right) \overset{d}{\to} N\left(0, \boldsymbol{\Sigma}^{-1}\boldsymbol{\Omega}\boldsymbol{\Sigma}^{-1}\right)\text{ as }n,T \to \infty.
\end{equation*}

\section{Simulations}
\label{Simulation}

This section reports simulation results to investigate the finite-sample performance of the QMLE. The data generating process (DGP) for the simulations is based on equation (\ref{m2-2-fd1}). We consider a setting with $m^{*} = 2$, $K = 2$, and $Q = 1$. The variables $\mathbf{X}_{t}$, and $\mathbf{X}_{t}^{\tau}$ are independently and identically distributed, drawn from $N\left( \mathbf{0}, \mathbf{I}_{n}\right)$. Similarly, each component of $\mathbf{c}_{0}$ and $\widetilde{\boldsymbol{\alpha}}_{t,0}$ is generated from $N\left(0, 1\right)$. For the disturbance term, $\operatorname{vec}\left( \mathbf{E}_{t}\right)$ follows $N\left(\mathbf{0},\left( \Sigma_{0} \otimes \mathbf{I}_{n}\right) \right)$, where $\Sigma_{0}=
\begin{bmatrix}
1 & 0.5 \\ 
0.5 & 1
\end{bmatrix}$.
Finally, $\mathbf{e}_{t}^{\tau}$ is generated from $N\left( \mathbf{0},\sigma_{0}^{2}\mathbf{I}_{n}\right)$, with $\sigma_{0}^{2} = 1$.

Our simulation design aligns with the empirical application in Section \ref{Application} regarding the sample size and the spatial weighting matrix $\mathbf{W}^{adj}$ there. That is, we consider $\left(n,T\right) =\left( 48,25\right)$. For each pair $(i,j)$, let $w_{ij}^{adj} = \frac{\tilde{w}_{ij}^{adj}}{\sum_{k=1}^{n}\tilde{w}_{ik}^{adj} }$ where 
\begin{equation}
\tilde{w}_{ij}^{adj} = \mathbf{1}\left\{ \left( i,j\right) \text{ are bordering
states}\right\} \text{ if }j\neq i;\text{ and }0\text{ if }j=i.
\label{m5-1-w}
\end{equation}
We choose $\delta = 0.9$. To have a stable DGP, we first generate the data with $30 + T$ periods and take the last $T$ periods
as our sample. For components in $\theta_{0}$, we consider $\Lambda_{0}=
\begin{bmatrix}
0.2 & 0.1 \\ 
0.1 & 0.2
\end{bmatrix}
$, $\boldsymbol{\rho}_{0}=
\begin{bmatrix}
0.2 & 0.1 \\ 
0.1 & 0.2
\end{bmatrix}
$, $P_{0}=
\begin{bmatrix}
0.2 & 0 \\ 
0 & 0.2
\end{bmatrix}%
$, $\Psi_{0}=
\begin{bmatrix}
1 & 0.2 \\ 
0.2 & 1
\end{bmatrix}
$, $\phi_{0}=\left(0.2, 0.2\right)^{\prime}$, 
$\Pi_{0}=
\begin{bmatrix}
1 & 0 \\ 
0 & -1
\end{bmatrix}
$, and $\beta_{0}=1$. In Section 4 of the supplement, we consider different parameter sets to check the robustness of the QMLE's finite sample performance. The simulation results under the alternative parameter sets show that the performance of the QMLE is quite similar to that in the main draft. Four criteria are reported for performance evaluation: (i) bias, (ii) standard deviation (SD), and (iii) the coverage probability of a nominal 95\% confidence interval (CP). Each simulation experiment consists of 300 sample repetitions.

Simulation results in Table \ref{simtable_1} are summarized as follows.
\begin{table}[htbp]
\caption{Simulation results}
\label{simtable_1}
\begin{center}
\footnotesize
\begin{tabular}{l|c|cccccccc}
\hline																			
	&		&	$\lambda_{11}$	&	$\lambda_{21}$	&	$\lambda_{12}$	&	$\lambda_{22}$	&	$\rho_{11}$	&	$\rho_{21}$	&	$\rho_{12}$	&	$\rho_{22}$	\\
\hline																			
Bias	&	$\hat{\theta}_{nmT}$	&	-0.0416	&	-0.0146	&	-0.0108	&	-0.0394	&	-0.0026	&	-0.0018	&	-0.0024	&	-0.0011	\\
	&	$\hat{\theta}_{nmT}^{c}$	&	-0.0068	&	-0.0014	&	0.0021	&	-0.0047	&	-0.0079	&	-0.0023	&	-0.0026	&	-0.0064	\\
SD	&	$\hat{\theta}_{nmT}$	&	0.0366	&	0.0499	&	0.0485	&	0.0398	&	0.0443	&	0.0557	&	0.0592	&	0.0428	\\
	&	$\hat{\theta}_{nmT}^{c}$	&	0.0375	&	0.0542	&	0.0529	&	0.0407	&	0.0436	&	0.0543	&	0.0576	&	0.0423	\\
CP	&	$\hat{\theta}_{nmT}$	&	0.8133	&	0.8967	&	0.9267	&	0.8167	&	0.8467	&	0.8733	&	0.8500	&	0.8600	\\
	&	$\hat{\theta}_{nmT}^{c}$	&	0.9500	&	0.8967	&	0.9167	&	0.9367	&	0.8600	&	0.8867	&	0.8667	&	0.8567	\\
\hline																			
	&		&	$p_{11}$	&	$p_{12}$	&	$p_{22}$	&	$\psi_{12}$	&	$\phi_{1}$	&	$\phi_{2}$	&	$\pi_{11}$	&	$\pi_{21}$	\\
\hline																			
Bias	&	$\hat{\theta}_{nmT}$	&	-0.0140	&	0.0038	&	-0.0137	&	0.0278	&	0.0004	&	0.0000	&	-0.0178	&	-0.0322	\\
	&	$\hat{\theta}_{nmT}^{c}$	&	-0.0071	&	0.0051	&	-0.0067	&	0.0258	&	0.0019	&	0.0015	&	-0.0121	&	-0.0299	\\
SD	&	$\hat{\theta}_{nmT}$	&	0.0211	&	0.0446	&	0.0208	&	0.2136	&	0.0224	&	0.0225	&	0.0521	&	0.2230	\\
	&	$\hat{\theta}_{nmT}^{c}$	&	0.0217	&	0.0458	&	0.0211	&	0.2121	&	0.0224	&	0.0228	&	0.0516	&	0.2227	\\
CP	&	$\hat{\theta}_{nmT}$	&	0.7833	&	0.8600	&	0.8200	&	0.8833	&	0.8933	&	0.8967	&	0.8933	&	0.8800	\\
	&	$\hat{\theta}_{nmT}^{c}$	&	0.8667	&	0.8700	&	0.8800	&	0.8800	&	0.9000	&	0.8967	&	0.9200	&	0.8967	\\
\hline																			
	&		&	$\pi_{12}$	&	$\pi_{22}$	&	$\beta$	&	$[\Sigma]_{11}$	&	$[\Sigma]_{12}$	&	$[\Sigma]_{22}$	&	$\sigma^{2}$	&		\\
\hline																			
Bias	&	$\hat{\theta}_{nmT}$	&	0.0313	&	0.0131	&	0.0028	&	-0.0276	&	0.0172	&	-0.0250	&	-0.0585	&		\\
	&	$\hat{\theta}_{nmT}^{c}$	&	0.0296	&	0.0078	&	0.0021	&	0.0394	&	0.0503	&	0.0424	&	-0.0024	&		\\
SD	&	$\hat{\theta}_{nmT}$	&	0.2279	&	0.0537	&	0.0289	&	0.1866	&	0.3857	&	0.1861	&	0.0407	&		\\
	&	$\hat{\theta}_{nmT}^{c}$	&	0.2281	&	0.0544	&	0.0290	&	0.2012	&	0.4107	&	0.2007	&	0.0430	&		\\
CP	&	$\hat{\theta}_{nmT}$	&	0.8867	&	0.8967	&	0.9500	&	0.7433	&	0.8767	&	0.7267	&	0.6800	&		\\
	&	$\hat{\theta}_{nmT}^{c}$	&	0.8900	&	0.8900	&	0.9500	&	0.8733	&	0.8667	&	0.8633	&	0.9233	&		\\
\hline															
	
\end{tabular}
\end{center}
\end{table}

($\Lambda = [\lambda_{l^\prime l}]$: Contemporary spatial interaction effects) We observe downward biases in all estimated elements of $\hat{\Lambda}_{nmT}$. Specifically, the biases in $\hat{\lambda}_{11, nmT}$ and $\hat{\lambda}_{22, nmT}$ are more significant than those in $\hat{\lambda}_{12, nmT}$ and $\hat{\lambda}_{21, nmT}$. Comparatively, the SDs of $\hat{\lambda}_{12, nmT}$ and $\hat{\lambda}_{21, nmT}$ are larger than those of $\hat{\lambda}_{11, nmT}$ and $\hat{\lambda}_{22, nmT}$. Upon correcting for the asymptotic biases, the magnitude of biases in $\hat{\Lambda}_{nmT}$ diminishes.

($\boldsymbol{\rho} = [\rho_{l^\prime l}]$: Diffusion Effects) We observe that the magnitudes of biases in $\boldsymbol{\hat{\rho}}_{nmT}$ are small relative to those of $\hat{\Lambda}_{nmT}$. Similar to the case of estimating $\Lambda_{0}$, the estimates $\hat{\rho}_{11,nmT}$ and $\hat{\rho}_{22,nmT}$ have smaller SDs compared to those of $\hat{\rho}_{12,nmT}$ and $\hat{\rho}_{21,nmT}$. In the case of $\boldsymbol{\hat{\rho}}_{nmT}$, their biases are significantly smaller than those for $\hat{\Lambda}_{nmT}$, and the bias correction does not result in significant changes in either the magnitudes of the biases or the CPs.

($P = [p_{l^\prime l}]$: Dynamic adjustment cost) In our findings, we observe downward biases in both $\hat{p}_{11, nmT}$ and $\hat{p}_{22, nmT}$, alongside a small upward bias in $\hat{p}_{12, nmT}$. Notably, the SD of $\hat{p}_{12, nmT}$ is larger compared to the SDs of $\hat{p}_{11, nmT}$ and $\hat{p}_{22, nmT}$. Upon applying bias correction techniques, the magnitudes of the biases in $\hat{p}_{11, nmT}$ and $\hat{p}_{22, nmT}$ are reduced. Concurrently, the CPs show an increase.

($\Psi = [\psi_{l^\prime l}]$: Cost of selecting activity level) We observe an upward bias in $\hat{\psi}_{12, nmT}$. The bias correction slightly reduces its upward bias.

($\phi$ : Effects of grants on local agents' decisions) We observe similar performance for $\hat{\phi}_{nmT}$ and $\hat{\phi}_{nmT}^{c}$. In the case of $\hat{\phi}_{nmT}$, the bias correction does not improve the magnitudes of biases and the CPs.

($\Pi = [\pi_{kl}]$: Effects of explanatory variables) We detect downward biases in \(\hat{\pi}_{11, nmT}\) and \(\hat{\pi}_{21, nmT}\) and upward biases in \(\hat{\pi}_{12, nmT}\) and \(\hat{\pi}_{22, nmT}\). We observe large SDs in \(\hat{\pi}_{12, nmT}\) and \(\hat{\pi}_{21, nmT}\) relative to those of \(\hat{\pi}_{11, nmT}\) and \(\hat{\pi}_{22, nmT}\).

($\beta$: Effect of an economic indicator $\mathbf{X}_{t}^{\tau}$) An upward bias is detected in $\hat{\beta}_{nmT}$. The bias correction yields a slight performance improvement.

($\Sigma$: Variance of $\mathbf{E}_{t}$) We observe that there exist downward biases in \( \left[\hat{\Sigma}_{nmT}\right]_{11} \) and \( \left[\hat{\Sigma}_{nmT}\right]_{22} \) and an upward bias in \( \left[ \hat{\Sigma}_{nmT}\right]_{12} \). The SD of \( \left[ \hat{\Sigma}_{nmT}\right]_{12} \) is larger than those of \( \left[ \hat{\Sigma}_{nmT}\right]_{11} \) and \( \left[ \hat{\Sigma}_{nmT}\right]_{22} \). The bias correction tends to improve the CPs of \( \hat{\Sigma}_{nmT} \). In terms of the average of biases, however, the bias correction for \( \hat{\Sigma}_{nmT} \) does not significantly reduce the biases. This is because the bias correction for \(\hat{\Sigma}_{nmT}\) is usually successful in reducing biases, but it sometimes generates outliers.

($\sigma^{2}$: Variance of $\mathbf{e}_{t}^{\tau}$) There exists a downward bias in \( \hat{\sigma}_{nmT}^{2} \). After the bias correction, the magnitude of the bias significantly decreases, and the CP improves.

In summary, the QMLE and its bias correction demonstrate reasonable performance in a finite sample. Additionally, we observe that the diagonal elements of a parameter matrix are generally estimated with greater precision compared to the off-diagonal elements.

\section{Application}
\label{Application}

\subsection{Background and data}
\label{Application_background}

\paragraph{Background.} In this section, we examine the relationship between the U.S. federal government (resource allocator) and the governments of the 48 U.S. states excluding Alaska and Hawaii (local agents), i.e., \(n=48\). The federal government provides intergovernmental revenue to state governments in the form of grants, and each state allocates these resources to two expenditure categories. In estimation, we consider the natural logarithms of per capita federal grants and the natural logarithms of per capita state government expenditures within these two specific sectors. Detailed descriptions are provided below:\footnote{The descriptions below are available from: \textsf{http://www.census.gov/govs/www/index.html}}
\begin{itemize}
    \item \textbf{Intergovernmental Revenue from the federal government}: These are funds directly provided to state governments from the federal government, encompassing formula-based grants, project-specific grants, shared taxes allocated under federal law, and contingent loans and advances.
    
    \item \textbf{Public Welfare}: This category includes expenditures for services such as poverty alleviation, emergency relief, refugee assistance, medical aid, housing assistance, and long-term health care.
    
    \item \textbf{Housing and Community Development}: This category covers funding for the construction, operation, and support of housing and redevelopment projects targeting both public and private housing as well as broader community development initiatives.
\end{itemize}
Hereafter, we denote per capita annual federal grants to states as $\text{Grant}$, per capita annual public welfare expenditures as $\text{PWE}$, and per capita annual housing and community development expenditure as $\text{HCDE}$. For each state $i$ and each year $t$, these are logged and represented as $g_{i,t} = \ln (\text{Grant}_{i,t})$, $y_{i,t,1} = \ln (\text{PWE}_{i,t})$, and $y_{i,t,2} = \ln (\text{HCDE}_{i,t})$.

Since these expenditure categories share the common goal of improving residents' living standards, we investigate whether these state-level expenditures are complementary or substitutable. The payoff functions,~(\ref{m2-2-fpayoff}) and~(\ref{m2-2-lpayoff}), differentiate between federal and state government incentives. An important aspect of state government behavior is their operation within an open economic environment, allowing resource flows and economic activities to freely cross state borders without stringent international border controls (\cite{Agrawaletal2022}). This openness, combined with non-cooperative decision-making, can lead to significant externalities arising from fiscal competition among states. Drawing from the frameworks of \cite{Brueckner2003} and \cite{Revelli2005}, we interpret the local agent's payoff~(\ref{m2-2-fpayoff}) as reflecting the state government's objective of maximizing the utility of a representative resident. Consequently, influences from other state governments captured in~(\ref{m2-2-fpayoff}) can be viewed as policy spillovers.

Another distinctive feature of the institutional environment is the existence of direct federal intergovernmental revenues to state governments, as highlighted by \citet{Agrawaletal2022}. These revenues encompass all non-repayable federal grants—such as formula-based, project-specific, and block grants—provided directly to the states without intermediate governmental routing. This institutional characteristic motivates our analysis. In particular, the terms \(\phi_{l}g_{i,t}y_{i,t,l}\) in~(\ref{m2-2-fpayoff}) capture the federal government's direct influence on state-level decisions. Given our assumption that each state government maximizes the utility of its own residents, the federal government's payoff function~(\ref{m2-2-lpayoff}) is interpreted as representing social welfare generated by the state two expenditures. In this context, federal grants are strategically allocated to maximize aggregate welfare. Our model formalizes the federal government’s altruism as the pursuit of this welfare-maximizing objective. Specifically, the federal government supplements state-level tax revenues by providing intergovernmental grants, thereby mitigating interstate externalities that arise when states optimize only their own objectives. Thus, the model highlights the role of federal transfers in improving the efficiency of decentralized decision-making~\cite[Ch.~6.3]{Agrawaletal2022}. Consequently, the system of equations in~\eqref{m2-2-fd1} incorporates policy reaction functions for state expenditures and equilibrium federal grants, thereby capturing the endogenous nature of federal funding in the presence of strategic interaction between federal and state governments.\footnote{\cite{Knight2002} develops a bargaining model addressing federal grant allocations and state spending policies on highways, presenting an identification strategy for estimating federal influence on state expenditures. We share the same spirit as \cite{Knight2002} in the aspect of endogenizing the federal government's decisions. Unlike Knight’s analysis, however, our model structurally incorporates both vertical and horizontal interactions and focuses on general intergovernmental revenue rather than grants for a specific function.}

\paragraph{Data.} The dataset covers the years 1992 to 2018, spanning \( T = 27 \) years across \( n = 48 \) states, yielding \( 1,296 \) observations. State financial, demographic, and economic data, along with federal grant information, are obtained from the U.S. Census Bureau. Additional macroeconomic variables—including GDP deflators, interest rates, consumption levels, and state income levels—are sourced from the Federal Reserve Bank of St. Louis. Specifically, \(\mathbf{X}_{t}\) comprises: (i) the logged per capita state tax revenue (\(x_{i,t,1}\)); (ii) state income growth rate (\(x_{i,t,2}\)); (iii) state population growth rate (\(x_{i,t,3}\)); and (iv) the growth rates of states' Gini coefficients (\(x_{i,t,4}\)).\footnote{The Gini coefficient measures income inequality within a region. For 1999–2018, data are from the U.S. Census Bureau. For 1992–1998, we use estimates provided by \cite{Frank2009}.} Growth rates rather than absolute levels are utilized for \(x_{i,t,2}\), \(x_{i,t,3}\), and \(x_{i,t,4}\) to maintain temporal stability. The variable \(x_{i,t,1}\) represents the formation of the second-type policy base, while \(x_{i,t,2}\), \(x_{i,t,3}\), and \(x_{i,t,4}\) reflect demographic and economic environment evolution. This second-type policy base ($x_{i,t,1}$) is mutually exclusive with the first-type base ($g_{i,t}$). Cyclical economic indicators (\(\mathbf{X}_{t}^{\tau}\)) are extracted using the Hodrick-Prescott decomposition of logged states' real per capita outputs. Note that all monetary values (federal grants, PWE, HCDE, and tax revenue) are inflation-adjusted using the GDP deflator (base year 2012). Consequently, we ensure the positivity of logged per capita amounts.

Table \ref{descstat_table} presents descriptive statistics. PWE represents the largest expenditure share (averaging 21.6\%), while HCDE accounts for only 0.5\%, despite functional similarities. Federal grants constitute approximately 37.7\% of total state revenue on average, highlighting their fiscal importance. State tax revenues ($x_{i,t,1}$) show higher variability than federal grants ($g_{i,t}$), whose variances are largely explained by fixed state and time effects (93.02\% compared to 84.39\% for tax revenues).

\begin{table}[htbp]
\caption{Descriptive statistics: 48 contiguous states in U.S.}
\label{descstat_table}
\begin{center}
\footnotesize
\begin{tabular}{lcc}
\hline									
Variables	&	Mean	&	STD	\\
\hline									
Public welfare per capita	&	1231.0142 	&	484.0848 	\\
\ (Proportion on total expenditure)	&	(0.2163) 	&	(0.0542) 	\\
$y_{i,t,1}$ & 7.0410 & 0.3921 \\
Housing and community development per capita	&	29.0410 	&	50.1910 	\\
\ (Proportion on total expenditure)	&	(0.0047) 	&	(0.0067) \\
$y_{i,t,2}$ & 2.6546 & 1.3286 \\
Total expenditure per capita	&	5633.4031 	&	1497.8889 	\\
Grants from the Federal government per capita	&	1529.3325 	&	556.5327 \\
$g_{i,t}$ & 7.2722 & 0.3449 \\

\hline

Tax revenue per capita	&	2491.8794 	&	711.7592 	\\
$x_{i,t,1}$ & 7.7864 & 0.2551 \\
$\frac{\text{Grant}}{\text{Grant + Tax} \times 100}$ (\%) & 37.7252 & 7.5234 \\
Personal income growth (\%)	&	1.0742 	&	5.0721 \\
Population growth (\%)	&	0.9358 	&	0.8563 	\\
Gini coefficient growth (\%)	&	0.3680 	&	1.6366 \\
Cyclical components of output per capita	&	0.0000 	&	0.0702 	\\
\hline										
						
\end{tabular}

\begin{tabular}{lcccc}
\hline
STD of the two resources & Level & Level (demeaned) & Logged & Logged (demeaned) \\
\hline
Grants from the Federal government per capita & 556.5327 &   185.4114 &   0.3449 &   0.0911 \\
Tax revenue per capita & 711.7592 &    346.4392 &   0.2551 &    0.1008 \\
\hline
\end{tabular}
\end{center}
\small \textit{Note}: Dollar amounts are per capita values. The first table reports the main descriptive statistics. The second table shows the standard deviations of the two resources for states' expenditures. For a variable $z$, ``demeaned" means $\tilde{z}_{it} = z_{it} - \bar{z}_{i.} - \bar{z}_{.t} + \bar{z}_{..}$, where $\bar{z}_{i.} = \frac{1}{T}\sum_{t=1}^{T}z_{it}$, $\bar{z}_{.t} = \frac{1}{n}\sum_{i=1}^{n}z_{it}$, and $\bar{z}_{..} = \frac{1}{nT}\sum_{i=1}^{n}\sum_{t=1}^{T}z_{it}$.
\end{table}

We define state interactions through spatial weighting matrices ($\mathbf{W}$), characterizing various neighbor concepts. As \cite{dePaulaetal2024} emphasize, fiscal competition neighbors can differ from geographic neighbors. We utilize five time-invariant, row-normalized spatial weighting matrices to explore expenditure competition:\footnote{Row-normalization ensures comparability across matrices. Unnormalized matrices like $\mathbf{W}^{adj}$ and $\mathbf{W}^{econ}$ originally represent binary adjacency. $\mathbf{W}^{mig}$ uses actual migration data, $\mathbf{W}^{nwh}$ is based on inverse differences in nonwhite population proportions, and $\mathbf{W}^{sim}$ relies on a comprehensive 100-point state similarity scale. Each row sumes to 1, ensuring weights between 0 and 1.} (i) adjacency matrix $\mathbf{W}^{adj}$ (see \eqref{m5-1-w}), (ii) economic neighbor matrix $\mathbf{W}^{econ}$ from \cite{dePaulaetal2024}, (iii) demographic similarity matrix $\mathbf{W}^{nwh}$ inspired by \cite{Caseetal1993}, (iv) comprehensive similarity matrix $\mathbf{W}^{sim}$ based on demographics, culture, politics, infrastructure, and geography;\footnote{This index is available at \texttt{https://objectivelists.com/state-similarity-index-distance-matrix/}.}, and (v) historical migration flow matrix $\mathbf{W}^{mig}$, as advocated by \cite{Figlioetal1999} and \cite{BAICKER2005529}. \cite{dePaulaetal2024} estimates $\mathbf{W}^{econ}$ to characterize the states' tax competition, but we employ this $\mathbf{W}^{econ}$ by recognizing this network as the states' economic relationships in their fiscal decisions. $\mathbf{W}^{adj}$ and $\mathbf{W}^{mig}$ are relevant to (i) geographic benefit spillovers and (ii) residents' welfare-motivated moves \citep{BAICKER2005529, SOLEOLLE200632}. Motivated by ``yardstick competitions'' (\cite{BesleyCase1995}), $\mathbf{W}^{econ}$, $\mathbf{W}^{nwh}$, and $\mathbf{W}^{sim}$ represent demographic or economic relationships among U.S. states beyond their geographic connectivities. In contrast to other specifications, $\mathbf{W}^{mig}$ captures these relationships through the actual movements of residents. More detailed descriptions of this issue can be found in Section 5 of the supplement.

To determine \(\delta\), we turn to the Euler equation from the consumption-based model and calculate it as \(\delta = \frac{1}{T} \sum_{t=1}^{T} \frac{Con_{t}}{Con_{t-1}} \frac{1}{1+i_{t}} = 0.9862\). Here, \(Con_{t}\) represents the U.S. real personal consumption expenditures, and \(i_{t}\) stands for the 10-year Treasury Constant Maturity Rate (long-run interest rate).

\subsection{Parameter estimates}
\label{Application_parameter}

\begin{table}[htbp]
\caption{Model estimation}
\label{estimation_table}
\begin{center}
\footnotesize
\begin{tabular}{l|c}
\hline			
$\lambda_{11}$, ($w_{ij}y_{j,t,1}$) $\mapsto$ ($y_{i,t,1}$)	&	0.0483 (0.0224)	\\
$\lambda_{21}$, ($w_{ij}y_{j,t,2}$) $\mapsto$ ($y_{i,t,1}$)	&	0.0000 (0.0180)	\\
$\lambda_{12}$, ($w_{ij}y_{j,t,1}$) $\mapsto$ ($y_{i,t,2}$)	&	-0.2828 (0.1088)	\\
$\lambda_{22}$, ($w_{ij}y_{j,t,2}$) $\mapsto$ ($y_{i,t,2}$)	&	-0.1633 (0.0128)	\\
$\rho_{11}$, ($w_{ij}y_{j,t-1,1}$) $\mapsto$ ($y_{i,t,1}$)	&	0.2779 (0.0365)	\\
$\rho_{21}$, ($w_{ij}y_{j,t-1,2}$) $\mapsto$ ($y_{i,t,1}$)	&	-0.0200 (0.0111)	\\
$\rho_{12}$, ($w_{ij}y_{j,t-1,1}$) $\mapsto$ ($y_{i,t,2}$)	&	0.0004 (0.1237)	\\
$\rho_{22}$, ($w_{ij}y_{j,t-1,2}$) $\mapsto$ ($y_{i,t,2}$)	&	0.0000 (0.0138)	\\
$p_{11}$, ($y_{i,t-1,1}$) $\mapsto$ ($y_{i,t,1}$)	&	0.0841 (0.0103)	\\
$p_{12}$, ($y_{j,t-1,k}$) $\mapsto$ ($y_{i,t,l}$)	&	-0.0854 (0.0157)	\\
$p_{22}$, ($y_{i,t-1,2}$) $\mapsto$ ($y_{i,t,2}$)	&	0.9580 (0.0021)	\\
$\psi_{12}$, ($y_{j,t,k}$) $\mapsto$ ($y_{i,t,l}$)	&	-0.0963 (0.0114)	\\
$\phi_{1}$, ($g_{i,t}$) $\mapsto$ ($y_{i,t,1}$)	&	0.0361 (0.0039)	\\
$\phi_{2}$, ($g_{i,t}$) $\mapsto$ ($y_{i,t,2}$)	&	0.0086 (0.0025)	\\
\hline			
$\pi_{11}$, Tax rev. ($x_{i,t,1}$) $\mapsto$ PWE ($y_{i,t,1}$)	&	0.2096 (0.0257)	\\
$\pi_{21}$, Personal income growth ($x_{i,t,2}$) $\mapsto$ PWE ($y_{i,t,1}$)	&	0.0010 (0.0010)	\\
$\pi_{31}$, Pop. growth ($x_{i,t,3}$) $\mapsto$ PWE ($y_{i,t,1}$)	&	-0.0489 (0.0092)	\\
$\pi_{41}$, Gini coeff. growth ($x_{i,t,4}$) $\mapsto$ PWE ($y_{i,t,1}$)	&	-0.0062 (0.0040)	\\
$\pi_{12}$, Tax rev. ($x_{i,t,1}$) $\mapsto$ HCDE ($y_{i,t,2}$)	&	-0.8614 (0.0228)	\\
$\pi_{22}$, Personal income growth ($x_{i,t,2}$) $\mapsto$ HCDE ($y_{i,t,2}$)	&	-0.0050 (0.0050)	\\
$\pi_{32}$, Pop. growth ($x_{i,t,3}$) $\mapsto$ HCDE ($y_{i,t,2}$)	&	0.2962 (0.0255)	\\
$\pi_{42}$, Gini coeff. growth ($x_{i,t,4}$) $\mapsto$ HCDE ($y_{i,t,2}$)	&	0.0493 (0.0260)	\\
\hline			
$\beta$	&	1.2436 (0.4767)	\\
$[\Sigma]_{11}$	&	0.0312 (0.0025)	\\
$[\Sigma]_{12}$	&	-0.0680 (0.0124)	\\
$[\Sigma]_{22}$	&	0.7354 (0.0505)	\\
$\sigma^{2}$	&	0.0041 (0.0002)	\\
\hline

No. of Obs.	&	1296	\\
Log-likelihood	&	492.1680	\\
\hline					
\end{tabular}

\begin{tabular}{l|ccccc}
\hline											
	&	 $\mathbf{W}^{adj}$	&	$\mathbf{W}^{econ}$	&	$\mathbf{W}^{nwh}$	&	$\mathbf{W}^{sim}$	&	$\mathbf{W}^{mig}$	\\
\hline											
Log-likelihood	&	492.1680	&	483.3810	&	462.6760	&	491.5320	&	429.5300	\\
Model probability	&	0.6538	&	0.0001	&	0.0000	&	0.3461	&	0.0000	\\
\hline											
								
\end{tabular}

\end{center}
\small
\textit{Note}: The first table reports the bias-corrected QMLEs, with their standard errors in parentheses. The second table presents the model probabilities, measured by the Akaike weight, along with the corresponding log-likelihood values.
\end{table}

Table~\ref{estimation_table} summarizes the estimation results. The Akaike weight guides our model selection, favoring the geographic network structure ($\mathbf{W}^{adj}$) due to evidence of benefit spillovers \citep{SOLEOLLE200632}: a fraction of the public good produced by one state government is utilized by residents in nearby states.\footnote{This criterion is based on the Akaike Information Criterion (AIC) as suggested by \cite{Akaike1973}. The Akaike weight maps each spatial weighting matrix to its model probability. Employing model selection criteria for appropriate spatial weighting matrices in SAR models has been justified by \cite{ZhangandYu2018} and \cite{JeongLee2024}.}

We begin by examining the signs and significance levels of each parameter estimate; the marginal effects will be discussed in Section~\ref{Application_marginaleffect}. The parameters of grant reliance, \(\phi_{1}\) and \(\phi_{2}\), indicate positive influences from \( g_{i,t} \) on \( y_{i,t,1} \) and \( y_{i,t,2} \). This is consistent with existing literature that reports a positive effect of \( g_{i,t} \) on various state expenditures such as state administration, health and human services, and highway \citep{Caseetal1993}, as well as Medicaid-related spending \citep{HanLee2016}.

We find strong evidence of spatial spillovers among the states' two expenditures. In detail, the estimates of $\lambda_{11}$ and $\rho_{11}$ are positive and significant. These estimation results provide empirical evidence of welfare spillovers from PWEs via geographic connections across states. Relevant empirical studies can be found in \cite{Figlioetal1999}, \cite{HanLee2016}, and \cite{Wang2018}. We observe that $\lambda_{22} < 0$, implying that the HCDEs of two neighboring states are strategic substitutes. We verify $p_{12} < 0$ and $\psi_{12} < 0$, indicating that the two types of expenditures within a state act as complements. Further, we estimate two negative cross-variable spillover effects ($\lambda_{12} < 0$). This result verifies the strategic substitutability between a state's PWE and its nearby states' HCDE. We find evidence of dynamic adjustment costs for selecting \(y_{i,t,1}\) and \(y_{i,t,2}\), as the estimates of $p_{11}$ and $p_{22}$ are both positive and significant.

For the effect of states' demographic and economic environments, we find a positive effect of a state's tax revenue on its PWE and a negative effect on HCDE. When a state's tax revenue increases, the state allocates its resources to more visible and politically salient programs like PWE instead of capital-intensive projects like HCDE. The state's income growth does not significantly influence these expenditures. The negative effect of population growth on PWE suggests economies of scale; as the population increases, the per capita cost of providing welfare services decreases. Conversely, the positive effect on HCDE indicates that population growth leads to increased demand for housing and community infrastructure, requiring more expenditure in this area.\footnote{This is in line with the findings of \cite{Caseetal1993}, who identified negative effects of population density on three specific types of expenditure: (i) state administration, (ii) health and human services, and (iii) education. The positive effect of \(x_{i,t,3}\) on \(y_{i,t,2}\) aligns with the findings of \cite{HanLee2016} in the context of Medicaid-related spending.} A rise in income inequality leads to increased HCDE. States may invest more in housing and community development to address disparities caused by inequality, such as inadequate housing and lack of community resources in lower-income areas. Additionally, the estimate of $\beta$ is found to be positive, suggesting that the autonomous transfer components of federal grants to states are procyclical.

\subsection{Equilibrium effects}
\label{Application_marginaleffect}

\subsubsection{Short-run and long-run effects}
\label{Application_marginaleffectdetail}

For interpretations, we compute the average equilibrium effects in levels. We focus on the characteristics for analysis whose relevant coefficients are significant at the 10\% significance level. Below, the ``Direct impact'' is computed based on $\mathsf{ADI}$ defined in Section~\ref{sec_me}, while the ``Total impact'' comes from $\mathsf{ADTSI}$ over the next ten years as described in Section~\ref{sec_drf}. For example, the average direct effect of a state's tax revenue per capita on its PWE is computed by $\frac{1}{nT}\sum_{i=1}^{n}\sum_{t=1}^{T}\frac{\partial \text{PWE}_{i,t}}{\partial \text{Tax rev}_{i,t} } = \frac{1}{nT}\sum_{i=1}^{n}\sum_{t=1}^{T}\frac{\partial y_{i,t,1}}{\partial x_{i,t,1}} \frac{\text{PWE}_{i,t}}{\text{Tax rev}_{i,t} }$. Their standard errors are reported in parentheses.

\begin{itemize}
    \item An increase of \$1,000 in a state's tax revenue per capita results in:
    \begin{itemize}
        \item \textbf{Direct impact}: An immediate increase of \$58.10 (7.92) in PWE and a decrease of \$4.68 (0.16) in HCDE.
        \item \textbf{Total impact}: PWE rises by \$1,007.35 (67.29) and HCDE decreases by \$64.22 (2.59) across all states.
    \end{itemize}
\end{itemize}

First, we compare the effects of increasing tax revenues with those of increasing federal grants. An increase of \$1,000 in a federal government grant per capita results in an immediate increase of \$28.38 (2.99) in PWE and an increase of \$0.14 (0.02) in HCDE.

Increases in both federal grants and tax revenues lead to a rise in PWE. However, the increase from federal grants is much less than that from tax revenue. Increasing state tax revenue leads to a decrease in HCDE, whereas federal grants have a slight increase in HCDE. Like PWE, the change in HCDE induced by federal grants is much less than that induced by tax revenue. This provides evidence of the limited flexibility of the federal grants, which is supported by the descriptive statistics for federal grants and states' tax revenues (recall the second table in Table~\ref{descstat_table}): federal grants are largely explained by the fixed-effect components compared to state tax revenues. Hence, the marginal effects of federal grants are significantly smaller than those of tax revenues. Second, we observe significant long-term positive effects on PWE and negative externalities on other states' HCDE resulting from increased tax revenue.

We also evaluate the effects of a state's population growth and the growth in its income inequality.

\begin{itemize}
    \item Effect of a 1\% increase in a state's population growth:
    \begin{itemize}
        \item \textbf{Direct impacts}: An immediate decrease of \$66.91 (18.00) in PWE and an increase of \$10.02 (0.86) in HCDE.
        
        \item \textbf{Total impacts}: In the long run, PWE decreases by \$685.91 (116.91), and HCDE increases by \$80.11 (6.43) across all states.
    \end{itemize}
    \item Effect of a 1\% increase in a state's growth in income inequality:
    \begin{itemize}
        \item \textbf{Direct impacts}: An immediate increase of \$0.51 (0.27) in HCDE.
        \item \textbf{Total impacts}: Over time, HCDE increases by \$0.31 (0.39) across all states.
    \end{itemize}
\end{itemize}

\subsubsection{Counterfactual simulation: effects of the federal government's interventions}

Lastly, by utilizing the estimates detailed in Section~\ref{Application_parameter}, we conduct counterfactual simulations to analyze the impact of the federal government's responsive interventions. We compare the DGPs from two scenarios of parameters: (i) Scenario 1: DGP from the estimated parameters in Table \ref{estimation_table} and (ii) Scenario 2: DGP from the same parameter setting with Scenario 1 but $\phi = \mathbf{0}$ ("$\widehat{\cdot}$" denotes the results from Scenario 1, while "$\widetilde{\cdot}$" stands for the results from Scenario 2). Under Scenario 1, the federal government reacts to the states' previous decisions and current demographic/economic characteristics. When $\phi = \mathbf{0}$ (Scenario 2), $\mathbf{g}_{t}^L = \mathbf{0}$ resulting in $\mathbf{g}_{t}^* = \mathbf{g}_{t}^{A} + \mathbf{R}_{0}^{-1}\mathbf{e}_{t}^\tau$. It indicates that the federal grants are determined solely by autonomous transfer components. In this case, both expenditures are unaffected by increases in federal grants.

\begin{table}[htbp]
\caption{Counterfactual simulation results/Variance decompositions for $\mathbf{g}_{t}$ (federal grant) and $\mathbf{x}_{t,1}$ (tax revenue)}
\label{cfactsim_table}
\begin{center}
\footnotesize
\begin{tabular}{lc}								
Part I.	&	$\widehat{\text{Scenario 1}} - \widetilde{\text{Scenario 2}}$ \\
\hline									
$\Delta$PWE & \$67.17 \\
$\Delta$HCDE & \$1.21 \\
\hline
$\Delta$Welfare (\%) & 7.27\% \\
\hline										
\end{tabular}

\begin{tabular}{lc | lc}					
Part II. & & & \\
\hline
$\text{VarShare}(\mathbf{g}^{L})$ & 1.24\% & $\text{VarShare}(\mathbf{x}_{1}^{L})$ & 12.15\% \\
$\text{VarShare}(\mathbf{g}^{A})$ & 94.93\% & $\text{VarShare}(\mathbf{x}_{1}^{A})$ & 85.17\% \\
$\text{VarShare}(\text{residuals})$ & 3.83\% & $\text{VarShare}(\text{residuals})$ & 2.66\% \\
\hline
\end{tabular}
\end{center}
\footnotesize \textit{Note}: The values in Part I are calculated as follows: $\Delta \text{PWE} = \frac{1}{nT}\sum_{i=1}^{n}\sum_{t=1}^{T}\left(\widehat{\text{PWE}}_{i,t} - \widetilde{\text{PWE}}_{i,t} \right)$, $\Delta \text{HCDE} = \frac{1}{nT}\sum_{i=1}^{n}\sum_{t=1}^{T}\left(\widehat{\text{HCDE}}_{i,t} - \widetilde{\text{HCDE}}_{i,t} \right)$, and $\Delta \text{Welfare} = \frac{\widehat{V}_{0} - \widetilde{V}_{0} }{\widetilde{V}_{0}} \times 100$, where $\widehat{V}_{0}$ and $\widetilde{V}_{0}$ denote respectively the resource allocator's value functions under the estimated and counterfactual parameters.

\ \ \ \ For Panel II, we employ $\widehat{\mathbb{Var}}(\mathbf{g})$ and $\widehat{\mathbb{Var}}(\mathbf{x}_1)$ denote the sample variances of $\lbrace g_{i,t} \rbrace_{i=1}^{n}\vert_{t=1}^{T}$ and $\lbrace x_{i,t,1} \rbrace_{i=1}^{n}\vert_{t=1}^{T}$. We observe $\widehat{\mathbb{Var}}(\mathbf{g}) = 0.1140$ and $\widehat{\mathbb{Var}}(\mathbf{x}_1) = 0.0643$. Then, we define $\text{VarShare}(\mathbf{g}^{L}) = \frac{\widehat{\mathbb{Var}}(\mathbf{g}^L) + \widehat{\mathbb{Cov}}(\mathbf{g}^L, \mathbf{g}^A)}{\widehat{\mathbb{Var}}(\mathbf{g})}\times 100$, $\text{VarShare}(\mathbf{g}^{A}) = \frac{\widehat{\mathbb{Var}}(\mathbf{g}^A) + \widehat{\mathbb{Cov}}(\mathbf{g}^L, \mathbf{g}^A)}{\widehat{\mathbb{Var}}(\mathbf{g})}\times 100$, $\text{VarShare}(\mathbf{x}_{1}^{L}) = \frac{\widehat{\mathbb{Var}}(\mathbf{x}_{1}^L) + \widehat{\mathbb{Cov}}(\mathbf{x}_{1}^L, \mathbf{x}_{1}^A)}{\widehat{\mathbb{Var}}(\mathbf{x}_{1})}\times 100$, and $\text{VarShare}(\mathbf{x}_{1}^{A}) = \frac{\widehat{\mathbb{Var}}(\mathbf{x}_{1}^A) + \widehat{\mathbb{Cov}}(\mathbf{x}_{1}^L, \mathbf{x}_{1}^A)}{\widehat{\mathbb{Var}}(\mathbf{x}_{1})}\times 100$. These quantities come from the variance-covariance breakdown with the equal-spilt rule. Note that $\widehat{\mathbb{Cov}}(\mathbf{g}^L, \mathbf{g}^A) = -0.0124$ and $\widehat{\mathbb{Cov}}(\mathbf{x}_{1}^L, \mathbf{x}_{1}^A) = 0.0057$.
\end{table}

The first part of Table \ref{cfactsim_table} summarizes the counterfactual simulation results. With responsive intervention, state governments increase their PWE by \$67.17 and their HCDE by \$1.21, corroborating the simulation results detailed in Case~3.3 of Table~\ref{equilactexample}. It means the responsive federal interventions lead to more states' efforts on their PWE and HCDE. Additionally, we evaluate the impact on social welfare by comparing the welfare estimates $\hat{V}_{0}(\mathbf{z}_{t}^{R})$ under the two scenarios. Our calculations show that responsive federal intervention increases social welfare by 7.27\%.

\paragraph{Discussion.}

These results indicate that the federal government's responsive intervention enhances social welfare by increasing states' PWE and HDCE. This finding suggests that federal intervention is more than just financial support; it strategically guides state policies to achieve sustained welfare improvements. However, the welfare gains are relatively modest compared to the federal grants' 37.7\% share in total state revenue. This aligns with Section~\ref{Application_marginaleffectdetail}, which highlighted the limited flexibility of federal grants relative to state tax revenues.

To investigate this issue further, we decompose the variance of $\mathbf{g}_{t}$ and compare it with the variance decomposition of $\mathbf{x}_{t,1}$ (the second part of Table \ref{cfactsim_table}). The contribution of the responsive part to the variance of $\lbrace g_{i,t} \rbrace_{i=1}^{n}\vert_{t=1}^{T}$ (denoted by $\text{VarShare}(\mathbf{g}^{L})$) is only 1.24\%, while that of the autonomous transfer part is $\text{VarShare}(\mathbf{g}^{A}) = 94.93\%$. For the logged states' tax revenues per capita, we consider a similar decomposition scheme: (i) $\mathbf{x}_{t,1}^{L} = \mathbf{A}_{1}^{x}\mathbf{x}_{t-1,1} + \mathbf{B}_{1}^{x}\text{vec}(\mathbf{Y}_{t-1}) + C_{1}^{x}\mathbf{g}_{t-1}$ (responsive part) and (ii) $\mathbf{x}_{t,1}^{A} = \mathbf{c}_{1}^{x} + \alpha_{t,1}^{x}$ (fixed effect part). In contrast, the tax revenue's responsive part explains 12.16\% of the variance of $\lbrace x_{i,t,1} \rbrace_{i=1}^{n}\vert_{t=1}^{T}$, which is much larger than the contribution of the responsive part for the federal grant ($\mathbf{g}^L$).

Hence, the high reliance on autonomous transfers, mainly consisting of fixed-effect components, may weaken the federal government's role in improving welfare via adjustment of states' policy spillovers. To make effective policy decisions, this result urges federal policymakers to consider more responsive strategy for the federal grant decisions to states' policies.

\section{Conclusion}

This paper presents a novel dynamic spatial/network interaction model aimed at elucidating (i) the resource allocations from the allocator to local agents and (ii) the decision-making processes of local agents regarding multiple activities. We establish a network game and characterize the activities at the Markov perfect Nash equilibrium (MPNE). We derive estimation equations to identify and estimate the key structural parameters (e.g., agents' payoff parameters). The MPNE also offers two essential insights: (i) the marginal (short-term) effects, and (ii) their long-term impacts.

To estimate the structural parameters, we adopt the quasi-maximum likelihood (QML) method and delve into the large sample properties of the QML estimator. In an empirical application, we explore policy interdependence among U.S. states, focusing on expenditures related to public welfare and housing and community development. Our findings highlight the positive influences of federal grants on both expenditures. We show significant spillovers across the states’ two expenditures. Further, these two expenditures function as complements.

Counterfactual simulations substantiate that federal resource allocations enhance social welfare by motivating state-level efforts on PWE and HDCE. However, the impact of federal grants is modest relative to their substantial share in total state revenues, which aligns with our findings on the high reliance on autonomous transfers. The heavy dependence on automatic transfers, primarily composed of fixed-effect elements, could diminish the federal government's ability to enhance welfare through adjusting the policy spillovers between states. First, the findings underscore the dual role of federal intervention as both financial support and a strategic guiding mechanism for state policies. Second, our findings suggest that policymakers should adopt a more responsive strategy for making federal grant decisions to make effective central government policy decisions.

\section*{Appendix A: A list of notations}

\setcounter{theorem}{0}
\setcounter{assumption}{0}
\setcounter{remark}{0}
\setcounter{definition}{0}
\renewcommand{\thetheorem}{A.\arabic{theorem}}
\renewcommand{\theassumption}{A.\arabic{assumption}}
\renewcommand{\theremark}{A.\arabic{remark}}
\renewcommand{\thedefinition}{A.\arabic{definition}}%

We list some frequently used notations and also define
some new ones, which will be used later:

\textbf{Value function components}\\
$\mathbb{Q}_{i}$, $\mathbb{L}_{i}$, and $c_{i}$ for $i = 1,\cdots,n$: Local agent $i$'s value function components,\\
$\mathbf{z}_{t}^{A} = \left( \operatorname{vec}\left( \mathbf{Y}_{t-1}\right)^{\prime}, \left(l_{m^{*}} \otimes \mathbf{g}_{t}^{\ast}\right)^{\prime}, \operatorname{vec}(\mathbf{X}_{t})', \operatorname{vec}\left( \mathbf{U}_{t}\right)^{\prime }\right)^{\prime} $: Local agents' state variables at time $t$, \\
$\Rightarrow V_{i}(\mathbf{z}_{t}^{A}) = \mathbf{z}_{t}^{A\prime} \mathbb{Q}_{i}\mathbf{z}_{t}^{A} + \mathbf{z}_{t}^{A\prime}\mathbb{L}_{i} + c_{i}$,\\
$\mathbb{Q}_{0}$, $\mathbb{L}_{0}$, and $c_{0} =$Resource allocator's value function components,\\
$\mathbf{z}_{t}^{R}
= \left( \operatorname{vec}\left( \mathbf{Y}_{t-1}\right)^{\prime}, \operatorname{vec}(\mathbf{X}_{t})^{\prime}, \operatorname{vec}\left( \mathbf{E}_{t}\right)^{\prime},\left( l_{m^{*}}\otimes \boldsymbol{\tau}_{t}\right)^\prime \right)^{\prime}$: Resource allocator's state variables at time $t$\\
$\Rightarrow V_{0}(\mathbf{z}_{t}^{R}) = \mathbf{z}_{t}^{R\prime} \mathbb{Q}_{0}\mathbf{z}_{t}^{R} + \mathbf{z}_{t}^{R\prime}\mathbb{L}_{0} + c_{0}$.\\
Note that $\overline{\mathbf{A}} = \mathbf{A} + \mathbf{A}^\prime$ for any square matrix $\mathbf{A}$.

\textbf{Econometric model}\\
$\mathbf{R} =
\begin{bmatrix}
    \mathbf{R}_{1:n} & -\left(\phi \otimes \mathbf{I}_{n}\right) \\
    \mathbf{0} & \mathbf{R}_{0}
\end{bmatrix}$, \\ 
$\boldsymbol{\Delta} =
\begin{bmatrix}
    \Sigma \otimes \mathbf{I}_{n} & \left( \Sigma \otimes \mathbf{I}_{n}\right)
C_{e}^{g\prime} \mathbf{R}_{0}^{\prime} \\ 
\mathbf{R}_{0}C_{e}^{g}\left( \Sigma \otimes \mathbf{I}_{n}\right) & 
\mathbf{R}_{0} C_{e}^{g}\left( \Sigma \otimes \mathbf{I}_{n}\right)
C_{e}^{g\prime} \mathbf{R}_{0}^{\prime} + \sigma^{2}\mathbf{I}_{n}
\end{bmatrix}$,
\\ 
$\mathbf{S} = \left(P + \Psi \right) \otimes \mathbf{I}_{n} - \left( \Lambda^{\prime} \otimes \mathbf{W} \right)$, and\\
$\mathbf{S}_{0} = \left(P + \Psi \right) \otimes \mathbf{I}_{n} - \left( \Lambda^{\prime} \otimes \left(\mathbf{W} + \mathbf{W}^{\prime}\right) \right)$.

\section*{Appendix B: Agents' value functions and MPNE activities}

\setcounter{theorem}{0}
\setcounter{assumption}{0}
\setcounter{remark}{0}
\setcounter{definition}{0}
\renewcommand{\thetheorem}{B.\arabic{theorem}}
\renewcommand{\theassumption}{B.\arabic{assumption}}
\renewcommand{\theremark}{B.\arabic{remark}}
\renewcommand{\thedefinition}{B.\arabic{definition}}%

\renewcommand{\thetable}{B.\arabic{table}}
\renewcommand{\thefigure}{B.\arabic{figure}}

As the first issue, we will derive the recursive forms of $\lbrace V_{i}(\cdot) \rbrace_{i=0}^{n}$. Some formulas and derivatives for this are relegated to Section 1 of the supplement file. Second, we will specify sufficient conditions for the uniqueness of the MPNE. Without loss of generality, we assume \( \boldsymbol{\tau} = \mathbf{0} \) to simply derive the forms of the value functions.

\subsection*{Components of $V_{i}\left( \cdot \right)$ for $i = 1,\cdots,n$}

We can rewrite the local agent payoff function (\ref{m2-2-fpayoff}), which is useful for deriving the closed-form value functions:
\begin{equation}
    U_{i}\left(\mathbf{Y}_{t}, \mathbf{z}_{t}^{A} \right) =     \operatorname{vec}(\mathbf{Y}_{t})'\mathbf{L}_{Z,i}\mathbf{z}_{t}^{A} 
    + \mathbf{z}_{t}^{A\prime}\mathbf{Q}_{Z,i}\mathbf{z}_{t}^{A}
    + \operatorname{vec}(\mathbf{Y}_{t})' \mathbf{Q}_{Y,i} \operatorname{vec}(\mathbf{Y}_{t}),
    \label{m2-2-fpayoff2}
\end{equation}
where 
\begin{equation*}
    \begin{split}
       \mathbf{L}_{Z,i} & = \left[\mathbf{K}_{i}^\prime, \diag(\lbrace \phi_{l} \rbrace) \otimes e_{n,i}e_{n,i}^{\prime}, \Pi^\prime \otimes e_{n,i}e_{n,i}^\prime, \mathbf{I}_{m^{*}} \otimes e_{n,i}e_{n,i}^\prime \right], \\
\mathbf{Q}_{Z,i} & =
        \begin{bmatrix}
            - \frac{1}{2}(\mathbf{I}_{m^{*}} \otimes e_{n,i})P (\mathbf{I}_{m^{*}} \otimes e_{n,i}^\prime) & \mathbf{0} \\
            \mathbf{0} & \mathbf{0}
        \end{bmatrix},\text{ and}\\
\mathbf{Q}_{Y,i} & = \left[ \left( \mathbf{I}_{m^{*}}\otimes w_{i.}^{\prime}\right)
\Lambda - \frac{1}{2}\left(\mathbf{I}_{m^{*}}\otimes e_{n,i}\right) \left(P + \Psi \right) \right] \left( \mathbf{I}_{m^{*}}\otimes e_{n,i}^{\prime}\right)
    \end{split}
\end{equation*}
with $\mathbf{K}_{i} = \left[ \left(\mathbf{I}_{m^{*}}\otimes w_{i.}^{\prime}\right) 
\boldsymbol{\rho} + \left( \mathbf{I}_{m^{*}} \otimes e_{n,i}\right) P\right] \left(\mathbf{I}_{m^{*}} \otimes e_{n,i}^\prime\right)$ and \( w_{i.} \) denotes the \( i \)th row of \( \mathbf{W} \) for $i = 1,\cdots,n$.

By the LQ payoff specification, the local agent \( i \)'s value function is also an LQ form:
\begin{equation*}
V_{i}\left( \mathbf{z}_{t}^{A} \right) 
= \mathbf{z}_{t}^{A\prime}\mathbb{Q}_{i}\mathbf{z}_{t}^{A}
+ \mathbf{z}_{t}^{A\prime}\mathbb{L}_{i} + c_{i},
\end{equation*}
where 
\begin{equation}
    \mathbb{Q}_{i} =
    \begin{bmatrix}
        \mathbf{Q}_{i} & \mathbf{L}_{i}^{g} & \mathbf{L}_{i}^{x} \text{diag}\left(\lbrace \boldsymbol{\Pi}_{k} \rbrace \right) & \mathbf{L}_{i}^{u} \\
        \mathbf{0} & \mathbf{Q}_{i}^{g}  &  \mathbf{L}_{i}^{g,x} \text{diag}\left(\lbrace \boldsymbol{\Pi}_{k} \rbrace \right) & \mathbf{L}_{i}^{g,u} \\
        \mathbf{0} & \mathbf{0} & \text{diag}\left(\lbrace \boldsymbol{\Pi}_{k}^\prime \rbrace \right) \mathbf{Q}_{i}^{x}\text{diag}\left(\lbrace \boldsymbol{\Pi}_{k} \rbrace \right) & \text{diag}\left(\lbrace \boldsymbol{\Pi}_{k}^\prime \rbrace \right) \mathbf{L}_{i}^{x,u} \\
        \mathbf{0} & \mathbf{0} & \mathbf{0} & \mathbf{Q}_{i}^{u}
    \end{bmatrix},
    \label{QF_form}
\end{equation}
\begin{equation}
    \mathbb{L}_{i} =
\begin{bmatrix}
\mathbf{L}_{i}^{c\prime}, & \mathbf{L}_{i}^{g,c\prime}, & \mathbf{L}_{1,i}^{x,c\prime}\boldsymbol{\Pi}_{1}, & \cdots, & \mathbf{L}_{K,i}^{x,c\prime}\boldsymbol{\Pi}_{K} & 
\mathbf{L}_{i}^{u,c\prime}
\end{bmatrix}
^{\prime},
\label{LF_form}
\end{equation}
and $c_{i}$ is a scalar. For the representations $\lbrace \mathbb{Q}_{i} \rbrace_{i=1}^{n}$ above, note the following block matrix representations:
\begin{equation*}
\begin{split}
    \mathbf{L}_{i}^{x} & = \left[\mathbf{L}_{i,1}^{x}, \cdots, \mathbf{L}_{i,K}^{x} \right],\text{ }\mathbf{L}_{i}^{g,x} = \left[\mathbf{L}_{i,1}^{g,x}, \cdots, \mathbf{L}_{i,K}^{g,x} \right], \text{ }
    \mathbf{Q}_{i}^{x} =
    \begin{bmatrix}
        \mathbf{Q}_{i,11}^{x} & \cdots & \mathbf{Q}_{i,1K}^{x} \\
        \vdots & \ddots & \vdots \\
        \mathbf{Q}_{i,K1}^{x} & \cdots & \mathbf{Q}_{i,KK}^{x}
    \end{bmatrix}, \text{ }\mathbf{L}_{i}^{x,u} = \left[\mathbf{L}_{i,1}^{x,u\prime},\cdots,\mathbf{L}_{i,K}^{x,u\prime} \right]^\prime, 
\end{split}
\end{equation*}
$\boldsymbol{\Pi} = \left[\boldsymbol{\Pi}_{1},\cdots,\boldsymbol{\Pi}_{K} \right]$, and $\diag\left(\lbrace \mathbf{\Pi}_{k} \rbrace \right)$ and $\diag\left(\lbrace \mathbf{\Pi}_{k}^\prime \rbrace \right)$ denote respectively diagonal block matrices with blocks $\boldsymbol{\Pi}_{k}$ and $\boldsymbol{\Pi}_{k}^\prime$.

The corresponding optimal decision vector is 
\begin{equation*}
\operatorname{vec}\left( \mathbf{Y}_{t}^{\ast }\right) = \mathbf{\bar{c}}^{y} + \mathbb{L}^{y}\mathbf{z}_{t}^{A},
\end{equation*}
where $\mathbb{L}^{y}
=
\begin{bmatrix}
\mathbf{A}^{y}, & \mathbf{B}^{y}, & \mathbf{C}_{1}^{y}, & \cdots, & 
\mathbf{C}_{K}^{y}, & \mathbf{C}_{u}^{y}
\end{bmatrix}
$. We will provide formulas for the components in $\mathbb{L}^{y}$ later. The state evolution can be represented by 
\begin{equation*}
\mathbb{E}_{t,2}\left(\mathbf{z}_{t+1}^{A}\right) = \mathbb{c}_{1:n} + \mathbb{A}_{1:n} \mathbf{z}_{t}^{A},
\end{equation*}
where $\mathbb{c}_{1:n} = \left[ \mathbf{\bar{c}}^{y\prime}, \mathbf{\bar{c}}^{g\ast \prime} + 
\mathbf{\bar{c}}^{y\prime} \mathbf{A}^{g\ast\prime},\cdots ,\mathbf{\bar{c}}^{y\prime}\mathbf{B}_{k}^{x\prime}
+ \mathbf{c}_{k}^{x\prime}, \cdots ,\operatorname{vec}\left( \boldsymbol{\eta} \right)^{\prime}\right]^{\prime}$,
\begin{equation}
    \mathbb{A}_{1:n} = \left[ \mathbb{L}^{y\prime}, \mathbb{L}^{y\prime}
\mathbf{A}^{y \ast \prime} + \mathbb{L}^{g \mapsto x \mapsto y \prime} + \mathbb{L}^{y \mapsto g\prime}, 
\mathbb{L}^{y\prime}\mathbf{B}_{1}^{x\prime} + \mathbb{L}_{1}^{g \mapsto x\prime} + \mathbb{L}_{1}^{x\prime},\cdots, 
\mathbb{L}^{y\prime}\mathbf{B}_{K}^{x\prime} + \mathbb{L}_{K}^{g \mapsto x\prime} + \mathbb{L}_{K}^{x\prime}
,\mathbf{0}\right]^{\prime},
\label{AF_form}
\end{equation}
\begin{equation*}
\mathbb{L}^{y \mapsto g} =
\begin{bmatrix}
\mathbf{0}, & \mathbf{0}, & 
\mathbf{C}_{1}^{g} \left( \mathbf{I}_{m^{*}}\otimes
\mathbf{A}_{1}^{x} \right) \boldsymbol{\Pi}_{1}, & \cdots , & \mathbf{C}_{K}^{g}\left(
\mathbf{I}_{m^{*}}\otimes \mathbf{A}_{K}^{x} \right) \boldsymbol{\Pi}_{K}, & \mathbf{0}
\end{bmatrix},
\end{equation*}
\begin{equation*}
    \mathbb{L}_{k}^{g \mapsto x}
    = \left[0, \mathbf{C}_{k}^{x}, \mathbf{0}, \cdots, \mathbf{0} \right]\text{ with }\mathbf{C}_{k}^{x} = \left[\frac{1}{m^{*}}C_{k}^{x}, \cdots, \frac{1}{m^{*}}C_{k}^{x} \right],
\end{equation*}
\begin{equation*}
    \mathbb{L}^{g \mapsto x \mapsto y}
    = \left[\mathbf{0}, \sum_{k=1}^{K}\mathbf{C}_{k}^{g}\boldsymbol{\Pi}_{k}\mathbf{C}_{k}^{x}, \mathbf{0}, \cdots, \mathbf{0} \right],\text{ }\mathbb{L}_{k}^{x}=
\begin{bmatrix}
\mathbf{0}, & \cdots , & \mathbf{A}_{k}^{x}, & \cdots , & \mathbf{0}
\end{bmatrix},
\end{equation*}
$\mathbf{\bar{c}}^{g\ast}
= \mathbf{\bar{c}}^{g}
+ \sum_{k=1}^{K}\mathbf{C}_{k}^{g}
\boldsymbol{\Pi}_{k}\mathbf{c}_{k}$, 
and $\mathbf{A}^{g\ast} = \mathbf{A}^{g} + \sum_{k=1}^{K}\mathbf{C}_{k}^{g}
\boldsymbol{\Pi}_{k}\mathbf{B}_{k}^{x}$.

Consequently, $\mathbb{Q}_{i}$ and $\mathbb{L}_{i}$ for $i = 1,\cdots,n$ satisfy the following fixed-point relations: 
\begin{equation}
\begin{split}
    \mathbb{Q}_{i} & =  \mathbb{L}^{y\prime} \mathbf{L}_{Z,i} + \mathbf{Q}_{Z,i}
+ \mathbb{L}^{y\prime} \mathbf{Q}_{Y,i}\mathbb{L}^{y}  + \delta \mathbb{A}_{1:n}^{\prime}\mathbb{Q}_{i} \mathbb{A}_{1:n}
\text{, and} \\
    \mathbb{L}_{i} & =
\left(\mathbf{L}_{Z,i}^{\prime} + \mathbb{L}^{y\prime}
\overline{\mathbf{Q}_{Y,i}}
\right) \mathbf{\bar{c}}^{y}
+ \delta \mathbb{A}_{1:n}^{\prime}\left( \overline{\mathbb{Q}_{i}}\mathbb{c}_{1:n}
+ \mathbb{L}_{i} \right). \label{fp_followers}
\end{split}
\end{equation}
For the above, note that
\begin{equation*}
\begin{split}
    U_{i}\left(\mathbf{Y}_{t}^{*}, \mathbf{z}_{t}^{A} \right) & = \mathbf{z}_{t}^{A\prime}
    \left(\mathbf{L}_{Z,i}^\prime +  \mathbb{L}^{y\prime}
\overline{\mathbf{Q}_{Y,i}}\right)\mathbf{\bar{c}}^{y}
+ \mathbf{z}_{t}^{A\prime}
\left( \mathbb{L}^{y\prime} \mathbf{L}_{Z,i} + \mathbf{Q}_{Z,i}
+ \mathbb{L}^{y\prime} \mathbf{Q}_{Y,i}\mathbb{L}^{y} \right)\mathbf{z}_{t}^{A}
+ \mathbf{\bar{c}}^{y\prime} \mathbf{Q}_{Y,i}\mathbf{\bar{c}}^{y},\\
\mathbb{E}_{t,2}\left(\mathbf{z}_{t+1}^{A\prime} \mathbb{Q}_{i} \mathbf{z}_{t+1}^{A} \right) & = 
\mathbf{z}_{t}^{A\prime} \mathbb{A}_{1:n}^\prime \mathbb{Q}_{i} \mathbb{A}_{1:n} \mathbf{z}_{t}^{A}
+ \mathbf{z}_{t}^{A\prime} \mathbb{A}_{1:n}^\prime \overline{\mathbb{Q}}_{i} \mathbb{c}_{1:n} + \mathbb{c}_{1:n}^\prime \mathbb{Q}_{i} \mathbb{c}_{1:n},\text{ and } 
\mathbb{E}_{t,2}\left(\mathbf{z}_{t+1}^{A\prime}\mathbb{L}_{i} \right) = \mathbf{z}_{t}^{A\prime} \mathbb{A}_{1:n}^\prime \mathbb{L}_{i} + \mathbb{c}_{1:n}^\prime \mathbb{L}_{i}.
\end{split}
\end{equation*}

By the first-order condition for choosing $y_{i,t,l}$ (see Section 2 in the supplement file for details),
\begin{equation}
e_{nm^{*},il}^\prime \mathbf{Q}_{1:n} = e_{nm^{*},il}^\prime \left(
\overline{\mathbf{Q}}_{i} 
+ \overline{\mathbf{L}_{i}^{g}\mathbf{A}^{g\ast}}
+ \mathbf{A}^{g\ast \prime}
\overline{\mathbf{Q}_{i}^{g}}
\mathbf{A}^{g\ast}  + \sum_{k}\overline{\left( \mathbf{L}_{i,k}^{x} + \mathbf{A}^{g\ast \prime
} \mathbf{L}_{i,k}^{g,x}\right) \boldsymbol{\Pi}_{k}
\mathbf{B}_{k}^{x}} + \sum_{k_{1}, k_{2}}\mathbf{B}_{k_{1}}^{x\prime}\boldsymbol{\Pi}_{k_{1}}^{\prime}\overline{\mathbf{Q}_{i,k_{1}k_{2}}^{x}}\boldsymbol{\Pi}_{k_{2}}\mathbf{B}_{k_{2}}^{x}
\right),
\label{BQF_form}
\end{equation}
and
\begin{equation}
   e_{nm^{*},il}^\prime \mathbf{L}_{1:n,k}
= e_{nm^{*},il}^\prime \left( 
\mathbf{L}_{i,k}^{x} + \mathbf{A}^{g\ast \prime}\mathbf{L}_{i,k}^{g,x}
+ \left(\mathbf{L}_{i}^{g} + \mathbf{A}^{g\ast \prime}
\overline{\mathbf{Q}_{i}^{g}}\right) \mathbf{C}_{k}^{g} 
+ \sum_{p}\mathbf{B}_{p}^{x\prime}\boldsymbol{\Pi}_{p}^{\prime }\left(\mathbf{L}_{i,p}^{g,x\prime}
\mathbf{C}_{k}^{g} + \overline{\mathbf{Q}_{i,pk}^{x}} \right)
\right),
\label{BLF_form}
\end{equation}
for $k=1,\cdots,K$. Here, $e_{nm^{*},il} = (e_{m^{*},l} \otimes e_{n,i})$ for $i=1,\cdots,n$ and $l=1,\cdots,m^{*}$.

Define $\mathbf{R}_{1:n} = \mathbf{S} - \delta\mathbf{Q}_{1:n}$ with $\mathbf{S} = \left( P + \Psi \right) \otimes \mathbf{I}_{n} - \left( \Lambda^{\prime} \otimes \mathbf{W} \right)$. Then, the first-order conditions yield the following equation system:
\begin{equation}
    \mathbf{R}_{1:n}\operatorname{vec}(\mathbf{Y}_{t}^{*}) = \ddot{\overline{\mathbf{c}}}^{y} + \ddot{\mathbb{L}}^{y}\mathbf{z}_{t}^{A},
    \label{follower_foc}
\end{equation}
where $\ddot{\overline{\mathbf{c}}}^{y}$ and $\ddot{\mathbb{L}}^{y}$ satisfy $\overline{\mathbf{c}}^{y} = \left(\mathbf{R}_{1:n} \right)^{-1}\ddot{\overline{\mathbf{c}}}^{y}$ and $\mathbb{L}^{y} = \left(\mathbf{R}_{1:n} \right)^{-1} \ddot{\mathbb{L}}^{y}$. If the invertibility of $\mathbf{R}_{1:n}$ is verified, the local agents' MPNE activities can be derived as 
\begin{equation*}
\operatorname{vec}\left( \mathbf{Y}_{t}^{\ast}\right) = \mathbf{\bar{c}}^{y} + \mathbb{L}^{y}\mathbf{z}_{t}^{A}
= \mathbf{\bar{c}}^{y}
+ \mathbf{A}^{y} \operatorname{vec}\left(\mathbf{Y}_{t-1}\right) + \mathbf{B}^{y}\left(l_{m^{*}}\otimes \mathbf{g}_{t}^{\ast}\right) + \sum_{k=1}^{K}\mathbf{C}_{k}^{y}
\boldsymbol{\Pi}_{k}\mathbf{x}_{t,k}
+ \mathbf{C}_{u}^{y}
\operatorname{vec}\left(\mathbf{U}_{t}\right),
\end{equation*}
where $\mathbf{R}_{1:n} = \mathbf{S} - \delta\mathbf{Q}_{1:n}$ with $\mathbf{S} = \left( P + \Psi \right) \otimes \mathbf{I}_{n} - \left( \Lambda^{\prime} \otimes \mathbf{W} \right)$, and $\mathbf{\bar{c}}^{y}
= \left( \mathbf{R}_{1:n}\right)^{-1}\mathbf{\bar{c}}^{y\ast}$. Here, we have the detailed formulas for the components in $\mathbb{L}^{y}$: $\mathbf{A}^{y} = \left(\mathbf{R}_{1:n}\right)^{-1}
\left[\left( P \otimes \mathbf{I}_{n}\right) + \left( \boldsymbol{\rho}^{\prime} \otimes \mathbf{W} \right) \right]$, $\mathbf{B}^{y} = \left( \mathbf{R}_{1:n} \right)^{-1}\left( \diag(\lbrace \phi_{l} \rbrace) \otimes \mathbf{I}_{n}\right)$, $\mathbf{C}_{k}^{y}
=\left(\mathbf{R}_{1:n} \right)^{-1}\left[ \mathbf{I}_{nm^{*}} + \delta \mathbf{L}_{1:n,k} \left( \mathbf{I}_{m^{*}}\otimes \mathbf{A}_{k}^{x}\right) \right]$ for $k=1,\cdots,K$, and $\mathbf{C}_{u}^{y} = \left( \mathbf{R}_{1:n} \right)^{-1}$ with $\mathbf{R}_{1:n} = \mathbf{S} - \delta \mathbf{Q}_{1:n}$. We do not provide the formula for the fixed-effect components ($\mathbf{\bar{c}}^{y\ast}$).

\subsection*{Components of $V_{0}\left( \cdot \right) $}

Given $\mathbf{z}_{t}^{R}$, the resource allocator's optimal activities can be derived from 
\begin{equation*}
V_{0}\left( \mathbf{z}_{t}^{R} \right) =
\mathbf{z}_{t}^{R\prime}\mathbb{Q}_{0}
\mathbf{z}_{t}^{R}
+ \mathbf{z}_{t}^{R\prime}\mathbb{L}_{0} + c_{0},
\end{equation*}
where 
\begin{equation}
    \mathbb{Q}_{0} =
    \begin{bmatrix}
        \mathbf{Q}_{0} & \mathbf{L}_{0}^{x}\text{diag}\left(\lbrace \boldsymbol{\Pi}_{k} \rbrace \right) & \mathbf{L}_{0}^{u} & \mathbf{L}_{0}^{\tau} \\
        \mathbf{0} & \text{diag}\left(\lbrace \boldsymbol{\Pi}_{k}^\prime \rbrace \right) \mathbf{Q}_{0}^{x} \text{diag}\left(\lbrace \boldsymbol{\Pi}_{k} \rbrace \right) & \text{diag}\left(\lbrace \boldsymbol{\Pi}_{k}^\prime \rbrace \right) \otimes \mathbf{L}_{0}^{x,u} & 
        \text{diag}\left(\lbrace \boldsymbol{\Pi}_{k}^\prime \rbrace \right) \mathbf{L}_{0}^{x,\tau} \\
        \mathbf{0} & \mathbf{0} & \mathbf{Q}_{0}^{u} & \mathbf{L}_{0}^{u,\tau} \\
        \mathbf{0} & \mathbf{0} & \mathbf{0} & \mathbf{Q}_{0}^{\tau}
    \end{bmatrix},
    \label{QL_form}
\end{equation}
\begin{equation}
    \mathbb{L}_{0} =
\begin{bmatrix}
\mathbf{L}_{0}^{c\prime}, & \mathbf{L}_{0,1}^{x,c\prime }\boldsymbol{\Pi}_{1}, & \cdots, & 
\mathbf{L}_{0,K}^{x,c\prime}\boldsymbol{\Pi}_{K}, & \mathbf{L}_{0}^{u,c\prime} & \mathbf{L}_{0}^{\tau,c\prime}
\end{bmatrix}
^{\prime},
\label{LL_form}
\end{equation}
and $c_{0}$ are scalars. For the representations $\mathbb{Q}_{0}$ above, note the following block matrix representations:
\begin{equation*}
\begin{split}
    \mathbf{L}_{0}^{x} & = \left[\mathbf{L}_{0,1}^{x},\cdots,\mathbf{L}_{0,K}^{x} \right],\text{ }\mathbf{Q}_{0}^{x} = 
        \begin{bmatrix}
        \mathbf{Q}_{0,11}^{x} & \cdots & \mathbf{Q}_{0,1K}^{x} \\
        \vdots & \ddots & \vdots \\
        \mathbf{Q}_{0,K1}^{x} & \cdots & \mathbf{Q}_{0,KK}^{x}
    \end{bmatrix},\text{ }
    \mathbf{L}_{0}^{x,u} = \left[\mathbf{L}_{0,1}^{x,u\prime}, \cdots, \mathbf{L}_{0,K}^{x,u\prime} \right]^\prime,\text{ }
    \mathbf{L}_{0}^{x,\tau} = \left[\mathbf{L}_{0,1}^{x,\tau\prime}, \cdots, \mathbf{L}_{0,K}^{x,\tau\prime} \right]^\prime.
\end{split}
\end{equation*}
The corresponding optimal grants are
\begin{equation*}
\left( l_{m}\otimes \mathbf{g}_{t}^{\ast}\right) = \mathbf{\bar{c}}^{g}
+ \mathbb{L}^{g}\mathbf{z}_{t}^{R}
\end{equation*}
with $\mathbb{L}^{g} =
\begin{bmatrix}
\mathbf{A}^{g}, & \cdots \mathbf{C}_{k}^{g} \boldsymbol{\Pi}_{k},\cdots, & \mathbf{C}_{e}^{g}, & \mathbf{B}^{g}
\end{bmatrix}$. We will provide formulas for the components in $\mathbb{L}^{g}$ later.

To derive the form of \(V_{0}(\mathbf{z}_{t}^{R})\), two expected values, \(\mathbb{E}_{t,1}(\mathbf{z}_{t}^{A})\) and \(\mathbb{E}_{t,1}(\mathbf{z}_{t+1}^{R})\), are required. Since the allocator faces uncertainty in the first stage regarding the local agents' decisions in the second stage, evaluating \(\mathbb{E}_{t,1}(\mathbf{z}_{t}^{A})\) is necessary. Additionally, \(\mathbb{E}_{t,1}(\mathbf{z}_{t+1}^{R})\) accounts for uncertainty in future economic environments.

First, note that 
\begin{equation*}
\mathbb{E}_{t,1}(\mathbf{z}_{t}^{A}) = \mathbb{c}_{1:n}^{R} + \mathbb{A}_{1:n}^{R}\mathbf{z}_{t}^{R},
\end{equation*}
where $\mathbb{c}_{1:n}^{R}
=
\begin{bmatrix}
\mathbf{0} \\ 
\mathbf{\bar{c}}^{g} \\ 
\mathbf{0} \\ 
\operatorname{vec} \left( \boldsymbol{\eta} \right)
\end{bmatrix}
$ and $\mathbb{A}_{1:n}^{R} =
\begin{bmatrix}
\mathbf{I}_{nm^{*}} & \mathbf{0} & \mathbf{0} & \mathbf{0} \\ 
\mathbf{A}^{g} & \cdots \mathbf{C}_{k}^{g} \boldsymbol{\Pi}_{k}\cdots  & 
\mathbf{C}_{e}^{g} & \mathbf{B}^{g} \\ 
\mathbf{0} & \mathbf{I}_{nK} & \mathbf{0} & \mathbf{0} \\ 
\mathbf{0} & \mathbf{0} & \mathbf{I}_{nm^{*}} & \mathbf{0}
\end{bmatrix}
$.

Second,
\begin{equation*}
\mathbb{E}_{t,1}(\mathbf{z}_{t+1}^{R})
= \mathbb{c}_{0} + \mathbb{A}_{0} \mathbf{z}_{t}^{R},
\end{equation*}
where $\mathbb{c}_{0} =
\begin{bmatrix}
\mathbf{c}^{R\prime}, & \mathbf{c}_{1}^{x\prime} + \mathbf{c}^{R\prime}\mathbf{B}_{1}^{x\prime} + \bar{\mathbf{c}}^{g\prime}, & \cdots, & \mathbf{c}_{K}^{x\prime} + \mathbf{c}^{R\prime}\mathbf{B}_{K}^{x\prime} + \bar{\mathbf{c}}^{g\prime}, & 
\mathbf{\bar{c}}^{y\prime}\mathbf{B}_{K}^{x\prime} + \mathbf{c}_{K}^{x\prime}, & \operatorname{vec}\left( \boldsymbol{\eta }\right)^{\prime}
\end{bmatrix}
^{\prime}$,
\begin{equation}
\mathbb{A}_{0} =
\begin{bmatrix}
\mathbf{A}^{R} & \mathbf{B}_{1}^{R}\boldsymbol{\Pi}_{1} & \cdots & \mathbf{B}_{K}^{R}\boldsymbol{\Pi}_{K} & \mathbf{B}_{u}^{R} & \mathbf{C}^{R}
\\ 
\mathbf{B}_{1}^{x}\mathbf{A}^{R} + \mathbf{C}_{1}^{x}\mathbf{A}^{g} & \mathbf{A}_{1}^{x} + (\mathbf{B}_{1}^{x}\mathbf{B}_{1}^{R} + \mathbf{C}_{1}^{x}\mathbf{C}_{1}^{g} )\boldsymbol{\Pi}_{1} & \cdots & (\mathbf{B}_{K}^{x}\mathbf{B}_{K}^{R} + \mathbf{C}_{K}^{x}\mathbf{C}_{K}^{g})\boldsymbol{\Pi}_{K} & \mathbf{B}_{1}^{x}\mathbf{B}_{u}^{R} + \mathbf{C}_{1}^{x}\mathbf{C}_{e}^{g} & \mathbf{B}_{1}^{x}\mathbf{C}^{R} + \mathbf{C}_{1}^{x}\mathbf{B}^{g} \\ 
\vdots & \vdots & \cdots & \ddots & \vdots & \vdots \\ 
\mathbf{B}_{K}^{x}\mathbf{A}^{R} + \mathbf{C}_{K}^{x}\mathbf{A}^{g} & (\mathbf{B}_{1}^{x}\mathbf{B}_{1}^{R} + \mathbf{C}_{1}^{x}\mathbf{C}_{1}^{g})\boldsymbol{\Pi}_{1} & \cdots & \mathbf{A}_{K}^{x} + (\mathbf{B}_{K}^{x}\mathbf{B}_{K}^{R} + \mathbf{C}_{K}^{x}\mathbf{C}_{K}^{g})\boldsymbol{\Pi}_{K} & \mathbf{B}_{K}^{x}\mathbf{B}_{u}^{R} + \mathbf{C}_{K}^{x}\mathbf{C}_{e}^{g} & 
\mathbf{B}_{K}^{x}\mathbf{C}^{R} + \mathbf{C}_{K}^{x}\mathbf{B}^{g} \\ 
\mathbf{0} & \mathbf{0} & \cdots & \mathbf{0} & \mathbf{0} & \mathbf{0}
\end{bmatrix},
\label{AL_form}
\end{equation}
$\mathbf{c}^{R} = \mathbf{\bar{c}}^{y} + \mathbf{B}^{y}\mathbf{\bar{c}}^{g}
+ \mathbf{C}_{u}^{y}\boldsymbol{\eta}$, 
$\mathbf{A}^{R} = \mathbf{A}^{y} + \mathbf{B}^{y}\mathbf{A}^{g}$, $\mathbf{B}_{k}^{R}
= \mathbf{C}_{k}^{y} + \mathbf{B}^{y}\mathbf{C}_{k}^{g}$ for $k=1,\cdots,K$, $\mathbf{B}_{u}^{R}
= \mathbf{C}_{u}^{y} + 
\mathbf{B}^{y}\mathbf{C}_{u}^{y}$, and $\mathbf{C}^{R} = \mathbf{B}^{y}\mathbf{B}^{g}$.

We provide the recursive formulas for $\mathbb{Q}_{0}$ and $\mathbb{L}_{0}$ as follow. Since $c_{0}$ is not
relevant to the allocator's optimization, we do not provide its formula. Since $V_{0}\left( \cdot \right)$ is based on the summation of local agents' payoffs, we have $\mathbb{Q}_{0} = \sum_{i=1}^{n}\mathbb{Q}_{0,i}$ and $\mathbb{L}_{0} = \sum_{i=1}^{n}\mathbb{L}_{0,i}$ with their $i$ components
derived recursively below: for $i=1,\cdots,n$, 
\begin{equation}
\begin{split}
    \mathbb{Q}_{0,i} = & \mathbb{A}_{1:n}^{R\prime}\left( \mathbb{L}^{y\prime} \mathbf{L}_{Z,i} 
    + \mathbf{Q}_{Z,i} + \mathbb{L}^{y\prime} \mathbf{Q}_{Y,i} \mathbb{L}^{y} \right) \mathbb{A}_{1:n}^{R}
    - \frac{1}{2m^{*}}\mathbb{L}^{g\prime}\left( \mathbf{I}_{m}\otimes e_{n,i}e_{n,i}^{\prime}\right) \mathbb{L}^{g}
    + \frac{1}{m}\left[ \mathbf{0}, \mathbf{I}_{nm^{*}}\right]^{\prime}
    \left(\mathbf{I}_{m}\otimes e_{n,i}e_{n,i}^{\prime}\right) \mathbb{L}^{g} \\
    & + \delta \mathbb{A}_{0}^{\prime}\mathbb{Q}_{0,i}\mathbb{A}_{0}
    \text{, and} \\
    \mathbb{L}_{0,i} = & \mathbb{A}_{1:n}^{R\prime}
    \left( \left( \overline{\mathbb{L}^{y\prime} \mathbf{L}_{Z,i} }  
    + \overline{\mathbf{Q}_{Z,i}} + \mathbb{L}^{y\prime} \overline{\mathbf{Q}_{Y,i}} \right)  \mathbb{c}_{1:n}^{L} + \left(\mathbf{L}_{Z,i}^\prime + 
    \mathbb{L}^{y\prime} \overline{\mathbf{Q}_{Y,i}} \right) \mathbf{\bar{c}}^{y}\right) \\
    & -\frac{1}{m^{*}}\left( \mathbb{L}^{g} 
    - \left[ \mathbf{0}, \mathbf{I}_{nm^{*}}\right]\right)^{\prime}\left( \mathbf{I}_{m^{*}}\otimes e_{n,i}e_{n,i}^{\prime}\right) 
    \mathbf{\bar{c}}^{g}
    + \delta \mathbb{A}_{0}^{\prime}\left(  
    \overline{\mathbb{Q}_{0,i}} \mathbb{c}_{0} + \mathbb{L}_{0,i} \right). \label{fp_leader}
\end{split}
\end{equation}
For the above, note that
\begin{equation*}
    \begin{split}
        \mathbb{E}_{t,1}\left(\operatorname{vec}(\mathbf{Y}_{t}^{*})^\prime \mathbf{L}_{Z,i} \mathbf{z}_{t}^{A} \right) & = \mathbf{z}_{t}^{R\prime} \mathbb{A}_{1:n}^{R\prime} \mathbb{L}^{y\prime} \mathbf{L}_{Z,i}\mathbb{A}_{1:n}^{R} \mathbf{z}_{t}^{R}
        + \mathbf{z}_{t}^{R\prime} \mathbb{A}_{1:n}^{R\prime}
        \left(\mathbf{L}_{Z,i}^\prime \bar{\mathbf{c}}^{y} + \overline{\mathbb{L}^{y\prime}\mathbf{L}_{Z,i}} \mathbb{c}_{1:n}^{R} \right) + \text{constant},\\
        \mathbb{E}_{t,1}\left(\mathbf{z}_{t}^{A\prime} \mathbf{Q}_{Z,i} \mathbf{z}_{t}^{A} \right) & =
        \mathbf{z}_{t}^{R\prime} \mathbb{A}_{1:n}^{R\prime} \mathbf{Q}_{Z,i} \mathbb{A}_{1:n}^{R} \mathbf{z}_{t}^{R}
        + \mathbf{z}_{t}^{R\prime} \mathbb{A}_{1:n}^{R\prime}
        \overline{\mathbf{Q}_{Z,i}} \mathbb{c}_{1:n}^{R},\text{ and}\\
        \mathbb{E}_{t,1}\left(\operatorname{vec}(\mathbf{Y}_{t}^{*})^\prime \mathbf{Q}_{Y,i} \operatorname{vec}(\mathbf{Y}_{t}^{*}) \right) & = \mathbf{z}_{t}^{R\prime} \mathbb{A}_{1:n}^{R\prime} \mathbb{L}^{y\prime} \mathbf{Q}_{Y,i} \mathbb{L}^{y}\mathbb{A}_{1:n}^{R}\mathbf{z}_{t}^{R} + \mathbf{z}_{t}^{R\prime} \mathbb{A}_{1:n}^{R\prime} \mathbb{L}^{y\prime}
        \overline{\mathbf{Q}_{Y,i}}\left(\mathbb{c}_{1:n}^{R} + \bar{\mathbf{c}}^{y} \right).
    \end{split}
\end{equation*}

By the first-order condition for choosing $g_{i,t}$ (see Section 2 in the supplement file), define $\mathbf{R}_{0} = 
\mathbf{I}_{n} - \mathbf{T}_{0}$, where
\begin{equation}
   \begin{split}
    \mathbf{T}_{0} =&  \overline{\left( \phi^\prime \otimes \mathbf{I}_{n}\right) \mathbf{B}^{y} \left(l_{m^{*}}\otimes \mathbf{I}_{n}\right)} 
 - \left( l_{m^{*}}^\prime \otimes \mathbf{I}_{n}\right) \mathbf{B}^{y\prime} \mathbf{S}_{0} \mathbf{B}^{y}
 \left(l_{m^{*}} \otimes \mathbf{I}_{n}\right) + \delta 
 \left(l_{m^{*}}^\prime \otimes \mathbf{I}_{n}\right) \mathbf{B}^{y\prime}
 \mathbf{Q}_{0}^{*} \mathbf{B}^{y} \left( l_{m^{*}}\otimes \mathbf{I}_{n}\right) \\
 & + \delta \sum_{k=1}^{K}\overline{ \left( l_{m^{*}}^{\prime} \otimes \mathbf{I}_{n}\right) 
\mathbf{B}^{y\prime}\mathbf{L}_{0,k}^{x} \boldsymbol{\Pi}_{k}C_{k}^{x} } 
+ \delta \sum_{k=1}^{K}\sum_{l=1}^{K}C_{k}^{x\prime} \boldsymbol{\Pi}_{k}^\prime \overline{\mathbf{Q}_{0,kl}^{x}}\boldsymbol{\Pi}_{l}C_{l}^{x} \\
& + \delta \sum_{k=1}^{K}\sum_{l=1}^{K}\left(
\left(l_{m^{*}}^\prime \otimes \mathbf{I}_{n}\right) \mathbf{B}^{y\prime}\mathbf{B}_{k}^{x\prime}\boldsymbol{\Pi}_{k}^\prime \overline{\mathbf{Q}_{0,kl}^{x}}\boldsymbol{\Pi}_{l} C_{l}^{x} + C_{l}^{x\prime} \boldsymbol{\Pi}_{l}^\prime \overline{\mathbf{Q}_{0,lk}^{x}} \boldsymbol{\Pi}_{k}\mathbf{B}_{k}^{x}\mathbf{B}^{y}\left(l_{m^{*}} \otimes \mathbf{I}_{n}\right)
\right),
\end{split}
 \label{T0_form}
\end{equation}
with $\mathbf{S}_{0} = \left(P + \Psi \right) \otimes \mathbf{I}_{n} - \left( \Lambda^{\prime}\otimes 
        \overline{\mathbf{W}} \right)$ and $\mathbf{Q}_{0}^{*} = \overline{\mathbf{Q}_{0}} + \sum_{k=1}^{K}\overline{\mathbf{L}_{0,k}\boldsymbol{\Pi}_{k}\mathbf{B}_{k}^{x}} + \sum_{k=1}^{K}\sum_{l=1}^{K}\mathbf{B}_{k}^{x\prime} \boldsymbol{\Pi}_{k}^\prime \overline{\mathbf{Q}_{0,kl}^{x}}\boldsymbol{\Pi}_{l}\mathbf{B}_{l}^{x}$. Then, we obtain the following system of equations from the first-order conditions:
\begin{equation}
    \left(\mathbf{I}_{m^{*}} \otimes \mathbf{R}_{0}  \right)\left(l_{m^{*}} \otimes \mathbf{g}_{t}^{*} \right) = \ddot{\overline{\mathbf{c}}}^{g} + \ddot{\mathbb{L}}^{g}\mathbf{z}_{t}^{R},
    \label{leader_foc}
\end{equation}
where $\ddot{\overline{\mathbf{c}}}^{g}$ and $\ddot{\mathbb{L}}^{g}$ satisfy $\overline{\mathbf{c}}^{g} =  \left(\mathbf{I}_{m^{*}} \otimes \mathbf{R}_{0}^{-1}  \right)\ddot{\overline{\mathbf{c}}}^{g}$ and $\mathbb{L}^{g} = \left(\mathbf{I}_{m^{*}} \otimes \mathbf{R}_{0}^{-1}  \right)\ddot{\mathbb{L}}^{g}$ if $\mathbf{R}_{0}$ is invertible.

If invertibility of $\mathbf{R}_{0}$ is guaranteed, we can derive the optimal grant at $t$:
\begin{equation*}
\mathbf{g}_{t}^{\ast} =
\bar{c}^{g} + A^{g} \operatorname{vec}\left(\mathbf{Y}_{t-1}\right) 
+ B^{g} \boldsymbol{\tau}_{t}
+ \sum_{k=1}^{K}C_{k}^{g} \boldsymbol{\Pi}_{k}\mathbf{x}_{t,k}
+ C_{e}^{g} \operatorname{vec}\left( \mathbf{E}_{t}\right).
\end{equation*}
Here, we have the detailed formulas for the components in $\mathbb{L}^{g}$. Note that $\mathbf{A}^{g} = l_{m^{*}} \otimes A^{g}$, $\mathbf{B}^{g} = \mathbf{I}_{m^{*}} \otimes B^{g}$, $\mathbf{C}_{e}^{g} = l_{m^{*}} \otimes C_{e}^{g}$, and $\mathbf{C}_{k}^{g} = l_{m^{*}} \otimes C_{k}^{g}$ for $k=1,\cdots,K$, where
\begin{equation*}
B^{g} = \mathbf{R}_{0}^{-1},
\end{equation*}
\begin{equation*}
\begin{split}
    \bar{A}^{g} = & B^g
    \left( 
    \begin{array}{l}
      (l_{m^*}^\prime \otimes \mathbf{I}_{n}) \mathbf{B}^{y\prime}\left( (P \otimes \mathbf{I}_{n}) + ( \boldsymbol{\rho}^\prime \otimes \mathbf{W}) \right) + ((\phi^\prime \otimes \mathbf{I}_{n}) - (l_{m^*}^\prime \otimes \mathbf{I}_{n}) \mathbf{B}^{y\prime}\mathbf{S}_{0} )\mathbf{A}^{y} + \delta (l_{m^*}^\prime \otimes \mathbf{I}_{n}) \mathbf{B}^{y\prime} \mathbf{Q}_{0}^* \mathbf{A}^{y}   \\
      + \delta \sum_{k=1}^{K}C_{k}^{x\prime}\boldsymbol{\Pi}_{k}^\prime \left(\mathbf{L}_{0,k}^{x\prime} + \sum_{l=1}^{K}\overline{\mathbf{Q}_{0,kl}^{x}}\boldsymbol{\Pi}_{l} \mathbf{B}_{l}^{x} \right)\mathbf{A}^{y}    
    \end{array}
    \right),
\end{split}
\end{equation*}
\begin{equation*}
    \begin{split}
        \bar{C}_{k}^{g} = & 
        B^g\left(
        \begin{array}{l}
           (l_{m^*}^\prime \otimes \mathbf{I}_{n} )\mathbf{B}^{y\prime} + ((\phi^\prime \otimes \mathbf{I}_{n}) - (l_{m^*}^\prime \otimes \mathbf{I}_{n}) \mathbf{B}^{y\prime}\mathbf{S}_{0})\mathbf{C}_{k}^{y} 
        + \delta (l_{m^*}^\prime \otimes \mathbf{I}_{n} )\mathbf{B}^{y\prime}\mathbf{Q}_{0}^{*}\mathbf{C}_{k}^{y} \\
        + \delta(l_{m^*}^\prime \otimes \mathbf{I}_{n} )\mathbf{B}^{y\prime}\left(\mathbf{L}_{0,k}^{x} + \sum_{l=1}^{K} \mathbf{B}_{l}^{x\prime} \boldsymbol{\Pi}_{l}^\prime \overline{\mathbf{Q}_{0,lk}^{x}} \right)(\mathbf{I}_{m^*} \otimes \mathbf{A}_{k}^{x}) \\
        + \delta \left(\sum_{l=1}^{K}C_{l}^{x\prime}\boldsymbol{\Pi}_{l}^\prime \left(\mathbf{L}_{0,l}^{x\prime} + \sum_{p=1}^{K}   \overline{\mathbf{Q}_{0,lp}^{x}} \boldsymbol{\Pi}_{p} \mathbf{B}_{p}^{x}\right)  \right)\mathbf{C}_{k}^{y}
        + \delta \sum_{l=1}^{K}C_{l}^{x\prime} \boldsymbol{\Pi}_{l}^\prime \overline{\mathbf{Q}_{0,lk}^{x}}(\mathbf{I}_{m^*} \otimes \mathbf{A}_{k}^{x})
        \end{array}
        \right),\text{ for }k = 1,\cdots,K,
    \end{split}
\end{equation*}
and
\begin{equation*}
    \begin{split}
        \bar{C}_{e}^{g} = & 
        B^g\left(
        \begin{array}{l}
           (l_{m^*}^\prime \otimes \mathbf{I}_{n} )\mathbf{B}^{y\prime} + ((\phi^\prime \otimes \mathbf{I}_{n}) - (l_{m^*}^\prime \otimes \mathbf{I}_{n}) \mathbf{B}^{y\prime}\mathbf{S}_{0})\mathbf{C}_{u}^{y} 
        + \delta (l_{m^*}^\prime \otimes \mathbf{I}_{n} )\mathbf{B}^{y\prime}\mathbf{Q}_{0}^{*}\mathbf{C}_{u}^{y} \\
           + \delta \left(\sum_{k=1}^{K}C_{k}^{x\prime}\boldsymbol{\Pi}_{k}^\prime \left(\mathbf{L}_{0,k}^{x\prime} + \sum_{l=1}^{K}   \overline{\mathbf{Q}_{0,kl}^{x}} \boldsymbol{\Pi}_{l} \mathbf{B}_{l}^{x}\right)  \right)\mathbf{C}_{u}^{y}   
        \end{array}
        \right).
    \end{split}
\end{equation*} 
We do not provide the formula for $\bar{c}^{g}$.

\subsection*{MPNE uniqueness}

We will specify a set of $(\theta_{P}, \theta_{E}, \delta, \mathbf{W})$, denoted by $\mathcal{M}$, such that $\operatorname{vec}(\mathbf{Y}_{t}^{\ast})$ and $\mathbf{g}_{t}^{\ast}$ are unique if $(\theta_{P}, \theta_{E}, \delta, \mathbf{W}) \in \mathcal{M}$. Further details are provided in the following paragraphs.

Consider the invertibility condition. By the first-order conditions, \eqref{follower_foc} and \eqref{leader_foc}, we need to guarantee the invertibility of the two matrices, $\mathbf{R}_{1:n}$ and $\mathbf{R}_{0}$. Assuming the diagonal dominance property of $P + \Psi$ leads to having the following representation:
\begin{equation*}
    \mathbf{R}_{1:n} = \left((P + \Psi) \otimes \mathbf{I}_{n} \right)\left(\mathbf{I}_{nm^{*}} - \mathbf{T}_{1:n} \right)
\end{equation*}
where $\mathbf{T}_{1:n} = (P + \Psi)^{-1}\Lambda^\prime \otimes \mathbf{W} + \delta\left((P + \Psi)^{-1} \otimes \mathbf{I}_{n} \right)\mathbf{Q}_{1:n}$. Similarly, we can represent $\mathbf{R}_{0}$ as $\mathbf{R}_{0} = \mathbf{I}_{n} - \mathbf{T}_{0}$, where $\mathbf{T}_{0}$ is defined in \eqref{T0_form}. Hence, we define
\begin{equation*}
    \mathcal{I} = \left\lbrace (\theta_{P}, \theta_{E}, \delta, \mathbf{W}): \Vert \mathbf{T}_{1:n} \Vert_{2} < 1 \text{ and }\Vert \mathbf{T}_{0} \Vert_{2} < 1 \right\rbrace.
\end{equation*}
If $(\theta_{P}, \theta_{E}, \delta, \mathbf{W}) \in \mathcal{I}$, $\mathbf{R}_{1:n}$ and $\mathbf{R}_{0}$ are invertible. Further, $\mathbf{R}_{1:n}^{-1} = \left(\sum_{k=0}^{\infty}\mathbf{T}_{1:n}^{k} \right) \cdot \left((P + \Psi)^{-1} \otimes \mathbf{I}_{n} \right)$ and $\mathbf{R}_{0} = \sum_{k=0}^{\infty}\mathbf{T}_{0}^{k}$. The restrictions imposed by $\mathcal{I}$ guarantee the unique representations of the optimal activities when the value functions are given. Its interpretation is provided in the remark below.

\begin{remark}
\label{remark_invertibility}
When $\delta = 0$, observe that 
\begin{equation*}
        \mathbf{R}_{1:n} = \mathbf{S} = ((P + \Psi) \otimes \mathbf{I}_{n}) \left(\mathbf{I}_{nm^{*}} - \mathbf{T}_{1:n} \right), \text{ and }
        \mathbf{R}_{0} = \mathbf{I}_{n} - \mathbf{T}_{0},
\end{equation*}
where $\mathbf{T}_{1:n} = (P + \Psi)^{-1} \Lambda^{\prime}\otimes \mathbf{W}$ and $\mathbf{T}_{0} = (\phi^{\prime} \otimes \mathbf{I}_{n})(\overline{\mathbf{S}^{-1\prime}} - \mathbf{S}^{-1\prime}\mathbf{S}_{0} \mathbf{S}^{-1})(\phi \otimes \mathbf{I}_{n})$ with $\mathbf{S}_{0} = \left(P + \Psi \right) \otimes \mathbf{I}_{n} - \left( \Lambda^{\prime} \otimes \overline{\mathbf{W}} \right)$.

First, consider the invertibility of $\mathbf{S} = \mathbf{R}_{1:n}$. A sufficient condition for the invertibility of $\mathbf{S}$ is 
\begin{equation*}
\Vert \mathbf{T}_{1:n} \Vert_{2} = \Vert (P + \Psi)^{-1}\Lambda^{\prime} \Vert_{2} \cdot \Vert \mathbf{W} \Vert_{2} < 1.
\end{equation*}
When $\left(\theta_{P}, \mathbf{W} \right)$ satisfies the condition above, we have a unique solution to equation \eqref{m2-2-example}:\footnote{When $\delta = 0$, $\theta_{E}$ and $\delta$ do not affect invertibility of $\mathbf{S}$.}
\begin{equation*}
    \operatorname{vec}(\mathbf{Y}_{t}^{*})
    = \mathbf{S}^{-1} \left(\left((P \otimes \mathbf{I}_{n}) + (\boldsymbol{\rho}^\prime \otimes \mathbf{W} ) \right)
    \operatorname{vec}(\mathbf{Y}_{t-1})
    + (\phi \otimes \mathbf{I}_{n})\mathbf{g}_{t}^{*}
    + \sum_{k=1}^{K}\boldsymbol{\Pi}_{k}\mathbf{x}_{t,k} 
    + \operatorname{vec}(\mathbf{U}_{t}) \right),
\end{equation*}
and $\mathbf{S}^{-1} = \left(\sum_{k=0}^{\infty}(\widetilde{\Lambda}^{\prime k} \otimes \mathbf{W}^{k}) \right) \cdot \left((P + \Psi)^{-1} \otimes \mathbf{I}_{n} \right)$ where $\widetilde{\Lambda} = \Lambda(P + \Psi)^{-1}$. Then, $\widetilde{\Lambda}^{\prime k} \otimes \mathbf{W}^{k}$ characterizes the $k$th ($k = 1,2,\cdots$) order network influences. We observe that $\mathbf{S}$ only contains signals from $j$ to $i$ due to the local agents' non-cooperative behaviors under myopia: $[\widetilde{\Lambda}^{k} ]_{l^\prime l} \left[\mathbf{W}^{k} \right]_{ij}$ in $\widetilde{\Lambda}^{\prime k} \otimes \mathbf{W}^{k}$ illustrates the $k$th order influence from $j$'s $l^\prime$th activity on $i$'s $l$th activity. By regarding $\widetilde{\Lambda}$ as a normalized interaction parameter matrix, this condition bears a similarity to the stability condition of the multivariate simultaneous equations spatial autoregressive models, as mentioned in Yang and Lee (\citeyear{YangLee2017}, Sec.2).

Second, note that \(\overline{\mathbf{S}^{-1}} - \mathbf{S}^{-1\prime}\mathbf{S}_{0} \mathbf{S}^{-1} = \left(\mathbf{I}_{nm^{*}} + \mathbf{S}^{-1\prime}(\Lambda^{\prime} \otimes \mathbf{W}^{\prime}) \right) \mathbf{S}^{-1}\) in $\mathbf{T}_{0}$ since \(\mathbf{S}_{0} = \mathbf{S} - (\Lambda^{\prime} \otimes \mathbf{W}^{\prime})\). A sufficient condition for the invertibility of \(\mathbf{R}_{0}\) is then
\begin{equation*}
    \Vert \mathbf{T}_{0} \Vert_{2} = \Vert (\phi^{\prime} \otimes \mathbf{I}_{n})\left(\mathbf{I}_{nm^{*}} + \mathbf{S}^{-1\prime}(\Lambda^{\prime} \otimes \mathbf{W}^{\prime}) \right)\mathbf{S}^{-1}(\phi \otimes \mathbf{I}_{n} ) \Vert_{2} < 1.
\end{equation*}
If \(m^{*}=1\) (i.e., single activity), \(\mathbf{R}_{0} = \mathbf{I}_{n} - \phi^{2}(\overline{\mathbf{S}^{-1}} - \mathbf{S}^{-1\prime}\mathbf{S}_{0}\mathbf{S}^{-1}) = \mathbf{I}_{n} - \phi^{2}\mathbf{S}^{-1\prime}\left(\mathbf{I}_{n} + \lambda \mathbf{W} \mathbf{S}^{-1} \right) = \mathbf{I}_{n} - \phi^{2} \left(\mathbf{S}\mathbf{S}^\prime \right)^{-1}\). Here, \(\mathbf{S} = \mathbf{I}_{n} - \lambda \mathbf{W}\) and \(\mathbf{S}_{0} = \mathbf{I}_{n} - \lambda \overline{\mathbf{W}}\), where \(\lambda\) denotes the spatial interaction parameter for the single activity case. Then, the invertibility condition can be characterized by
\begin{equation*}
\mathcal{I} = \lbrace (\lambda, \phi, \mathbf{W}): \phi^{2}\Vert ( \mathbf{S} \mathbf{S}^{\prime} )^{-1} \Vert_{2} < 1 \rbrace
\end{equation*}
since the invertibility of \(\mathbf{S}\) for the second stage should be imposed to define \(\mathbf{S}^{-1}\). We observe that \(\phi^{2}\Vert ( \mathbf{S} \mathbf{S}^{\prime} )^{-1} \Vert_{2}\) is an increasing function of \(\lambda\) and \(\phi\) and shows larger values with large column sums of \(\mathbf{W}\) when \(\mathbf{W}\) is row-normalized (see Sec.1.3.2 in the supplement file). Thus, \(\mathcal{I}\) restricts the parameter space and network, implying the model's stable contemporaneous network spillovers. The resulting network for the allocator's decision is \((\mathbf{S}\mathbf{S}^{\prime} )^{-1}\). If $w_{ij}$ illustrates the signal from $j$ to $i$, its implication is that \(\mathbf{S}^{-1}\) contains signals of all orders from \(j\) to \(i\) (i.e., \(j \mapsto \cdots \mapsto i\)), while \(\mathbf{S}^{-1\prime}\) includes those from \(i\) to \(j\) (i.e., \(i \mapsto \cdots \mapsto j\)) if \(\lambda \neq 0\). The benevolent allocator considers both directions of signals, even for myopia.$\blacksquare$
\end{remark}

The next condition specifies the uniqueness of the value functions. For this, stability is required. The stability condition enables equivalence between the recursive forms of value functions (\eqref{leadervalue} and \eqref{m2-2-flpv}) and their infinite sum representations. To characterize the stability condition, we define
\begin{equation*}
    \begin{split}
    \widetilde{\mathbb{Q}}_{i} = & \mathbb{L}^{y\prime} \mathbf{L}_{Z,i} 
+ \mathbf{Q}_{Z,i}
+ \mathbb{L}^{y\prime} \mathbf{Q}_{Y,i}\mathbb{L}^{y}, \\
\widetilde{\mathbb{L}}_{i} = &
\left(\mathbf{L}_{Z,i}^\prime + \mathbb{L}^{y\prime} \overline{\mathbf{Q}_{Y,i}}  \right) \mathbf{\bar{c}}^{y}, \\
\widetilde{\mathbb{Q}}_{0,i} = & \mathbb{A}_{1:n}^{R\prime }\left( \mathbb{L}^{y\prime} \mathbf{L}_{Z,i} 
    + \mathbf{Q}_{Z,i} + \mathbb{L}^{y\prime} \mathbf{Q}_{Y,i}\mathbb{L}^{y} \right) \mathbb{A}_{1:n}^{R} \\
    & - \frac{1}{2m^{*}}\mathbb{L}^{g\prime}\left( \mathbf{I}_{m^{*}}\otimes e_{n,i}e_{n,i}^{\prime} \right) \mathbb{L}^{g}
    + \frac{1}{m^{*}}\left[ \mathbf{0},\mathbf{I}_{nm^{*}}\right]^{\prime}
    \left(\mathbf{I}_{m^{*}}\otimes e_{n,i}e_{n,i}^{\prime}\right) \mathbb{L}^{g}
    \text{, and} \\
    \widetilde{\mathbb{L}}_{0,i} = & \mathbb{A}_{1:n}^{R\prime}
    \left( \left( \overline{\mathbb{L}^{y\prime}\mathbf{L}_{Z,i} } 
    + \overline{\mathbf{Q}_{Z,i}}
    + \mathbb{L}^{y\prime}\overline{\mathbf{Q}_{Y,1}} \right)  \mathbb{c}_{1:n}^{R} 
    + \left(\mathbf{L}_{Z,i}^\prime + 
    \mathbb{L}^{y\prime} \overline{\mathbf{Q}_{Y,i}} \right) \mathbf{\bar{c}}^{y}\right)  -\frac{1}{m^{*}}\left( \mathbb{L}^{g} 
    - \left[ \mathbf{0}, \mathbf{I}_{nm^{*}}\right]\right)^{\prime }\left( \mathbf{I}_{m^{*}}\otimes e_{n,i}e_{n,i}^{\prime}\right) 
    \mathbf{\bar{c}}^{g}.
    \end{split}
\end{equation*}
Those components constitute the agents' per-period payoffs.

Consequently, we have the following recursive representations:
\begin{equation*}
    \begin{split}
        \mathbb{Q}_{i} & = \widetilde{\mathbb{Q}}_{i} + \delta \mathbb{A}_{1:n}^\prime \mathbb{Q}_{i}\mathbb{A}_{1:n},\\
        \mathbb{L}_{i} & = \widetilde{\mathbb{L}}_{i} + \delta
        \mathbb{A}_{1:n}^\prime (\overline{\mathbb{Q}}_{i} \mathbb{c}_{1:n} + \mathbb{L}_{i}),\\
        \mathbb{Q}_{0,i} & = \widetilde{\mathbb{Q}}_{0,i} + \delta \mathbb{A}_{0}^\prime \mathbb{Q}_{0,i}\mathbb{A}_{0},\text{ and}\\
        \mathbb{L}_{0,i} & = \widetilde{\mathbb{L}}_{0,i} + \delta
        \mathbb{A}_{0}^\prime (\overline{\mathbb{Q}}_{0,i} \mathbb{c}_{0} + \mathbb{L}_{0,i}),
    \end{split}
\end{equation*}
where $\mathbb{Q}_{0,i}$ and $\mathbb{L}_{0,i}$ for $i = 1,\cdots,n$ satisfy $\mathbb{Q}_{0} = \sum_{i=1}^{n}\mathbb{Q}_{0,i}$ and $\mathbb{L}_{0} = \sum_{i=1}^{n}\mathbb{L}_{0,i}$, and $\mathbb{A}_{1:n}$ and $\mathbb{A}_{0}$ govern the dynamics of the state variables, i.e., $\mathbb{E}_{t,2}(\mathbf{z}_{t+1}^{A}) = \mathbb{c}_{1:n} + \mathbb{A}_{1:n}\mathbf{z}_{t}^{A}$ and $\mathbb{E}_{t,1}(\mathbf{z}_{t+1}^{R}) = \mathbb{c}_{0} + \mathbb{A}_{0}\mathbf{z}_{t}^{R}$.

We observe that the equations for $\mathbb{Q}_{i}$ and $\mathbb{Q}_{0,i}$ follow the discrete \textit{Lyapunov equation}. By Theorem 7.3.2 in Datta (\citeyear{DATTA2004201}, Ch.7), the necessary and sufficient condition for uniqueness of $\mathbb{Q}_{i}$ is $\Vert \sqrt{\delta}\mathbb{A}_{1:n} \Vert_{2} < 1$ by achieving the stable dynamic system ($\Vert \sqrt{\delta}\mathbb{A}_{0} \Vert_{2} < 1$ for uniqueness of $\mathbb{Q}_{0,i}$). If $\Vert \sqrt{\delta}\mathbb{A}_{1:n} \Vert_{2} < 1$ and $\Vert \sqrt{\delta}\mathbb{A}_{0} \Vert_{2} < 1$ are satisfied, we obtain
\begin{equation*}
    \begin{split}
        \mathbb{Q}_{i} & = \widetilde{\mathbb{Q}}_{i} + \delta \mathbb{A}_{1:n}^\prime \mathbb{Q}_{i}\mathbb{A}_{1:n} = \sum_{k=0}^{\infty}\delta^{k} \mathbb{A}_{1:n}^{k\prime} \widetilde{\mathbb{Q}}_{i}\mathbb{A}_{1:n}^{k},\text{ and }\\
        \mathbb{Q}_{0,i} & = \widetilde{\mathbb{Q}}_{0,i} + \delta \mathbb{A}_{0}^\prime \mathbb{Q}_{0,i}\mathbb{A}_{0}
        = \sum_{k=0}^{\infty}\delta^{k} \mathbb{A}_{0}^{k\prime} \widetilde{\mathbb{Q}}_{0,i}\mathbb{A}_{0}^{k}.
    \end{split}
\end{equation*}
To achieve uniqueness of $\mathbb{L}_{i}$ and $\mathbb{L}_{0,i}$, we need to impose stronger conditions, $\Vert \mathbb{A}_{1:n} \Vert_{2} < 1$ and $\Vert \mathbb{A}_{0} \Vert_{2} < 1$ to achieve well-definedness of $\sum_{j=0}^{\infty}\mathbb{A}_{1:n}^{j}$ and $\sum_{j=0}^{\infty}\mathbb{A}_{0}^{j}$. When $\sum_{j=0}^{\infty}\mathbb{A}_{1:n}^{j}$ and $\sum_{j=0}^{\infty}\mathbb{A}_{0}^{j}$ are well-defined, we have
\begin{equation*}
    \begin{split}
        \mathbb{L}_{i} & = \widetilde{\mathbb{L}}_{i} + \delta
        \mathbb{A}_{1:n}^\prime (\overline{\mathbb{Q}}_{i} \mathbb{c}_{1:n} + \mathbb{L}_{i}) 
        =\sum_{k=0}^{\infty}\delta^{k}(\mathbb{A}_{1:n}^\prime)^{k} \widetilde{\mathbb{L}}_{i} + \sum_{k=1}^{\infty}\delta^{k}(\mathbb{A}_{1:n}^\prime)^{k} \overline{\widetilde{\mathbb{Q}}}_{i}\left(\sum_{j=0}^{\infty}\mathbb{A}_{1:n}^{j} \right)\mathbb{c}_{1:n},\text{ and} \\
        \mathbb{L}_{0,i} & = \widetilde{\mathbb{L}}_{0,i} + \delta
        \mathbb{A}_{0}^\prime (\overline{\mathbb{Q}}_{0,i} \mathbb{c}_{0} + \mathbb{L}_{0,i}) = \sum_{k=0}^{\infty}\delta^{k}(\mathbb{A}_{0}^\prime)^{k} \widetilde{\mathbb{L}}_{0,i} + \sum_{k=1}^{\infty}\delta^{k}(\mathbb{A}_{0}^\prime)^{k} \overline{\widetilde{\mathbb{Q}}}_{0,i}\left(\sum_{j=0}^{\infty}\mathbb{A}_{0}^{j} \right)\mathbb{c}_{0}.
    \end{split}
\end{equation*}
As a result, we define
\begin{equation*}
    \mathcal{S} = \left\lbrace (\theta_{P}, \theta_{E}, \delta, \mathbf{W}): \Vert \mathbb{A}_{1:n} \Vert_{2} < 1 \text{ and }\Vert \mathbb{A}_{0} \Vert_{2} < 1 \right\rbrace.
\end{equation*}
If $(\theta_{P}, \theta_{E}, \delta, \mathbf{W}) \in \mathcal{S}$, $\mathbb{Q}_{i}$, $\mathbb{Q}_{0,i}$, $\mathbb{L}_{i}$ and $\mathbb{L}_{0,i}$ for $i = 1,\cdots,n$ are unique.

Finally, we conclude that the uniqueness of the MPNE is guaranteed by the following set $\mathcal{M} = \mathcal{I} \cap \mathcal{S}$:
\begin{equation*}
    \mathcal{M} = \left\lbrace (\theta_{P}, \theta_{E}, \delta, \mathbf{W}): \Vert \mathbf{T}_{1:n} \Vert_{2} < 1,\text{ } \Vert \mathbf{T}_{0} \Vert_{2} < 1,\text{ } \Vert \mathbb{A}_{1:n} \Vert_{2} < 1,\text{ and } \Vert \mathbb{A}_{0} \Vert_{2} < 1 \right\rbrace.
\end{equation*}

\subsection*{Numerical experiments on the linear-quadratic payoff}

The main purpose of this subsection is to conduct a numerical sensitivity analysis justifying the linear-quadratic (LQ) payoff specification. We consider a simplified two-stage static game model. State variables are defined as follows:

\begin{itemize}
\item Resource allocator's state variables: $\mathbf{x} = (x_{1}, \dots, x_{n})^\prime$ and $\boldsymbol{\tau} = (\tau_{1}, \dots, \tau_{n})^\prime$,
\item Local agents' state variables: $\mathbf{g} = (g_{1}, \dots, g_{n})^\prime$ and $\mathbf{x}$.
\end{itemize}

Local agent $i$'s payoff function for activity $y_{i}$ is specified by:
\begin{equation}
\begin{split}
    U_{i}\left(\mathbf{y}, \mathbf{g}, \mathbf{x} \right)  
    & = \left(\pi x_{i} + \phi g_{i} + \lambda \sum_{j=1}^{n}w_{ij}y_{j} \right)y_{i} - \frac{1}{2}y_{i}^2 + \nu \cdot y_{i}\cdot F\left(\pi x_{i} + \phi g_{i} + \lambda \sum_{j=1}^{n}w_{ij}y_{j} \right) \\
    & = \phi \mathbf{y}^\prime e_{n,i}e_{n,i}^\prime \mathbf{g}
    + \pi \mathbf{y}^\prime e_{n,i}e_{n,i}^\prime \mathbf{x}
    + \mathbf{y}^\prime \left(\lambda e_{n,i}w_{i.} - \frac{1}{2}e_{n,i}e_{n,i}^\prime \right) \mathbf{y}
    + \nu \cdot \mathbf{y}^\prime e_{n,i}e_{n,i}^\prime
    \tilde{\mathbf{F}}(\mathbf{z})
\end{split}
\label{agent_simu}
\end{equation}
where $\phi$, $\pi$, and $\lambda$ are the main parameters of the simplified model and $\tilde{\mathbf{F}}(\mathbf{z})  =(F(z_{1}),\dots,F(z_{n}))^{\prime}$. The function $F(\cdot)$ captures the nonlinear interactions among agents, and $\nu \geq 0$ controls the degree of nonlinearity. Following \cite{XuandLee2015a}, we set $F(z) = \frac{1}{1 + e^{-z}}$ (logistic CDF) for numerical simulations. When $\nu = 0$, \eqref{agent_simu} reduces to an LQ payoff.

The allocator's corresponding payoff is:
\begin{equation}
U_{0}(\mathbf{g}, \mathbf{y}, \mathbf{x}, \boldsymbol{\tau}) = \sum_{i=1}^{n}U_{i}(\mathbf{y}, \mathbf{g}, \mathbf{x}) + \sum_{i=1}^{n}\tau_{i}g_{i} - \frac{1}{2}\sum_{i=1}^{n}g_{i}^2.
\label{ra_simu}
\end{equation}

The strategic interaction unfolds in two sequential stages:
\begin{itemize}
    \item \textbf{Stage 1.} Given $\mathbf{x}$ and $\boldsymbol{\tau}$, the allocator chooses $\mathbf{g} = \left(g_{1}, \cdots, g_{n} \right)^\prime$ to maximize $U_{0}(\mathbf{g}, \mathbf{y}, \mathbf{x}, \boldsymbol{\tau})$. The allocator expects the local agents' best-response function $\mathbf{y}^{*}(\mathbf{g}, \mathbf{x}) = 
 \left(y_{1}^{*}(\mathbf{g}, \mathbf{x}), \cdots, y_{n}^{*}(\mathbf{g}, \mathbf{x}) \right)^\prime$ given $\mathbf{g}$ and $\mathbf{x}$. Formally, $\mathbf{g}^{*}(\mathbf{x}, \boldsymbol{\tau}) = \text{argmax}_{\mathbf{g}}U_{0}(\mathbf{g}, \mathbf{y}^{*}(\mathbf{g}, \mathbf{x}), \mathbf{x}, \boldsymbol{\tau})$.

    \item \textbf{Stage 2.} After observing the grant allocations, $n$ local agents simultaneously choose their activities.
    
    Hence, $y_{i}^{*}(\mathbf{y}_{-i}, \mathbf{g}, \mathbf{x}) = \text{argmax}_{y_{i}} U_{i}\left(y_{i}, \mathbf{y}_{-i}, \mathbf{g}, \mathbf{x} \right)$ where $\mathbf{y}_{-i} = \left(y_{1}, \cdots, y_{i-1}, y_{i+1}, \cdots, y_{n} \right)^\prime$ is a vector of local agents' activities except for $i$'s one.

    \item[$\Rightarrow$] When a unique Nash equilibrium exists, $\mathbf{y}^{*}(\mathbf{g}, \mathbf{x}) = \left(y_{1}^*(\mathbf{g}, \mathbf{x}), \cdots,  y_{n}^*(\mathbf{g}, \mathbf{x})\right)^\prime$ satisfies the fixed point relation $y_{i}^{*}(\mathbf{g}, \mathbf{x}) = y_{i}^{*}(\mathbf{y}_{-i}^{*}(\mathbf{g}, \mathbf{x}), \mathbf{g}, \mathbf{x})$.
\end{itemize}

The first-order conditions of \textbf{Stage 2} yield:
\begin{equation}
    \mathbf{y} = \lambda \mathbf{W}\mathbf{y} + \nu \tilde{\mathbf{F}}(\mathbf{z}) + \phi \mathbf{g} + \pi \mathbf{x}.
    \label{foc_stage2}
\end{equation}
A sufficient condition for the uniqueness of Nash equilibrium at the second stage is $\vert \lambda \vert \Vert \mathbf{W} \Vert_{\infty} (1 + \frac{\nu}{4}) < 1$ by Assumption 3 in \cite{XuandLee2015a}.\footnote{ To see this, note that
\begin{equation*}
    \tilde{\mathbf{F}}(\mathbf{z}) = \tilde{\mathbf{F}}(\mathbf{0}) + \phi \tilde{\mathbf{f}}(\mathbf{\bar{z}})\mathbf{g} + \pi \tilde{\mathbf{f}}(\mathbf{\bar{z}})\mathbf{x}
    + \lambda \tilde{\mathbf{f}}(\mathbf{\bar{z}}) \mathbf{y},
\end{equation*}
where $\tilde{\mathbf{f}}(\mathbf{z}) = \text{diag}\left\lbrace f\left(z_1 \right), \cdots, f\left(z_n \right) \right\rbrace$ with $f(z) = \frac{e^{-z}}{1 + e^{-z}}$ and $\mathbf{\bar{z}}$ lies between $\mathbf{z}$ and $\mathbf{0}$. Then, equation \eqref{foc_stage2} can be written as
\begin{equation*}
    \left(\mathbf{I}_{n} - \lambda\left(\mathbf{I}_{n} + \nu \tilde{\mathbf{f}}(\mathbf{\bar{z}}) \right)\mathbf{W} \right)\mathbf{y}
    = \phi \left(\mathbf{I}_{n} + \nu \tilde{\mathbf{f}}(\mathbf{\bar{z}}) \right) \mathbf{g}
    + \pi \left(\mathbf{I}_{n} + \nu \tilde{\mathbf{f}}(\mathbf{\bar{z}}) \right) \mathbf{x}
    + \tilde{\mathbf{F}}(\mathbf{0}).
\end{equation*}
Then, a sufficient condition for invertibility of $\mathbf{I}_{n} - \lambda\left(\mathbf{I}_{n} + \nu \tilde{\mathbf{f}}(\mathbf{\bar{z}}) \right)\mathbf{W}$ is $\vert \lambda \vert \Vert \mathbf{W} \Vert_{\infty} (1 + \frac{\nu}{4}) < 1$ since $\sup_{z}F^\prime (z) = \frac{1}{4}$ when $F(z) = \frac{1}{1 + e^{-z}}$.} Under this condition, there is a unique $\mathbf{y}^{*}\left(\mathbf{g}, \mathbf{x} \right)$ satisfying \eqref{foc_stage2}.

The allocator’s first-order condition is:
\begin{equation}
\begin{split}
    \mathbf{0} = & \phi \mathbf{y}^{*}\left(\mathbf{g}, \mathbf{x} \right) + \phi \frac{\partial \mathbf{y}^{*}\left(\mathbf{g}, \mathbf{x} \right)}{\partial \mathbf{g}^\prime}\mathbf{g}
    + \pi  \frac{\partial \mathbf{y}^{*}\left(\mathbf{g}, \mathbf{x} \right)}{\partial \mathbf{g}^\prime}\mathbf{x}
    + \frac{\partial \mathbf{y}^{*}\left(\mathbf{g}, \mathbf{x} \right)^\prime}{\partial \mathbf{g}}\left(\lambda (\mathbf{W} + \mathbf{W}^\prime) - \mathbf{I}_{n} \right)\mathbf{y}^{*}\left(\mathbf{g}, \mathbf{x} \right) \\
    & + \nu \frac{\partial \mathbf{y}^{*}\left(\mathbf{g}, \mathbf{x} \right)^\prime}{\partial \mathbf{g}}\tilde{\mathbf{F}}(\mathbf{z})
    + \nu \tilde{\mathbf{f}}(\mathbf{z})
    \left(\phi \mathbf{I}_{n} + \lambda \frac{\partial \mathbf{y}^{*}\left(\mathbf{g}, \mathbf{x} \right)^\prime}{\partial \mathbf{g}}\mathbf{W}^\prime \right)\mathbf{y}^{*}\left(\mathbf{g}, \mathbf{x} \right)
    + \boldsymbol{\tau}
    - \mathbf{g}.
\end{split}
\label{foc_stage1}
\end{equation}
Equation \eqref{foc_stage1} takes an implicit function form: $\mathcal{G}\left(\mathbf{g}, \mathbf{x}, \boldsymbol{\tau} \right) = \mathcal{G}\left(\mathbf{g}, \mathbf{y}^{*}\left(\mathbf{g}, \mathbf{x} \right), \mathbf{x}, \boldsymbol{\tau} \right) = \mathbf{0}$. From \eqref{foc_stage1}, 
\begin{equation}
\begin{split}
  \frac{\partial \mathcal{G}\left(\mathbf{g}, \mathbf{x}, \boldsymbol{\tau} \right)}{\partial \mathbf{g}^\prime} 
    & = - \mathbf{I}_{n} + \widetilde{\mathbf{T}}_{0}(\mathbf{g}, \mathbf{x})
\end{split}
\label{soc_stage1}
\end{equation}
where
\begin{equation*}
    \widetilde{\mathbf{T}}_{0}(\mathbf{g}, \mathbf{x}) =
    \frac{\partial}{\partial \mathbf{g}^\prime}\left(
    \begin{array}{cc}
        \phi \mathbf{y}^{*}\left(\mathbf{g}, \mathbf{x} \right) + \phi \frac{\partial \mathbf{y}^{*}}{\partial \mathbf{g}^\prime}\left(\mathbf{g}, \mathbf{x} \right)\mathbf{g}
    + \pi  \frac{\partial \mathbf{y}^{*}}{\partial \mathbf{g}^\prime}\left(\mathbf{g}, \mathbf{x} \right)\mathbf{x}
    + \frac{\partial \mathbf{y}^{*\prime}}{\partial \mathbf{g}}\left(\mathbf{g}, \mathbf{x} \right)\left(\lambda (\mathbf{W} + \mathbf{W}^\prime) - \mathbf{I}_{n} \right)\mathbf{y}^{*}\left(\mathbf{g}, \mathbf{x} \right) \\
          + \nu \frac{\partial \mathbf{y}^{*\prime}}{\partial \mathbf{g}}\left(\mathbf{g}, \mathbf{x} \right)\tilde{\mathbf{F}}(\mathbf{g}, \mathbf{y}^{*}\left(\mathbf{g}, \mathbf{x} \right), \mathbf{x})
    + \nu \tilde{\mathbf{f}}(\mathbf{g}, \mathbf{y}^{*}\left(\mathbf{g}, \mathbf{x} \right), \mathbf{x})
    \left(\phi \mathbf{I}_{n} + \lambda \frac{\partial \mathbf{y}^{*\prime}}{\partial \mathbf{g}}\left(\mathbf{g}, \mathbf{x} \right)\mathbf{W}^\prime \right)\mathbf{y}^{*}\left(\mathbf{g}, \mathbf{x} \right) 
    \end{array}
    \right).
\end{equation*}

Fix $(\mathbf{x}, \boldsymbol{\tau})$ for the following analysis. Given $(\mathbf{x}, \boldsymbol{\tau})$, let $\mathbf{g}^{*}(\mathbf{x}, \boldsymbol{\tau})$ be a solution satisfying $\mathcal{G}\left(\mathbf{g}, \mathbf{x}, \boldsymbol{\tau} \right) = \mathbf{0}$. Note that $\mathcal{G}(\cdot, \cdot, \cdot)$ is an infinitely differentiable function with respect to its arguments, and $\frac{\partial \mathcal{G}\left(\mathbf{g}, \mathbf{x}, \boldsymbol{\tau} \right)}{\partial \mathbf{g}^\prime}$ in \eqref{soc_stage1} is invertible at $(\mathbf{g}^{*}(\mathbf{x}, \boldsymbol{\tau}), \mathbf{x}, \boldsymbol{\tau})$ if $\Vert \widetilde{\mathbf{T}}_{0}(\mathbf{g}, \mathbf{x})\textbf{} \Vert_2 < 1$. Then, by the implicit function theorem, the optimal grant vector $\mathbf{g}^{*}(\mathbf{x}, \boldsymbol{\tau})$ is unique.

The uniqueness of the subgame-perfect Nash equilibrium (SPNE) can be characterized by the sufficient conditions for contraction:
\begin{equation*}
|\lambda|\Vert\mathbf{W}\Vert_{\infty}\left(1 + \frac{\nu}{4}\right) < 1, \quad \text{and} \quad \Vert\widetilde{\mathbf{T}}_{0}(\mathbf{g}, \mathbf{x})\Vert_{\infty} < 1.
\end{equation*}
Under these bounds, the joint mapping
$(\mathbf{g},\mathbf{y})\mapsto(\mathbf{g}^{*},\mathbf{y}^{*})$ is a
contraction, guaranteeing a unique SPNE.

\paragraph{Numerical simulations.} For numerical experiments, we employ empirical network structure $\mathbf{W}^{adj}$ and set $x_{i} = \frac{1}{T}\sum_{t=1}^{T}y_{i,t,1}$ and $\tau_{i} = \frac{1}{T}\sum_{t=1}^{T}g_{i,t}$ for $i = 1,\cdots,n$ from our application. Note that the averages of $x_{i}$ and $\tau_{i}$ are respectively 7.0410 and 7.2722. Under these settings, we consider two scenarios of policy dependence:
\begin{itemize}
\item Scenario 1: Strong dependence ($\lambda = 0.3$, $\phi = 0.3$, $\pi = 0.6$).
\item Scenario 2: Moderate dependence ($\lambda = 0.2$, $\phi = 0.2$, $\pi = 0.4$).
\end{itemize}

We measure deviations from the LQ payoff ($\nu=0$) using:
\begin{equation*}
d^{y}(\nu) = \max_{i}\left| y_{i}^{*}(\nu)-y_{i}^{*}(0) \right|,\text{ and } d^{g}(\nu)=\max_{i}\left| g_{i}^{*}(\nu)-g_{i}^{*}(0) \right|,
\end{equation*}
where $y_{i}^*(\nu)$ and $g_{i}^*(\nu)$ denote the optimal decisions at evaluated each $\nu$.

\begin{figure}[!htbp]
\begin{center}
\caption{Deviations from the LQ payoff}
\label{fig_devlq_casedeath}
\begin{subfigure}[t]{0.495\textwidth}
   \caption{Scenario 1: strong dependence}
  \includegraphics[width=\linewidth]{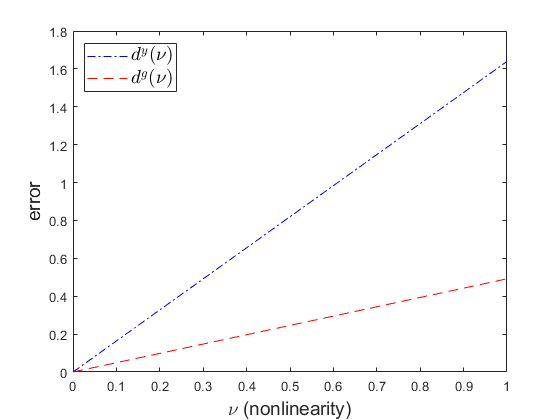}
\end{subfigure}
\begin{subfigure}[t]{0.495\textwidth}
   \caption{Scenario 2: moderate dependence}
  \includegraphics[width=\linewidth]{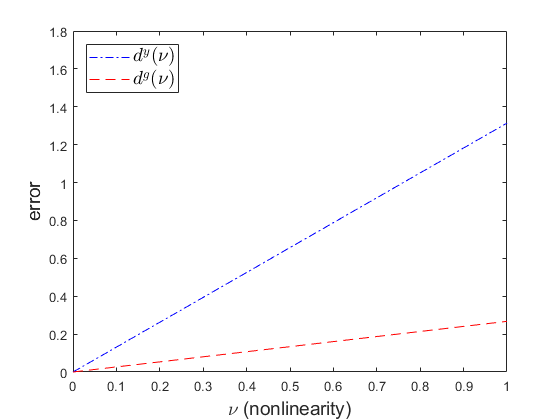} 
\end{subfigure}
\end{center}
\footnotesize
\textit{Note}: Average baseline values for Scenario 1 are $10.4993$ (activity) and $10.4220$ (allocation), with maximum deviations at $\nu=1$ of $d^{y}(1)=1.6393$ and $d^{g}(1)=0.4919$. For Scenario 2, averages are $5.6185$ (activity) and $8.3969$ (allocation), with $d^{y}(1)=1.3139$ and $d^{g}(1)=0.2674$.
\end{figure}

Figure~\ref{fig_devlq_casedeath} demonstrates that deviations increase approximately linearly with $\nu$. Deviations in activities ($d^{y}$) are consistently larger than in grant allocations ($d^{g}$). Even at maximal nonlinearity ($\nu=1$), deviations remain moderate. Specifically, deviations are under 10\% and 5\% when $\nu \leq 0.3$. These results imply that if the payoff moderately deviates from the LQ specification and satisfies the contraction mapping condition, equilibrium activities and grants closely approximate those under the simpler LQ payoff.

\section*{Appendix C: Large sample properties}
\label{AppendixB_chap3}

\setcounter{theorem}{0}
\setcounter{assumption}{0}
\setcounter{remark}{0}
\setcounter{definition}{0}
\renewcommand{\thetheorem}{C.\arabic{theorem}}
\renewcommand{\theassumption}{C.\arabic{assumption}}
\renewcommand{\theremark}{C.\arabic{remark}}
\renewcommand{\thedefinition}{C.\arabic{definition}}

\renewcommand{\thetable}{C.\arabic{table}}
\renewcommand{\thefigure}{C.\arabic{figure}}

\subsection*{Consistency}

In this subsection, we provide proof of the consistency of the QMLE in Theorem \ref{thmconsistency}. For this, we study the structure of $\frac{1}{nT}\ell_{nmT}^{c}\left( \theta \right) - \mathbb{E}\left(\frac{1}{nT}\ell_{nmT}^{c}(\theta) \right)$. By (\ref{m3-3-cll_chap3}),
\begin{equation*}
\frac{1}{nT}\ell_{nmT}^{c}\left( \theta \right) - \mathbb{E}\left(\frac{1}{nT}\ell_{nmT}^{c}(\theta) \right) =
- \frac{1}{2nT}\left\{ 
\begin{array}{c}
\mathbf{v}_{nmT}^{\prime}\left( \theta_{P}\right) \left( \mathbf{J}_{T}\otimes
\mathbf{J}_{n,m}\boldsymbol{\Delta}^{-1}\left( \theta_{P}, \Sigma, \sigma^{2} \right) \mathbf{J}_{n,m}\right) \mathbf{v}_{nmT}\left( \theta_{P}\right) \\ 
-\mathbb{E}\left( \mathbf{v}_{nmT}^{\prime}\left( \theta_{P}\right) \left(
\mathbf{J}_{T}\otimes \mathbf{J}_{n,m}\boldsymbol{\Delta}^{-1}\left( \theta_{P}, \Sigma, \sigma^{2} \right) \mathbf{J}_{n,m}\right) 
\mathbf{v}_{nmT}\left( \theta_{P}\right) \right)
\end{array}
\right\},
\end{equation*}
where $\mathbf{v}_{nmT}(\theta_{P}) = \left(\mathbf{v}_{1}^\prime (\theta_{P}), \cdots, \mathbf{v}_{T}^\prime(\theta_{P}) \right)^\prime$. 

As
\begin{equation*}
    \begin{split}
        & \left( \operatorname{vec}\left( \mathbf{Y}_{1}\right)^\prime, \mathbf{g}_{1}^{\prime},\cdots,
\operatorname{vec}\left( \mathbf{Y}_{T}\right)^\prime,
\mathbf{g}_{T}^\prime \right)^\prime \\
= & \left( \mathbf{I}_{T}\otimes \mathbf{R}^{-1}\right) \left( \left(
\mathbf{I}_{T}\otimes \mathbf{G}
\right) \mathbf{y}_{nmT,-1} + \sum_{k=1}^{K}\left(\mathbf{I}_{T}\otimes 
\boldsymbol{\Gamma}_{k}
\boldsymbol{\Pi}_{k,0}\right) \mathbf{x}_{nmT,k}
+ \sum_{q=1}^{Q}\beta_{q,0} \mathbf{x}_{nmT,q}^{\tau}\right) + \left(\mathbf{I}_{T}\otimes \mathbf{R}^{-1}\right) \left( \mathbf{c}_{nmT,0}
+ \widetilde{\boldsymbol{\alpha}}_{nmT,0} + \mathbf{v}_{nmT}\right),
    \end{split}
\end{equation*}
\begin{equation}
\begin{split}
    & \left( \mathbf{J}_{T}\otimes \mathbf{J}_{n,m} \boldsymbol{\Delta}^{-\frac{1}{2}}\left( \theta
\right) \mathbf{J}_{n,m}\right) \mathbf{v}_{nmT}\left( \theta \right) \\
= & \left( \mathbf{J}_{T}\otimes \mathbf{J}_{n,m}\boldsymbol{\Delta}^{-\frac{1}{2}}\left( \theta \right) \mathbf{J}_{n,m}\right) \left\{ 
\begin{array}{c}
\left( \mathbf{I}_{T}\otimes \mathbf{F}^{y}\left( \theta_{P}\right) \right) 
\mathbf{y}_{nmT,-1}
+ \sum_{k=1}^{K}\left( \mathbf{I}_{T}\otimes \mathbf{F}_{k}^{x}\left( \theta_{P}\right) \right) \mathbf{x}_{nmT,k} \\ 
+\sum_{q=1}^{Q}\left( \mathbf{I}_{T}\otimes \mathbf{F}_{q}^{\tau}\left( \theta_{P}, \beta_{q}\right) \right) \mathbf{x}_{nmT,q}^{\tau} 
+ \left( \mathbf{I}_{T}\otimes \mathbf{R} \left( \theta_{P}\right) \mathbf{R}^{-1}\right) \left( \widetilde{\boldsymbol{\alpha}}_{nmT,0}
+ \mathbf{v}_{nmT}\right)
\end{array}
\right\},
\label{transuform}
\end{split}
\end{equation}
where 
\begin{itemize}
    \item $\mathbf{G}(\theta_{P}) = \begin{bmatrix}
    (P \otimes \mathbf{I}_{n}) + (\boldsymbol{\rho}^\prime \otimes \mathbf{W})\\
    \mathbf{R}_{0}(\theta_{P})A^{g}(\theta_{P})
\end{bmatrix}$, $\mathbf{G} = \mathbf{G}(\theta_{P,0})$, $\boldsymbol{\Gamma}_{k}(\theta_{P}) = \begin{bmatrix}
    \mathbf{I}_{nm^{*}} + \delta\mathbf{L}_{1:n,k}(\theta_{P}) \\
    \mathbf{R}_{0}(\theta_{P})C_{k}^{g}(\theta_{P})
\end{bmatrix}
$, $\boldsymbol{\Gamma}_{k} = \boldsymbol{\Gamma}_{k}(\theta_{P,0})$,
   \item  $\mathbf{y}_{nmT,-1} = \left(\operatorname{vec}(\mathbf{Y}_{0})^\prime,\cdots,\operatorname{vec}(\mathbf{Y}_{T-1})^\prime \right)^\prime$,
   \item $\mathbf{x}_{nmT,k} = \left(\mathbf{x}_{1,k}^\prime, \cdots, \mathbf{x}_{T,k}^\prime \right)^\prime$ for $k = 1,\cdots,K$, $\mathbf{x}_{nmT,q}^{\tau} = \left(l_{m}^\prime \otimes \mathbf{x}_{1,q}^{\tau\prime}, \cdots, l_{m}^\prime \otimes \mathbf{x}_{T,q}^{\tau\prime} \right)^\prime$ for $q = 1,\cdots,Q$,
   \item $\mathbf{c}_{nmT,0} = l_{T}\otimes \mathbf{c}_{0}$, $\widetilde{\boldsymbol{\alpha}}_{nmT,0} = \left(\widetilde{\boldsymbol{\alpha}}_{1,0}^\prime \otimes l_{n}^\prime, \cdots, \widetilde{\boldsymbol{\alpha}}_{T,0}^\prime \otimes l_{n}^\prime \right)^\prime$, and $\mathbf{v}_{nmT} = \left(\mathbf{v}_{1}^\prime, \cdots, \mathbf{v}_{T}^\prime \right)^\prime$.
\end{itemize}
Also, recall that $\mathbf{F}^{y}\left(\theta_{P}\right) 
= \mathbf{R}\left(\theta_{P}\right) \mathbf{R}^{-1}\mathbf{G}
- \mathbf{G}(\theta_{P})$, $\mathbf{F}_{k}^{x}\left( \theta_{P}\right) 
=  \mathbf{R}\left( \theta_{P}\right) \mathbf{R}^{-1}
\boldsymbol{\Gamma}_{k}
\boldsymbol{\Pi}_{k,0}
- \boldsymbol{\Gamma}_{k}(\theta_{P})
\boldsymbol{\Pi}_{k}$ for $k = 1,\cdots,K$, and $\mathbf{F}_{q}^{\tau}\left( \theta_{P}, \beta_{q}\right) = \beta_{q,0}
\mathbf{R} \left( \theta_{P}\right) \mathbf{R}^{-1}-\beta_{q}\mathbf{I}_{nm}$ for $q = 1,\cdots,Q$. The projector $\mathbf{J}_{T}\otimes \mathbf{I}_{n}$ eliminates $\mathbf{c}_{nmT,0}$.

The first step for proving consistency involves verifying that \(\sup_{\theta \in \Theta} \left| \frac{1}{nT}\ell_{nmT}^{c}(\theta) - \mathbb{E}\left(\frac{1}{nT}\ell_{nmT}^{c}(\theta) \right) \right| \overset{p}{\to} 0\) as \(n,T \rightarrow \infty\). The second step requires showing the uniform stochastic equicontinuity of \(\left\{ \mathbb{E}\left(\frac{1}{nT} \ell_{nmT}^{c}(\theta)\right) \right\}\) over \(\theta \in \Theta\). Both of these steps are based on the decomposition (\ref{transuform}). Detailed proofs and explanations can be found in Section 3.3 of the supplementary file.

\subsubsection*{Identification}

Last, we need to ensure that the true parameters are identifiable. We will use the information inequality, as detailed in \cite{Rothenberg1971}. For this purpose, we reproduce two definitions from \cite{Rothenberg1971}.

\begin{definition}
\label{Defiden} For each $\theta \in \Theta$, $\mathcal{L}_{nmT}\left( \theta |\left\{ \mathbf{Y}_{t}, \mathbf{g}_{t}\right\}_{t=1}^{T}\right)$ denotes the density function with data $\left\{\mathbf{Y}_{t}, \mathbf{g}_{t}\right\}_{t=1}^{T}$.

(i) $\theta^{\prime}$ and $\theta^{\prime \prime}$ in $\Theta$ are observationally equivalent if $\mathcal{L}_{nmT}\left( \theta^{\prime}|\left\{
\mathbf{Y}_{t}, \mathbf{g}_{t}\right\}_{t=1}^{T} \right) = \mathcal{L}_{nmT} \left(\theta^{\prime \prime }|\left\{\mathbf{Y}_{t}, \mathbf{g}_{t}\right\}_{t=1}^{T}\right) $ a.e.

(ii) $\theta _{0}\in \Theta$ is identifiable if and only if there is no observationally equivalent $\theta \in \Theta \backslash \left\{ \theta_{0}\right\}$.
\end{definition}

We apply Definition \ref{Defiden} to $\ell_{nmT}^{c}\left( \theta
\right) $ since the information inequality argument is valid for the concentrated log-likelihood function for SAR models even if the likelihood is only a quasi-likelihood (\cite{LeeYu2016}). Then, identification of $\theta_{0}$ is based on (i) $\mathbb{E}\left( \ell_{nmT}^{c}\left( \theta \right) \right)
\leq \mathbb{E}\left( \ell_{nmT}^{c}\left( \theta_{0}\right) \right)$, (ii) $\ell_{nmT}^{c}\left( \theta \right) = \ell_{nmT}^{c}\left( \theta_{0}\right) $ a.e. in $\left\{\mathbf{Y}_{t}, \mathbf{g}_{t}\right\}_{t=1}^{T}$ if and only if $\mathbb{E}\left(
\ell_{nmT}^{c}\left( \theta \right) \right) = \mathbb{E}\left( \ell_{nmT}^{c}\left(\theta_{0}\right) \right)$, and (iii) $\ell_{nmT}^{c}\left(\theta \right) = \ell_{nmT}^{c}\left( \theta_{0}\right) $ a.e., gives $\theta = \theta_{0}$.

To derive the identification condition for $\theta_{0}$, consider the decomposition:
\begin{eqnarray*}
\left( \mathbf{J}_{T}\otimes \mathbf{J}_{n,m} \boldsymbol{\Delta}^{-\frac{1}{2}}\left(
\theta_{P}, \Sigma, \sigma^{2} \right) \mathbf{J}_{n,m}\right) \mathbf{v}_{nmT}\left( \theta \right) 
&=& \left(\mathbf{J}_{T} \otimes \mathbf{J}_{n,m}\boldsymbol{\Delta}^{-\frac{1}{2}}\left( \theta_{P}, \Sigma, \sigma^{2}
\right) \mathbf{J}_{n,m}\right) \left( \underset{=\text{Part of }\theta_{P,0}}{\underbrace{\boldsymbol{\Xi}_{nmT}\left( \theta_{P}\right) }}
+ \underset{=\text{Part of }\beta_{0}}{\underbrace{\sum_{q=1}^{Q}(\beta_{q,0} - \beta_{q})\mathbf{x}_{nmT,q}^{\tau} }}\right) \\
&& + \underset{=\text{Part of the error structure}}{\underbrace{\left(\mathbf{J}_{T} \otimes \mathbf{J}_{n,m} \boldsymbol{\Delta}^{-\frac{1}{2}}\left( \theta_{P}, \Sigma, \sigma^{2}
\right) \mathbf{J}_{n,m}\mathbf{R}\left( \theta_{P}\right) \mathbf{R}^{-1}\right) \mathbf{v}_{nmT}}}.
\end{eqnarray*}
The term $\boldsymbol{\Xi}_{nmT}\left( \theta_{P}\right) $ indicates a potential misspecification error for $\theta_{P,0}$, as $\left( \mathbf{J}_{T}\otimes \mathbf{J}_{n,m}\boldsymbol{\Delta}^{-\frac{1}{2}}\left( \theta_{P,0}, \Sigma, \sigma^{2}\right) \mathbf{J}_{n,m}\right) \boldsymbol{\Xi}_{nmT}\left( \theta_{P,0}\right)
= 0$ for any $\left( \Sigma, \sigma^{2}\right)$.

Then, we have 
\begin{equation*}
    \begin{split}
     & \frac{1}{nT}\mathbb{E}\left[ \mathbf{v}_{nmT}^{\prime }\left( \theta \right)
     \left( \mathbf{J}_{T}\otimes \mathbf{J}_{n,m} 
     \boldsymbol{\Delta}^{-1}\left( \theta_{P}, \Sigma, \sigma^{2} \right)
     \mathbf{J}_{n,m}\right) \mathbf{v}_{nmT}\left( \theta \right) \right] \\
     = &  \frac{1}{nT}\mathbb{E}\left[ \boldsymbol{\Xi}_{nmT}^{\ast}
     \left( \theta_{P}, \beta \right)^{\prime}
     \left( \mathbf{J}_{T}\otimes \mathbf{J}_{n,m} 
     \boldsymbol{\Delta}^{-1}\left( \theta_{P}, \Sigma, \sigma^{2} \right) \mathbf{J}_{n,m}\right) \boldsymbol{\Xi}_{nmT}^{\ast}
     \left(\theta_{P},\beta \right) \right] \\
     & +\frac{1}{n}\operatorname{tr}\left( \mathbf{R}^{-1\prime}\mathbf{R}
     \left(\theta_{P}\right)^{\prime}
     \boldsymbol{\Delta}^{-1}
     \left( \theta_{P}, \Sigma, \sigma^{2} \right) \mathbf{R}
     \left( \theta_{P}\right) \mathbf{R}^{-1}
     \boldsymbol{\Delta}\right) + o\left( 1\right),
    \end{split}
\end{equation*}
where the following relation holds because\footnote{
For an $nm \times nm$ matrix $A_{nm}$ such that $\frac{1}{n}A_{nm} = O\left(1\right)$ and $\max\left\{\left\Vert A_{nm}\right\Vert_{1}, \left\Vert A_{nm}\right\Vert_{\infty }\right\} < \infty$ uniformly in $n$, 
\begin{equation*}
\frac{1}{nm}\operatorname{tr}\left( \mathbf{J}_{T}\otimes \mathbf{J}_{n,m}A_{nm}\right) = \frac{1}{nm}\operatorname{tr}\left( \mathbf{J}_{T}\right) \operatorname{tr}\left( \mathbf{J}_{n,m}A_{nm}\right) = \frac{1}{n}\operatorname{tr}\left(
A_{nm}\right) + o\left( 1\right) .
\end{equation*}
} 
\begin{eqnarray*}
&& \frac{1}{nT}\operatorname{tr}\left( \mathbf{J}_{T} \otimes 
\left( \mathbf{R}^{-1\prime}
\mathbf{R}\left( \theta_{P}\right)^{\prime}
\mathbf{J}_{n,m}\boldsymbol{\Delta}^{-1}
\left( \theta_{P}, \Sigma, \sigma^{2} \right) \mathbf{J}_{n,m}
\mathbf{R}\left( \theta
_{P}\right) \mathbf{R}^{-1}
\boldsymbol{\Delta} \right) \right) \\
&=&\frac{1}{nT} \operatorname{tr}\left( \mathbf{J}_{T}\right) \operatorname{tr}\left( \mathbf{R}^{-1\prime}
\mathbf{R}\left( \theta_{P}\right)^{\prime}\mathbf{J}_{n,m}
\boldsymbol{\Delta}^{-1}\left( \theta_{P}, \Sigma, \sigma^{2} \right) \mathbf{J}_{n,m}\mathbf{R}\left( \theta_{P}\right) \mathbf{R}^{-1} \boldsymbol{\Delta} \right) \\
&=&\frac{1}{n}\operatorname{tr}\left( \mathbf{R}^{-1\prime}\mathbf{R}
\left(\theta_{P}\right)^{\prime}\boldsymbol{\Delta}^{-1}\left( \theta_{P}, \Sigma, \sigma^{2} \right) \mathbf{R}\left( \theta_{P}\right) \mathbf{R}^{-1}
\boldsymbol{\Delta} \right) + o\left( 1\right).
\end{eqnarray*}

The identification condition is derived by the following relation: for arbitrary $\varepsilon >0$,
\begin{equation*}
\underset{n,T \to \infty}{\lim \sup}\max_{\theta \in \mathcal{N}^{c}\left( \theta_{0}, \varepsilon \right)}
\left[ \mathbb{E}\left(\frac{1}{nT}\ell_{nmT}^{c}(\theta) \right) -  \mathbb{E}\left(\frac{1}{nT}\ell_{nmT}^{c}(\theta_{0}) \right) \right] < 0,
\end{equation*}
where $\mathcal{N}^{c}\left( \theta_{0}, \varepsilon \right)$ is the complement of an open neighborhood centered at $\theta_{0}$ with radius $\varepsilon > 0$. For $\mathbb{E}\left(\frac{1}{nT}\ell_{nmT}^{c}(\theta_{0}) \right)$, observe that 
\begin{equation*}
\frac{1}{2nT}\mathbb{E}\left[ \mathbf{v}_{nmT}^{\prime}\left( \mathbf{J}_{T}\otimes
\mathbf{J}_{n,m}\boldsymbol{\Delta}^{-1}\mathbf{J}_{n,m}\right) \mathbf{v}_{nmT}\right] =
\frac{m}{2} + o\left( 1\right).
\end{equation*}

Then, 
\begin{eqnarray*}
\mathbb{E}\left(\frac{1}{nT}\ell_{nmT}^{c}(\theta) \right) - \mathbb{E}\left(\frac{1}{nT}\ell_{nmT}^{c}(\theta_{0}) \right) 
&=& \frac{1}{n}\left( \ln \left\vert \mathbf{R} \left( \theta_{P}\right) \right\vert
- \ln \left\vert \mathbf{R} \right\vert \right) 
- \frac{1}{2n}\left( \ln \left\vert \boldsymbol{\Delta} \left( \theta_{P}, \Sigma, \sigma^{2} \right) \right\vert 
- \ln \left\vert \boldsymbol{\Delta} \right\vert \right)  \\
&& - \left( \frac{1}{2nT}\mathbb{E}\left[\mathbf{v}_{nmT}^{\prime}
\left( \theta_{P}\right) \left( \mathbf{J}_{T}\otimes \mathbf{J}_{n,m}\boldsymbol{\Delta}^{-1}
\left(\theta_{P}, \Sigma, \sigma^{2} \right) \mathbf{J}_{n,m}\right) \mathbf{v}_{nmT}\left( \theta_{P}\right) \right]
- \frac{m}{2}\right) + o\left( 1\right)  \notag \\
&=&\frac{1}{2}\left[ \frac{1}{n}\ln \left\vert \mathbf{M} \left( \theta_{P}, \Sigma, \sigma^{2} \right) \right\vert 
- \frac{1}{n}\left( \operatorname{tr}\left( \mathbf{M}\left(
\theta_{P}, \Sigma, \sigma^{2} \right) \right) - nm \right) \right]  \notag \\
&&-\frac{1}{2nT}\mathbb{E}\left[ \boldsymbol{\Xi}_{nmT}^{\ast}\left( \theta_{P}, \beta \right)^{\prime}\left( \mathbf{J}_{T}\otimes \mathbf{J}_{n,m}\boldsymbol{\Delta}^{-1}
\left( \theta_{P}, \Sigma, \sigma^{2} \right) \mathbf{J}_{n,m}\right) 
\boldsymbol{\Xi}_{nmT}^{\ast}\left(\theta_{P},\beta \right) \right] 
+ o\left( 1\right),  \notag
\end{eqnarray*}
where $\mathbf{M}\left( \theta_{P}, \Sigma, \sigma^{2} \right) = \boldsymbol{\Delta}^{\frac{1}{2}}\mathbf{R}^{-1\prime}\mathbf{R}\left( \theta_{P}\right)^{\prime}\boldsymbol{\Delta}^{-1}\left( \theta_{P}, \Sigma, \sigma^{2} \right) \mathbf{R} \left( \theta_{P}\right) \mathbf{R}^{-1}\boldsymbol{\Delta}^{\frac{1}{2}}$, which is positive definite because $\mathbf{R}\left(
\theta_{P}\right) \mathbf{R}^{-1}$ for $\theta_{P} \in \Theta_{P}$
is nonsingular.

Observe that 
\begin{equation*}
\frac{1}{n}\ln \left\vert \mathbf{M}\left( \theta_{P}, \Sigma, \sigma^{2} \right) \right\vert - \frac{1}{n}\left( \operatorname{tr}\left( \mathbf{M}\left( \theta_{P}, \Sigma, \sigma^{2} \right) \right)
-nm\right) = \frac{1}{n}\left[ \sum_{i}\left( \ln \varphi_{i}\left( \mathbf{M}\left(\theta_{P}, \Sigma, \sigma^{2} \right) \right) -
\left( \varphi_{i}\left( \mathbf{M}\left(\theta_{P}, \Sigma, \sigma^{2} \right) \right) - 1\right) \right) \right] \leq 0
\end{equation*}
by the arithmetic-geometric mean relationship and $\mathbb{E}\left[ \boldsymbol{\Xi}_{nmT}^{\ast}\left( \theta_{P}, \beta \right)
^{\prime}\left( \mathbf{J}_{T} \otimes \mathbf{J}_{n,m}\boldsymbol{\Delta}^{-1}\left( \theta_{P}, \Sigma, \sigma^{2} \right) \mathbf{J}_{n,m}\right) \boldsymbol{\Xi}_{nmT}^{\ast}\left( \theta_{P}, \beta \right) \right] \geq 0$ since $\mathbf{J}_{T} \otimes \mathbf{J}_{n,m}\boldsymbol{\Delta}^{-1}\left( \theta_{P}, \Sigma, \sigma^{2} \right) \mathbf{J}_{n,m}$ is nonnegative definite. Hence, we can obtain two identification conditions.
First, $\lim_{n\to \infty}\frac{1}{n}\sum_{i}\left[ \ln \varphi_{i}\left( \mathbf{M}\left(\theta_{P}, \Sigma, \sigma^{2} \right) \right) - \left( \varphi_{i}\left(\mathbf{M}\left(\theta_{P}, \Sigma, \sigma^{2} \right) \right) - 1\right) \right] < 0$ for $(\theta_{P}, \Sigma, \sigma^{2}) \neq (\theta_{P,0}, \Sigma_{0}, \sigma_{0}^{2})$. For
identification of $\theta_{P,0}$, second, we need to have $\mathbb{E}\left[\boldsymbol{\Xi}_{nmT}^{\ast}\left( \theta_{P}, \beta \right)^{\prime}\left(\mathbf{J}_{T} \otimes \mathbf{J}_{n,m}\boldsymbol{\Delta}^{-1}\left( \theta_{P}, \Sigma, \sigma^{2} \right)
\mathbf{J}_{n,m}\right) \boldsymbol{\Xi}_{nmT}^{\ast}\left( \theta_{P}, \beta \right) \right] > 0$ for $\theta_{P} \neq \theta_{P,0}$. To identify $\beta_{0}$, we additionally require $\plim_{n,T\to \infty}\frac{1}{nT} \mathbf{X}_{nmT}^{\tau \prime}\left(\mathbf{J}_{T}\otimes \mathbf{J}_{n,m}\boldsymbol{\Delta}^{-1} \mathbf{J}_{n,m}\right) \mathbf{X}_{nmT}^{\tau} > 0$.

When $\theta_{P,0}$ and $\beta_{0}$ are identified, it remains to consider identification of $\Sigma_{0}$ and $\sigma_{0}^{2}$.
Note that 
\begin{eqnarray*}
\mathbf{M}\left( \theta_{P,0}, \Sigma, \sigma^{2}\right) &=&
\boldsymbol{\Delta}^{-1}\left( \theta_{P,0}, \Sigma, \sigma^{2}\right) \boldsymbol{\Delta} \\
&=&
\begin{bmatrix}
\begin{array}{c}
\left( \Sigma^{-1} \otimes \mathbf{I}_{n}\right) + \frac{1}{\sigma^{2}}\bar{C}_{e}^{g\prime}
\bar{C}_{e}^{g}
\end{array}
& -\frac{1}{\sigma^{2}}\bar{C}_{e}^{g\prime} \\ 
-\frac{1}{\sigma^{2}}\bar{C}_{e}^{g} & \frac{1}{\sigma^{2}}\mathbf{I}_{n}
\end{bmatrix}
\begin{bmatrix}
\Sigma_{0}\otimes \mathbf{I}_{n} & \left( \Sigma_{0}\otimes \mathbf{I}_{n}\right) \bar{C}_{e}^{g\prime} \\ 
\bar{C}_{e}^{g}\left( \Sigma_{0} \otimes \mathbf{I}_{n}\right) & 
\begin{array}{c}
\bar{C}_{e}^{g}\left( \Sigma_{0} \otimes \mathbf{I}_{n}\right) \bar{C}_{e}^{g\prime} + \sigma_{0}^{2}\mathbf{I}_{n}
\end{array}
\end{bmatrix}
\\
&=&
\begin{bmatrix}
\left( \Sigma^{-1}\Sigma_{0} \otimes \mathbf{I}_{n}\right) & \left( \left(\Sigma^{-1}\Sigma_{0} \otimes \mathbf{I}_{n}\right) -\frac{\sigma_{0}^{2}}{\sigma^{2}} \mathbf{I}_{nm^{*}}\right) \bar{C}_{e}^{g\prime} \\ 
\mathbf{0} & \frac{\sigma_{0}^{2}}{\sigma^{2}}\mathbf{I}_{n}
\end{bmatrix},
\end{eqnarray*}
where $\bar{C}_{e}^{g} = \mathbf{R}_{0}C_{e}^{g}$, by the formula for an inverse of a block matrix. Since $\Sigma$ and $\Sigma_{0}$ are positive definite, it is not possible to have $\Sigma^{-1}\Sigma_{0} = \mathbf{I}_{m^{*}}$ if $\Sigma \neq \Sigma_{0}$.

\subsection*{Asymptotic distribution of the MLE}

The primary focus of this subsection is to investigate the asymptotic distribution of \( \frac{1}{\sqrt{nT}}\frac{\partial \ell_{nmT}^{c}(\theta_0)}{\partial \theta} \). Here is the form of $\widetilde{\mathbf{s}}_{nmT}^{\theta_{j}} = \mathbf{s}_{nmT}^{\theta_{j}} - \textbf{bias}_{1,nmT}^{\theta_{j}} - \textbf{bias}_{2,nmT}^{\theta_{j}}$:
\begin{equation}
    \begin{split}
        \mathbf{s}_{nmT}^{\theta_{j}}  = & 
        \frac{1}{\sqrt{nT}}\sum_{t=1}^{T}\left[ 
\mathbf{D}_{y}^{\theta_{j}} \operatorname{vec}\left( \mathbf{Y}_{t-1}\right)
+ \sum_{k=1}^{K}\mathbf{D}_{k}^{\theta_{j}}\mathbf{x}_{t,k} 
+ \mathbf{D}_{\tau,t}^{\theta_{j}}\right]^{\prime} \mathbf{J}_{n,m} \boldsymbol{\Delta}^{-1} \mathbf{J}_{n,m}\mathbf{v}_{t} \\
& +\frac{1}{\sqrt{nT}}\sum_{t=1}^{T}\left[ 
\begin{array}{c}
\mathbf{v}_{t}^{\prime} \left( \mathbf{D}_{q}^{\theta_{j}\prime
}\mathbf{J}_{n,m} + \mathbf{J}_{n,m}\mathbf{D}_{\Delta}^{\theta_{j}}\right) \boldsymbol{\Delta}^{-1}\mathbf{J}_{n,m}\mathbf{v}_{t} \\ 
- \operatorname{tr}\left( \left( \mathbf{D}_{q}^{\theta_{j}\prime }\mathbf{J}_{n,m} + \mathbf{J}_{n,m}
\mathbf{D}_{\Delta}^{\theta_{j}}\right) \boldsymbol{\Delta }^{-1}\mathbf{J}_{n,m}\boldsymbol{\Delta} \right)
\end{array}
\right], \label{m4-3-LQform}
    \end{split}
\end{equation}
\begin{equation*}
 \textbf{bias}_{1,nmT}^{\theta_{j}} =
 \frac{1}{\sqrt{nT}}
 \left[ 
\begin{array}{c}
\sum_{s=0}^{T-1}\operatorname{vec}\left(\mathbf{Y}_{s}\right)^{\prime }\mathbf{D}_{y}^{\theta_{j}\prime}\mathbf{J}_{n,m}
\boldsymbol{\Delta}^{-1}\mathbf{J}_{n,m}
\overline{\mathbf{v}}_{nmT} \\ 
+\sum_{k=1}^{K}\sum_{s=1}^{T}\mathbf{x}_{s,k}^{\prime}\mathbf{D}_{k}^{\theta_{j}\prime} \mathbf{J}_{n,m}
\boldsymbol{\Delta}^{-1} \mathbf{J}_{n,m} \overline{\mathbf{v}}_{nmT}
\end{array}
\right] \\
+ \sqrt{\frac{T}{n}}\overline{\mathbf{v}}_{nmT}^{\prime}
\left( \mathbf{D}_{q}^{\theta_{j}\prime}
\mathbf{J}_{n,m} + \mathbf{J}_{n,m} \mathbf{D}_{\Delta}^{\theta_{j}}\right) \boldsymbol{\Delta}^{-1}\mathbf{J}_{n,m}\overline{\mathbf{v}}_{nmT},
\end{equation*}
and
\begin{equation*}
\textbf{bias}_{2,nmT}^{\theta_{j}} = \sqrt{\frac{T}{n}}\left[ \operatorname{tr}\left( 
\mathbf{D}_{q}^{\theta_{j}} + \mathbf{D}_{\Delta}^{\theta_{j}}\right)
- \operatorname{tr}\left( \left( \mathbf{D}_{q}^{\theta_{j}\prime} \mathbf{J}_{n,m} + \mathbf{J}_{n,m}\mathbf{D}_{\Delta}^{\theta_{j}}\right) \boldsymbol{\Delta}^{-1} \mathbf{J}_{n,m}
\boldsymbol{\Delta} \right) \right].
\end{equation*}
Here, $\mathbf{D}_{y}^{\theta_{j}}$, $\left\{ \mathbf{D}_{k}^{\theta_{j}}\right\}_{k=1}^{K}$, $\mathbf{D}_{q}^{\theta_{j}}$, and $\mathbf{D}_{\Delta}^{\theta_{j}}$ are coefficient matrices, $\mathbf{D}_{\tau,t}^{\theta_{j}}$ denotes a strictly exogenous component generated by $\mathbf{X}_{t}^\tau$, and $\overline{\mathbf{v}}_{nmT} = \frac{1}{T}\sum_{t=1}^{T}\mathbf{v}_{t}$. The exact forms of $\mathbf{D}_{y}^{\theta_{j}}$, $\left\{ \mathbf{D}_{k}^{\theta_{j}}\right\}_{k=1}^{K}$, $\mathbf{D}_{q}^{\theta_{j}}$, and $\mathbf{D}_{\Delta}^{\theta_{j}}$ are relegated to the next subsection.

The detailed derivation of the asymptotic distribution of $\mathbf{s}_{nmT}^{\theta_{j}}$ can be found in Section 3.5 of the supplement file. The below shows the forms of $\mathbf{a}_{nm,1}^{\theta_{j}}$ and $\mathbf{a}_{nm,2}^{\theta_{j}}$:
\begin{equation*}
\begin{split}
    \mathbf{a}_{nm,1}^{\theta_{j}} 
= &\frac{1}{n}\sum_{s=t+1}^{\infty}\mathbb{E}
\left[ \left(\mathbf{D}_{y}^{\theta_{j}}\operatorname{vec}\left(\mathbf{Y}_{s-1}\right) + \sum_{k=1}^{K}\mathbf{D}_{k}^{\theta_{j}}\mathbf{x}_{s,k}\right)^{\prime} \mathbf{J}_{n,m} 
\boldsymbol{\Delta}^{-1} 
\mathbf{J}_{n,m}
\mathbf{v}_{t}\right] \\
& + \frac{1}{n}\operatorname{tr}\left( \left( \mathbf{D}_{q}^{\theta_{j}\prime} \mathbf{J}_{n,m} + \mathbf{J}_{n,m} \mathbf{D}_{\Delta}^{\theta_{j}}\right) \boldsymbol{\Delta}^{-1}
\mathbf{J}_{n,m}\boldsymbol{\Delta} \right),\text{ and}\\
    \mathbf{a}_{nm,2}^{\theta_{j}} = & \operatorname{tr}\left( \mathbf{D}_{q}^{\theta_{j}}
+ \mathbf{D}_{\Delta}^{\theta_{j}}\right) - \operatorname{tr}\left( \left(\mathbf{D}_{q,nm}^{\theta_{j}\prime} \mathbf{J}_{n,m} + \mathbf{J}_{n,m}\mathbf{D}_{\Delta}^{\theta_{j}}\right) 
\boldsymbol{\Delta}^{-1}\mathbf{J}_{n,m}\boldsymbol{\Delta} \right).
\end{split}
\end{equation*}

\subsubsection*{For the asymptotic bias formula}

Note that 
\begin{equation*}
\theta  = \left( \operatorname{vec}\left(\Lambda\right)^\prime,
\operatorname{vec}\left(\boldsymbol{\rho} \right)^\prime,
\operatorname{vec}\left(\mathring{P}\right)^\prime,
\operatorname{vec}\left(\mathring{\Psi}\right)^\prime,
\phi^\prime,
\operatorname{vec}\left(\Pi \right)^\prime, \beta^\prime,
\operatorname{vec}\left( \mathring{\Sigma} \right)^\prime , \sigma ^{2}\right)^\prime,
\end{equation*}
where $\operatorname{vec}\left( \mathring{P} \right) $, $\operatorname{vec}\left( \mathring{\Psi} \right)$, and $\operatorname{vec}\left(\mathring{\Sigma} \right)$ denote
respectively distinct elements of $P$, $\Psi$, and $\Sigma$. The total number of free parameters, i.e., the dimension of $\theta$, is $2m^{*2} + 3m^{*} + \frac{3m^{*}(m^{*}-1)}{2} + m^{*}K + Q + 1$. In this subsection, we define $A_{\theta_{j}}(\theta) = \frac{\partial A(\theta)}{\partial \theta_{j}}$, where $A(\theta)$ is a matrix which is a function of $\theta \in \Theta$ and $A_{\theta_{j}} = A_{\theta_{j}}(\theta_{0})$.

For the components $\theta_{j}$ with $j=1,\cdots,2m^{*2}+2m^{*}+m^{*}(m^{*}-1) + m^{*}K$,
i.e., the parameters before $\beta$, 
\begin{eqnarray*}
\frac{\partial \ell_{nmT}^{c}\left( \theta_{0}\right)}{\partial \theta_{j}}
&=&\left[ \left( \mathbf{I}_{T} \otimes \mathbf{D}_{y}^{\theta_{j}}\right) \mathbf{y}_{nmT,-1} + \sum_{k=1}^{K}\left( \mathbf{I}_{T}\otimes \mathbf{D}_{k}^{\theta_{j}}\right) \mathbf{x}_{nmT,k}
+ \left( \mathbf{I}_{T}\otimes - \mathbf{R}_{\theta_{j}}
\mathbf{R}^{-1}\right) \left( \sum_{q=1}^{Q}\beta_{q,0}\mathbf{x}_{nmT,q}^{\tau} + \widetilde{\boldsymbol{\alpha}}_{nmT,0}\right) \right]^{\prime} \\
&&\times \left( \mathbf{J}_{T}\otimes \mathbf{J}_{n,m}
\boldsymbol{\Delta}^{-1} \mathbf{J}_{n,m}\right) \mathbf{v}_{nmT} \\
&& + \left[ 
\begin{array}{c}
\mathbf{v}_{nmT}^{\prime}\left( \mathbf{J}_{T}\otimes \left( -\mathbf{R}^{-1\prime} \mathbf{R}_{\theta_{j}}^{\prime} \mathbf{J}_{n,m} + \frac{1}{2}\mathbf{J}_{n,m}
\boldsymbol{\Delta}^{-1} 
\boldsymbol{\Delta}_{\theta_{j}}\right)\boldsymbol{\Delta}^{-1} \mathbf{J}_{n,m}\right) \mathbf{v}_{nmT} \\ 
- \operatorname{tr}\left( \mathbf{I}_{T}\otimes \left( -\mathbf{R}_{\theta_{j}}
\mathbf{R}^{-1} + \frac{1}{2}\boldsymbol{\Delta}^{-1}
\boldsymbol{\Delta}_{\theta_{j}}\right) \right)
\end{array}
\right],
\end{eqnarray*}
where $\mathbf{D}_{y}^{\theta_{j}} = - \left( \mathbf{R}_{\theta_{j}}
\mathbf{R}^{-1}\mathbf{G} - \mathbf{G}_{\theta_{j}}\right)$,
and $\mathbf{D}_{k}^{\theta_{j}}
= - \mathbf{R}_{\theta_{j}}\mathbf{R}^{-1}
\boldsymbol{\Gamma}_{k}\boldsymbol{\Pi}_{k,0}
+ \left[ \boldsymbol{\Gamma}_{k} \boldsymbol{\Pi}_{k}\right]_{\theta_{j}}$ for $k=1,\cdots,K$.\footnote{Recall that $\mathbf{R}_{\theta_{j}}$, $\boldsymbol{\Delta}_{\theta
_{j}}$, $\mathbf{G}_{\theta_{j}}$, and $\left[\boldsymbol{\Gamma }_{k}\boldsymbol{\Pi}_{k}\right]_{\theta_{j}}$ are respectively the derivatives of $\mathbf{R}\left( \theta_{P}\right)$, $\boldsymbol{\Delta}\left( \theta \right)$, $\mathbf{G}\left(\theta_{P}\right)$, and $\boldsymbol{\Gamma}_{k}\left( \theta_{P}\right)\boldsymbol{\Pi}_{k}$ with respect to $\theta_{j}$ evaluated at $\theta_{0}$.} For $\beta$, $\frac{\partial
\ell_{nmT}^{c}\left( \theta_{0}\right) }{\partial \beta } =
\mathbf{X}_{nmT}^{\tau\prime}\left( \mathbf{J}_{T}\otimes \mathbf{J}_{n,m}\boldsymbol{\Delta}^{-1} \mathbf{J}_{n,m}\right) \mathbf{v}_{nmT}$.

For the variance parameters, define $\boldsymbol{\sigma}_{\Sigma} = \left( \operatorname{vec}\left( \mathring{\Sigma} \right)^{\prime}, \sigma^{2}\right)^{\prime}$. Then, 
\begin{equation*}
\frac{\partial \ell_{nmT}^{c}\left( \theta_{0}\right)}{\partial \left[ \boldsymbol{\sigma}_{\Sigma}\right]_{l}}
= -\frac{T}{2}\operatorname{tr}\left( \boldsymbol{\Delta}^{-1}
\boldsymbol{\Delta}_{\left[ \boldsymbol{\sigma}_{\Sigma}\right]_{l}}\right) + \frac{1}{2}\mathbf{v}_{nmT}^{\prime} \left( \mathbf{J}_{T}\otimes \mathbf{J}_{n,m} \boldsymbol{\Delta}^{-1}\boldsymbol{\Delta}_{\left[ \boldsymbol{\sigma}_{\Sigma}\right]_{l}} \boldsymbol{\Delta}^{-1} \mathbf{J}_{n,m}\right) \mathbf{v}_{nmT},
\end{equation*}
where $\boldsymbol{\Delta}_{\left[ \boldsymbol{\sigma }_{\Sigma}\right]_{l}}$ denotes the derivative of $\boldsymbol{\Delta}\left( \theta \right)$
with respect to the $l$th-element of $\boldsymbol{\sigma}_{\Sigma}$.

\clearpage
\bibliographystyle{apalike}
\bibliography{main}

\end{document}